\documentclass[prb,twocolumn,notitlepage,superscriptaddress]{revtex4-2}
\usepackage{amsmath,amsfonts,amssymb,amsthm,epsfig,array}
\usepackage{dsfont}
\usepackage{slashed}
\usepackage{graphics}
\usepackage{float}
\usepackage{verbatim}
\usepackage{color}
\usepackage[dvipsnames]{xcolor}
\usepackage[hidelinks,colorlinks,linkcolor=blue,
citecolor=blue,urlcolor=blue]{hyperref}
\usepackage[titletoc,title]{appendix}
\usepackage{multirow}
\usepackage{bibentry}
\usepackage{tabularx}
\usepackage[mathscr]{euscript}
\usepackage{mathtools}
\usepackage{hhline}
\usepackage[normalem]{ulem}

\newcommand{\bsl}[1]{\boldsymbol{#1}}

\renewcommand{\mod}{\,\mathrm{mod}\,}

\newcommand{\bra}[1]{\langle #1|}
\newcommand{\ket}[1]{|#1 \rangle}

\newcommand{\ii}{\mathrm{i}}

\newcommand{\dsZ}{\mathbb{Z}}

\newcommand{\dsR}{\mathbb{R}}

\newcommand{\dsC}{\mathbb{C}}
\newcommand{\Tr}{\mathop{\mathrm{Tr}}}

\renewcommand{\O}{\mathop{\mathrm{O}}}

\newcommand{\U}{\mathrm{U}}

\newcommand{\eqnref}[1]{Eq.\,\eqref{#1}}
\newcommand{\figref}[1]{Fig.\,\ref{#1}}
\newcommand{\tabref}[1]{Tab.\,\ref{#1}}
\newcommand{\secref}[1]{Sec.\,\ref{#1}}
\newcommand{\appref}[1]{Appendix.\,\ref{#1}}
\newcommand{\refcite}[1]{Ref.\,\cite{#1}}

\newcommand{\mat}[1]{\left(\begin{matrix}#1\end{matrix}\right)}

\newcommand{\eq}[1]{\begin{equation} #1 \end{equation}}
\newcommand{\eql}[1]{\begin{widetext}\begin{align}\begin{split} #1 \end{split}\end{align}\end{widetext}}
\newcommand{\eqa}[1]{\begin{align}\begin{split} #1 \end{split}\end{align}}
\newcommand{\ave}[1]{\left\langle #1 \right\rangle}

\usepackage{environ}
\NewEnviron{eqs}{%
\begin{equation}\begin{split}
    \BODY
\end{split}\end{equation}
}

\let\oldAA\AA
\renewcommand{\AA}{\text{\normalfont\oldAA}}
\newcommand{\sgn}[1]{\text{sgn}(#1)}
\newcommand{\ie}{{\emph{i.e.}}}
\newcommand{\eg}{{\emph{e.g.}}}
\newcommand{\TR}{\mathcal{T}}

\newcommand{\A}{\mathcal{A}}

\newcommand{\C}{\mathcal{C}}

\newcommand{\E}{\mathcal{E}}

\renewcommand{\P}{\mathcal{P}}

\newcommand{\Q}{\mathcal{Q}}
\newcommand{\meV}{\text{meV}}

\newcommand{\ch}{\text{Ch}}

\newcommand{\K}{\mathrm{K}}
\newcommand{\KM}{\mathrm{K}_{\mathrm{M}}}
\newcommand{\GM}{\Gamma_{\mathrm{M}}}
\newcommand{\MM}{\mathrm{M}_{\mathrm{M}}}

\newcommand{\MBZ}{\text{MBZ}}
\newcommand{\Chern}{\text{Chern}}

\newcommand{\VH}{\text{VH}}

\newtheorem{proposition}{Proposition}
\newcommand{\propref}[1]{Prop.\,\ref{#1}}

\begin{document}
\title{Magic-angle twisted symmetric trilayer graphene as a topological heavy-fermion problem}
\author{Jiabin Yu}
\affiliation{Condensed Matter Theory Center and Joint Quantum Institute, Department of Physics, University of Maryland, College Park, MD 20742, USA}
\affiliation{Department of Physics,
Princeton University,
Princeton, NJ 08544, USA}
\author{Ming Xie}
\affiliation{Condensed Matter Theory Center and Joint Quantum Institute, Department of Physics, University of Maryland, College Park, MD 20742, USA}
\author{B. Andrei Bernevig}
\affiliation{Department of Physics,
Princeton University,
Princeton, NJ 08544, USA}
\affiliation{Donostia International Physics Center, P. Manuel de Lardizabal 4, 20018 Donostia-San Sebastian, Spain}
\affiliation{IKERBASQUE, Basque Foundation for Science, Bilbao, Spain}
\author{Sankar Das Sarma}
\affiliation{Condensed Matter Theory Center and Joint Quantum Institute, Department of Physics, University of Maryland, College Park, MD 20742, USA}

\begin{abstract}
Recently, [\emph{Song and Bernevig, Phys. Rev. Lett. 129, 047601 (2022)}] reformulated magic-angle twisted bilayer graphene as a topological heavy fermion problem, and used this reformulation to provide a deeper understanding for the correlated phases at integer fillings. 
In this work, we generalize this heavy-fermion paradigm to magic-angle twisted symmetric trilayer graphene, and propose a low-energy $f-c-d$ model that reformulates magic-angle twisted symmetric trilayer graphene as heavy localized $f$ modes coupled to itinerant topological semimetalic $c$ modes and itinerant Dirac $d$ modes.
Our $f-c-d$ model well reproduces the single-particle band structure of magic-angle twisted symmetric trilayer graphene at low energies for displacement field $\E\in[0,300]$meV.
By performing Hartree-Fock calculations with the $f-c-d$ model for $\nu=0,-1,-2$ electrons per Moir\'e unit cell, we reproduce all the correlated ground states obtained from the previous numerical Hartree-Fock calculations with the Bistritzer-MacDonald-type model, and we find additional new correlated ground states at high displacement field.
Based on the numerical results, we propose a simple rule for the ground states at high displacement fields by using the $f-c-d$ model, and provide analytical derivation for the rule at charge neutrality.
We also provide analytical symmetry arguments for the (nearly-)degenerate energies of the high-$\E$ ground states at all the integer fillings of interest, and make experimental predictions of which charge-neutral states are stabilized in magnetic fields.
Our $f-c-d$ model provides a new perspective for understanding the correlated phenomena in magic-angle twisted symmetric trilayer graphene, suggesting that the heavy fermion paradigm of [\emph{Song and Bernevig, Phys. Rev. Lett. 129, 047601 (2022)}] should be the generic underpinning of correlated physics in multilayer moire graphene structures.
\end{abstract}

\maketitle

\section{Introduction}
Magic-angle twisted bilayer graphene (MATBG)~\cite{Bistritzer2011BMModel} hosts superconductivity~\cite{Cao2018TBGSC,Yankowitz2019SCMATBG,Lu2019SCMATBG,Stepanov2019SCCorrTBG,Saito2020SCCorrTBG,arora_2020,Cao2021NematicSCMATBG,deVries2021JosephsonMATBG,Oh2021NodalSCMATBG,Dibattista2021NodalSCMATBG,Tian2021ExpTBGSW} and various other interaction-induced phenomena~\cite{Cao2018TBGMott,sharpe_emergent_2019,liu2021tuning,serlin_QAH_2019,xie2019spectroscopic,choi_imaging_2019,kerelsky_2019_stm, jiang_charge_2019, polshyn_linear_2019, cao_strange_2020, wong_cascade_2020, zondiner_cascade_2020,  nuckolls_chern_2020, choi2020tracing, saito2020,das2020symmetry, wu_chern_2020,park2020flavour, saito2020isospin,rozen2020entropic, lu2020fingerprints,Efetov11222022MATBGReentrant}.
In the last several years, models have been constructed (in the real space~\cite{po_origin_2018,kang_symmetry_2018,kang_strong_2019,koshino_maximally_2018,po_faithful_2019,vafek2021lattice,zou2018,xux2018,yuan2018model}, in the momentum space~\cite{Biao-TBG4,TBG5,TBG6,zhang2021momentum,bultinck_ground_2020,cea_band_2020,zhang_HF_2020,hofmann2021fermionic}, or phenomenologically~\cite{Efimkin2018TBG,wux2018b,xu2018topological,thomson2018triangular,classen2019competing,eugenio2020dmrg,repellin_EDDMRG_2020,fernandes_nematic_2020}) to understand the physics observed in  MATBG, among other research efforts~\cite{tarnopolsky_origin_2019, liu2019pseudo,song_all_2019,hejazi_multiple_2019,padhi2018doped,lian2020, hejazi_landau_2019, padhi2020transport,ochi_possible_2018,guinea2018, venderbos2018, you2019,  wu_collective_2020, Lian2019TBG,Wu2018TBG-BCS, isobe2018unconventional,liu2018chiral, bultinck2020, zhang2019nearly, liu2019quantum, dodaro2018phases, gonzalez2019kohn, seo_ferro_2019, hejazi2020hybrid, khalaf_charged_2020, xie_superfluid_2020,julku_superfluid_2020, hu2019_superfluid, kang_nonabelian_2020, soejima2020efficient, pixley2019, knig2020spin, christos2020superconductivity,lewandowski2020pairing,Kwan2020Twisted, Parameswaran2021Exciton, xie_HF_2020,liu2020theories, liu2020nematic, daliao_VBO_2019,daliao2020correlation, kennes2018strong, huang2020deconstructing, huang2019antiferromagnetically,guo2018pairing, ledwith2020, abouelkomsan2020,repellin_FCI_2020, vafek2020hidden, Wilson2020TBG, wang2020chiral, TBG1,Song-TBG2,Biao-TBG3, cha2021strange,Chichinadze2020Nematic,sheffer2021chiral,kang2021cascades,Chou05022021SCMATBG, Calder2020Interactions,Thomson2021Recovery,Yu2022EOCPTBG,DGG20220919TBGStrainMag,Xie20220929TBGPhaseDia,He20220929TBGReconstruct}. 
Recently a physically-relevant and symmetry-preserving model that separates the correct energy scales and is convenient
for studying the correlated phenomena was  proposed in \refcite{Song20211110MATBGHF}. It is called the topological heavy-fermion model. 
At the single-particle level, the model proposed in \refcite{Song20211110MATBGHF} consists of localized heavy $f$ modes (of $p_x\pm i p_y$ symmetry) and itinerant $c$ modes, where the nearly flat bands in MATBG are given by coupling $f$ and $c$ modes (mainly around $\Gamma_{\text{M}}$).
The model is topological because the $c$ modes are anomalous in one valley (when the normal-state particle-hole symmetry is imposed exactly) and have a double-vortex dispersion akin to that in one of the valleys of \emph{untwisted} bilayer graphene, but at the $\GM$ point. 
Using the topological heavy-fermion model, \refcite{Song20211110MATBGHF} finds that the filling of the system is governed by the heavy fermions, which in a  Hartree-Fock calculation polarize. The Hartree-Fock calculation can be done efficiently for the correlated states at integer fillings, and a simple rule for the stability of the correlated ground-states can be derived analytically for those correlated states. Furthermore, the hope is that, using the differentiation of degrees of freedom in local and itinerant, progress can be made in the hard physics at non-integer filling, as well as at nonzero temperature. 
Recently, the heavy-fermion picture has been used to construct Kondo lattice model in MATBG~\cite{Chou20221128MATBGKondo,Hu2023KondoMATBG,Hu2023KondoMATBGStrain,Zhou2023KondoMATBG}, and has been generalized to twisted $(M + N)$-layer graphenes~\cite{Dai20220920HFTBGTMG} and to a variant of the kagome lattice~\cite{Si20220921HeavyFermion}.

Motivated by \refcite{Song20211110MATBGHF}, 
in this work, we generalize the topological heavy-fermion picture to magic-angle twisted symmetric trilayer graphene (MATSTG)
~\cite{Vishwanath20190129MATMG,MacDonald20190729TTGSP,Luskin20191113TTGSPRelax,Sachdev20191202TSTGSP,Kaxiras20200131TMGSP1,Kaxiras20200131TMGSP2,Lado20200505TSTGSPandSC,Luskin20200531TTGSP,MacDonald20201012TSTGMirrorBreaking,Wu20201226TSTGSP,BAB20210211TSTGI,Stauber20210316TSTGInt,Choi20210330TBGTMGEPC,Jung20210404TTGSP,Klebl20210420TSTGInt,Senthil20210428TSTGInt,MacDonald20210428TSTGSC,BAB20210628TSTGII,Mora20210628TSTGVHS,Guinea20210629TSTG,Sachdev20210603TSTGInt,Yuan20210805TSTGPressurePlasmon,Huang20211027TSTGSC,Liu20211111ATMG,Vishwanath20211122AMTG,Mathias20211216TSTGSC,Konig20211229TSTGInt,Ni20220401TTGRelax,Samajdar20220513TSTGPhonons,Jung20220619TMG,Liu20220906TSTGSPExp,Shin20221230ATMG,Zhang20221102MATTGExpEleAni,Vishwanath2021MATBGReview,Tian20211220TBGTMGReview}, which has also been experimentally confirmed to host correlated insulating states and superconductivity~\cite{Park20201026TSTGSCEXP,Kim20201204TSTGSCExp,Cao20210304TSTGSCExp,Dean20210627TSTGExp,NadjPerge20210920TSTGSCExp,Liu20210928TSTGSCExp,Kim20211026Asseumbly2DExp,NadjPerge20211217TMGSCExp,Efetov20220414TSTGExpAndTheory}.
Specifically, we first follow \refcite{Song20211110MATBGHF} to construct the heavy $f$ and itinerant $c$ modes, and then generalize the framework to include the nonzero displacement field $\E$, which couples $f$ and $c$ electrons to the relativistic Dirac ($d$) modes. 
The resultant single-particle $f-c-d$ model can reproduce almost identically the band structure of the Bistritzer-MacDonald-type (BM-type) model~\cite{BAB20210211TSTGI} in the energy window $[-50 \meV, 50 \meV]$ and for displacement field $\E\in[0,300]$meV.
We find that the $f$ modes dominate the low-energy single-particle physics for $\E\in[0,300]$meV.

The interaction in the $f-c-d$ model is obtained by projecting the Coulomb interaction to the $f-c-d$ basis.
Using this model, we perform the self-consistent Hartree-Fock calculation for the correlated states at fillings $\nu=0,-1,-2$ per Moir\'e unit cell.
The numerical results of our Hartree-Fock calculation are generally consistent with the previous numerical results in \refcite{Klebl20210420TSTGInt,Sachdev20210603TSTGInt,BAB20210628TSTGII,Guinea20210629TSTG,Vishwanath20211122AMTG,Efetov20220414TSTGExpAndTheory}, where a phase transition to states with zero intervalley coherence at all $\nu=0,-1,-2$ fillings exists when increasing the displacement field.
Nevertheless, we find \emph{more} additional correlated ground states than found in the previous literature at high displacement fields.
We further perform analytical one-shot Hartree-Fock analysis at the considered integer fillings.
At $\nu=0$, we provide analytical understanding of the loss of intervalley coherence for high displacement field, and derive a simple rule for the ground states at high fields.
The same rule is also derived for $\nu=-1,-2$ under an unrealistic approximation, but the rule turns out to be consistent with the self-consistent calculation for $\nu=-1,-2$.
We also find a symmetry reason for the similar energies of the ground states at high displacement fields at all $\nu=0,-1,-2$.
Finally, we discuss the experimental implication of our results.

The rest of the paper is organized as follows.
In \secref{sec:BM_TSTG}, we review the BM-type model for MATSTG.
In \secref{sec:fcd}, we build the heavy fermion $f-c-d$ model for MATSTG.
In \secref{sec:NHF_cal}, we perform numerical Hartree-Fock calculations with the $f-c-d$ model for $\nu=0,-1,-2$.
In \secref{sec:ana}, we perform analytical one-shot Hartree-Fock analysis for the correlated states with $\nu=0,-1,-2$.
In \secref{sec:conclusion}, we conclude the paper and discuss the experimental predictions.
A series of appendices provide all the technical details of our theory.

\section{Review: Interacting BM-type Model for MATSTG}
\label{sec:BM_TSTG}

In this section, we review the interacting BM-type model of MATSTG, which has been theoretically studied in \refcite{Vishwanath20190129MATMG,MacDonald20190729TTGSP,Luskin20191113TTGSPRelax,Sachdev20191202TSTGSP,Kaxiras20200131TMGSP1,Kaxiras20200131TMGSP2,Lado20200505TSTGSPandSC,Luskin20200531TTGSP,MacDonald20201012TSTGMirrorBreaking,Wu20201226TSTGSP,BAB20210211TSTGI,Stauber20210316TSTGInt,Choi20210330TBGTMGEPC,Jung20210404TTGSP,Klebl20210420TSTGInt,Senthil20210428TSTGInt,MacDonald20210428TSTGSC,BAB20210628TSTGII,Mora20210628TSTGVHS,Guinea20210629TSTG,Sachdev20210603TSTGInt,Yuan20210805TSTGPressurePlasmon,Huang20211027TSTGSC,Liu20211111ATMG,Vishwanath20211122AMTG,Mathias20211216TSTGSC,Konig20211229TSTGInt,Ni20220401TTGRelax,Samajdar20220513TSTGPhonons,Jung20220619TMG}.
Here the interaction is the Coulomb interaction screened by a top gate and a bottom gate, where the sample is placed in the middle of the two gates.
%
%
We will only review the contents that are essential for our later discussions and are specific to our theory presented in this work; a more complete and detailed discussion can be found in \refcite{BAB20210211TSTGI}.

\subsection{Single-Particle BM-type Model}
\label{sec:SP_BM}

In this part, we review the BM-type model for MATSTG following \refcite{Vishwanath20190129MATMG,BAB20210211TSTGI}.

MATSTG is constructed from a AAA-stacking trilayer graphene by rotating the graphene layers alternatively, \ie, rotating the top ($l=3$) and bottom ($l=1$) layers by $-\theta/2$ and rotating middle ($l=2$) layer by $\theta/2$, where $\theta>0$ corresponds to the counterclockwise rotation and $l=1,2,3$ is the layer index.
We label the lattice constant and the Fermi velocity of the monolayer graphene as $a_{G} = 2.46\ \AA$ and $v_0 = 5944\ \text{meV}\cdot \AA$, respectively.
We refer to the unit system in which $\AA$ is the length unit and $\meV$ is the energy unit as the experimental unit system (EUS), since this unit system is convenient for the comparison to the experiments.
However, EUS is not the most convenient unit system for the theoretical study of MATSTG. 
The most convenient unit system is the following simplified unit system in which
\eq{
\label{eq:unit_system}
\hbar = 1\ ,\ \epsilon_0 = 1\ ,\ k_\theta = 1 \ ,\ v_0 = 1\ ,
}
where $k_\theta =  \frac{4\pi}{3 a_{G} }2 \sin(\frac{\theta}{2})$ and $\epsilon_0$ is the vacuum permittivity.
Throughout the entire work, we will use \eqnref{eq:unit_system} unless otherwise (e.g., EUS) is specified.

With the unit system specified by \eqnref{eq:unit_system}, the single-particle BM-type model for MATSTG reads
\eq{
\label{eq:H_0}
H_0 = H_{0,+} + H_{0,-}\ .
}
Here ``$+$'' and ``$-$'' label two graphene valleys, which are related by time-reversal (TR) symmetry as
\eq{
H_{0,-} = \TR H_{0,+} \TR^{-1}\ .
}
Specifically, $H_{0,+}$ reads
\eql{
\label{eq:H0+_r}
H_{0,+}=\int d^2r\ \mat{\psi^\dagger_{+,\bsl{r},1} & \psi^\dagger_{+,\bsl{r},2} & \psi^\dagger_{+,\bsl{r},3} }  
\mat{
-\ii \bsl{\sigma}\cdot \bsl{\nabla}-\frac{\E}{2} & T(\bsl{r}) & \\
T^\dagger(\bsl{r}) & -\ii \bsl{\sigma}\cdot \bsl{\nabla} &  T^\dagger(\bsl{r})\\
 & T(\bsl{r}) & -\ii \bsl{\sigma}\cdot \bsl{\nabla}+\frac{\E}{2} 
} \otimes s_0\ 
\mat{ \psi_{+,\bsl{r},1} \\ \psi_{+,\bsl{r},2} \\ \psi_{+,\bsl{r},3} }  \ ,
}
where $\psi^\dagger_{+,\bsl{r},l} = (\psi^\dagger_{+,\bsl{r},l,A,\uparrow},\psi^\dagger_{+,\bsl{r},l,A,\downarrow},\psi^\dagger_{+,\bsl{r},l,B,\uparrow},\psi^\dagger_{+,\bsl{r},l,B,\downarrow})$ is the vector of creation operators for the $+$ valley and the $l$th layer, $\bsl{\sigma}=(\sigma_x,\sigma_y)$, and $\sigma_{0,x,y,z}$ and $s_{0,x,y,z}$ label the Pauli matrices for the sublattice index $\sigma=A/B$ and the spin index $s=\uparrow/\downarrow$, respectively.
The expression of $H_{0,-}$ can be obtained via
\eq{
\TR \psi^\dagger_{+,\bsl{r},l} \TR^{-1} = \psi^\dagger_{-,\bsl{r},l}\sigma_0 \ii s_y. 
}

In \eqnref{eq:H0+_r}, we assume that the twist angle $\theta$ is small enough such that the kinetic terms of order $O(\theta)$ can be safely neglected.
Moreover, $T(\bsl{r}) = \sum_{j=1,2,3} T_j e^{\ii \bsl{r} \cdot \bsl{q}_j}$
stands for the interlayer hopping between neighbouring layers with 
\eqa{
\label{eq:q}
& \bsl{q}_1 =(0,1)^T \\
& \bsl{q}_2 =(-\frac{\sqrt{3}}{2},- \frac{1}{2})^T \\
& \bsl{q}_3 =(\frac{\sqrt{3}}{2}, - \frac{1}{2})^T\ ,
}
and 
\eq{
T_j = w_0 \sigma_0 + w_1 \left[\cos(\frac{2\pi}{3} (j-1)) \sigma_x + \sin (\frac{2\pi}{3}(j-1) )\sigma_y \right]\ .
}
$w_0$ and $w_1$ are the AA and AB interlayer tunnellings, respectively, with $w_0=88$meV and $w_1=110$meV in EUS.
The values of $w_{0,1}$ in the unit system specified by \eqnref{eq:unit_system} depend on $\theta$.

\begin{figure}[t]
    \centering
    \includegraphics[width=0.5\columnwidth]{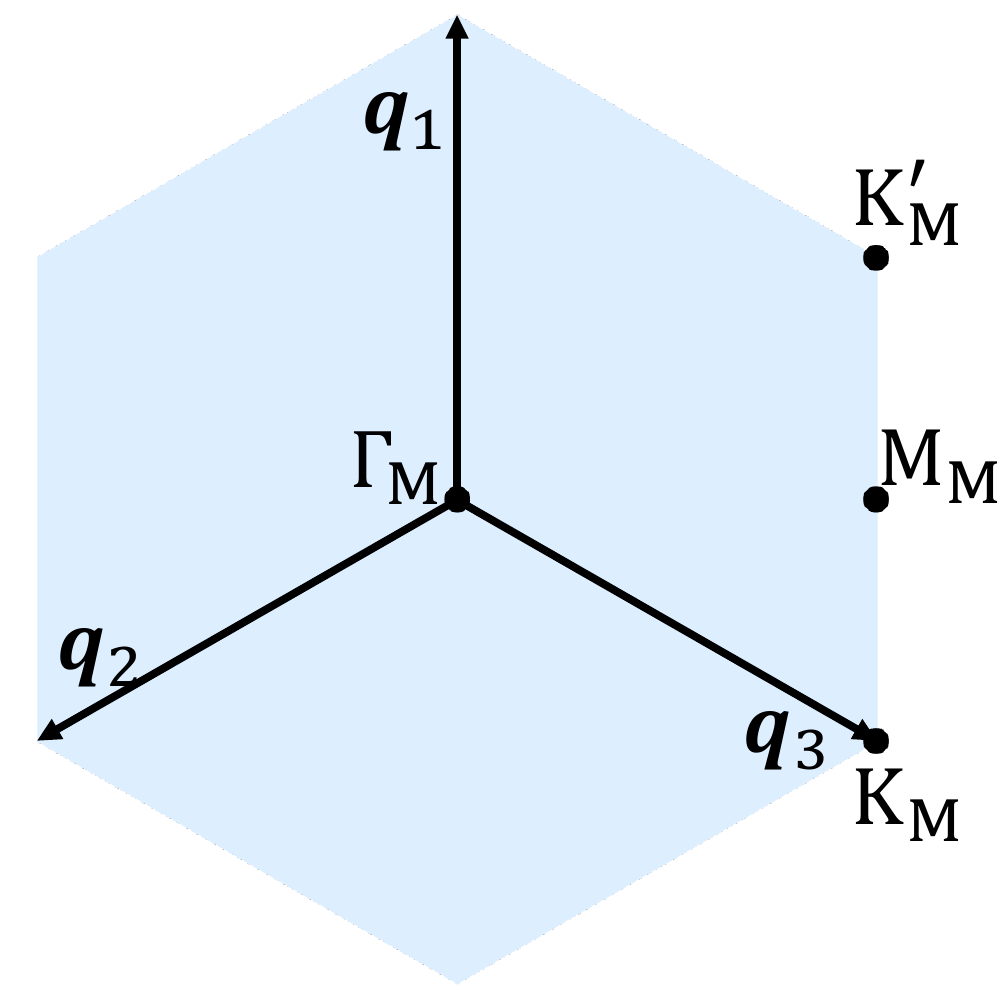}
    \caption{This figure shows MBZ, as well as the $\bsl{q}_{1,2,3}$ and various high-symmetry points.
    Note that $\K'_{\text{M}}$ is equivalent to $-\KM$.
    }
    \label{fig:MBZ}
\end{figure}

In \eqnref{eq:H0+_r}, $\E$ is the energy difference generated by the displacement field, \ie, the external electric field perpendicular to MATSTG.
When the displacement field is zero ($\E=0$), $H_0$ has a mirror symmetry $m_z$ with mirror plane lying in the middle layer, which is represented as
\eq{
m_z \psi^\dagger_{\eta,\bsl{r},l} m_z^{-1} = -\psi^\dagger_{\eta,\bsl{r},4-l}
}
with $\eta=\pm$ the graphene valley index.
Here the extra minus sign comes from the fact that $\psi^\dagger_{\eta,\bsl{r},l}$ are constructed from the $p_z$ orbital of graphene.
In fact, MATSTG is called symmetric owing to the presence of $m_z$ symmetry for $\E=0$.
$m_z$ allows us to recombine $\psi^\dagger_{\eta,\bsl{r},l}$ into a $m_z$-odd sector 
\eqa{
\label{eq:psi_tilde_r}
\left\{
\begin{array}{l}
 \widetilde{\psi}_{\eta,\bsl{r},t}^\dagger=\frac{1}{\sqrt{2}} (\psi^\dagger_{\eta,\bsl{r},3}+\psi^\dagger_{\eta,\bsl{r},1})\\
 \widetilde{\psi}_{\eta,\bsl{r},b}^\dagger= \psi^\dagger_{\eta,\bsl{r},2}
\end{array}\right.\ ,
}
and a $m_z$-even sector
\eq{
\label{eq:d_r}
d_{\eta,\bsl{r}}^\dagger = \frac{1}{\sqrt{2}} (\psi^\dagger_{\eta,\bsl{r},3}-\psi^\dagger_{\eta,\bsl{r},1})\ .
}
With the recombination, $H_{0,\eta}$ can be split into three parts
\eq{
H_{0,\eta} = H_{0,TBG,\eta} + H_{0,D,\eta} + H_{0,\E,\eta}\ ,
}
where
\eq{
\label{eq:H0TBG+_r}
H_{0,TBG,+}=\int d^2r\ \widetilde{\psi}^\dagger_{+,\bsl{r}}
\mat{
-\ii \bsl{\sigma}\cdot \bsl{\nabla} & \sqrt{2} T(\bsl{r}) & \\
\sqrt{2}  T^\dagger(\bsl{r}) & -\ii \bsl{\sigma}\cdot \bsl{\nabla} }\otimes s_0\ 
\widetilde{\psi}_{+,\bsl{r}}
}
is equivalent to the (valley $+$) BM model of the ordinary TBG with $w_0\rightarrow \sqrt{2} w_0$ and $w_1\rightarrow \sqrt{2} w_1$,
\eq{
\label{eq:H0D+_r}
H_{0,D,+}= \int d^2r\ d^\dagger_{+,\bsl{r}} (-\ii) \bsl{\sigma}\cdot \bsl{\nabla} d_{+,\bsl{r}}
}
is just a Dirac cone, the displacement field term
\eq{
\label{eq:H0E+_r}
 H_{0,\E,+} =\int d^2r\ \frac{\E}{2} \widetilde{\psi}_{+,\bsl{r},t}^\dagger d_{+,\bsl{r}} + h.c.
}
becomes the coupling between the TBG modes and the Dirac modes, and 
\eqa{
& H_{0,TBG,-}  = \TR H_{0,TBG,+}  \TR^{-1} \\
& H_{0,D,-} = \TR H_{0,D,+}  \TR^{-1} \\
& H_{0,\E,-} = \TR H_{0,\E,+}  \TR^{-1}\ .
}
It is clear that $H_{0,TBG,\eta}$ and $H_{0,D,\eta}$ commute with $m_z$, while $H_{0,\E,\eta}$ anticommutes with $m_z$.

$H_0$ has Moir\'e lattice translation symmetry, which is represented as
\eqa{
\label{eq:psi_tilder_r_translation}
& T_{\bsl{R}} \widetilde{\psi}_{\eta,\bsl{r},\widetilde{l}}^\dagger T_{\bsl{R}}^{-1} = \widetilde{\psi}_{\eta,\bsl{r}+\bsl{R},\widetilde{l}}^\dagger\ e^{-\eta\ii \bsl{K}_{\widetilde{l}}\cdot\bsl{R}} \\
& T_{\bsl{R}} d_{\eta,\bsl{r}}^\dagger T_{\bsl{R}}^{-1} = d_{\eta,\bsl{r}+\bsl{R}}^\dagger\ e^{-\eta\ii \bsl{K}_{t}\cdot\bsl{R}}\ ,
}
where $\widetilde{l}=t,b$ labels the ``layer" of the TBG part,  and $\bsl{R}$ is the Moir\'e lattice vector with two primitive Moir\'e vectors being $\bsl{a}_{\mathrm{M},1} = \frac{4\pi}{3}(0,-1)^T$ and $\bsl{a}_{\mathrm{M},2} = \frac{4\pi}{3}(\frac{\sqrt{3}}{2},\frac{1}{2})^T$.
In particular, $\bsl{K}_{t} $ and $\bsl{K}_{b}$ in \eqnref{eq:psi_tilder_r_translation} arise from the graphene valley as shown in \appref{app:sym_rep}, which read
\eq{
\label{eq:K_tb}
\bsl{K}_{t} = \frac{1}{2}(\cot(\theta/2),-1)^T\ ,\ \bsl{K}_{b} = \frac{1}{2}(\cot(\theta/2),1)^T \ .
}

To exploit the Moir\'e lattice translational symmetry of $H_0$, it is better to transform the Hamiltonian to the momentum space.
To do so, we first transform the basis to the momentum space as 
\eqa{
\label{eq:psi_tilde_d_p}
& \widetilde{\psi}_{\eta,\bsl{p},\widetilde{l}}^\dagger = \frac{1}{\sqrt{\A}}\int d^2 r e^{\ii \bsl{p}\cdot\bsl{r}} \widetilde{\psi}_{\eta,\bsl{r},\widetilde{l}}^\dagger \\
& d_{\eta,\bsl{p}}^\dagger = \frac{1}{\sqrt{\A}}\int d^2 r e^{\ii \bsl{p}\cdot\bsl{r}} d_{\eta,\bsl{r}}^\dagger\ ,
}
where $\bsl{p}\in\dsR^2$ and $\A$ is the area of MATSTG. 
Then, we define 
\eq{
\Q = \Q_+ \cup \Q_-\ ,\ 
\Q_\pm = \bsl{b}_{\mathrm{M},1} \dsZ +\bsl{b}_{\mathrm{M},2} \dsZ \pm \bsl{q}_1
}
with 
\eqa{ 
\label{eq:bM}
& \bsl{b}_{\mathrm{M},1}= \bsl{q}_3 - \bsl{q}_1 = (\frac{\sqrt{3}}{2}, -\frac{3}{2})^T\\
& \bsl{b}_{\mathrm{M},2}= \bsl{q}_3 - \bsl{q}_2 = ( \sqrt{3} , 0)^T
}forming the basis of the Moir\'e reciprocal lattice.
Finally, we define
\eqa{
\label{eq:BM_basis_k_Q}
& \widetilde{\psi}^\dagger_{\eta,\bsl{k},\bsl{Q}} = \widetilde{\psi}^\dagger_{\eta,\bsl{k}-\bsl{Q}, \widetilde{l}^{\eta}_{\bsl{Q}}}\text{ with $\bsl{Q}\in \Q$ } \\
& d^\dagger_{\eta,\bsl{k},\bsl{Q}} = d^\dagger_{\eta,\bsl{k}-\bsl{Q}} \text{ with $\bsl{Q}\in \Q_\eta$ }
}
with $\widetilde{l}^{\eta}_{\bsl{Q}}=t$ for $\bsl{Q}\in\Q_\eta$ and $\widetilde{l}^{\eta}_{\bsl{Q}}=b$ for $\bsl{Q}\in\Q_{-\eta}$.
With \eqnref{eq:BM_basis_k_Q}, the Hamiltonian becomes
\eqa{
\label{eq:H_0TBG_k}
H_{0,TBG,\eta} & = \sum_{\bsl{k}\in\MBZ} \sum_{\bsl{Q},\bsl{Q}'\in \Q} \widetilde{\psi}^\dagger_{\eta,\bsl{k},\bsl{Q}} \\
& \left[ h_{\eta,\bsl{Q}}^D(\bsl{k}) \delta_{\bsl{Q}\bsl{Q}'} + \sqrt{2} h_{\eta,\bsl{Q}\bsl{Q}'}^I \right] s_0 \widetilde{\psi}_{\eta,\bsl{k},\bsl{Q}'}\ ,
}
\eq{
\label{eq:H_0D_k}
 H_{0,D,\eta} = \sum_{\bsl{k}\in\MBZ} \sum_{\bsl{Q}\in \Q_\eta} d^\dagger_{\eta,\bsl{k},\bsl{Q}}  h_{\eta,\bsl{Q}}^D(\bsl{k})  s_0  d_{\eta,\bsl{k},\bsl{Q}}\ ,
}
and
\eq{
\label{eq:H_0E_k}
 H_{0,\E,\eta} = \sum_{\bsl{k}\in\MBZ} \sum_{\bsl{Q}\in \Q_\eta} \frac{\E}{2}  \widetilde{\psi}^\dagger_{\eta,\bsl{k},\bsl{Q}} d_{\eta,\bsl{k},\bsl{Q}} +h.c.\ .
}
Here $h_{+,\bsl{Q}}^D(\bsl{k})= (\bsl{k}-\bsl{Q})\cdot\bsl{\sigma}$, $h_{+,\bsl{Q}\bsl{Q}'}^I=\sum_{j} T_j (\delta_{\bsl{Q},\bsl{Q}'+\bsl{q}_j} + \delta_{\bsl{Q}',\bsl{Q}+\bsl{q}_j})$, $h_{-,\bsl{Q}}^D(\bsl{k})=[h_{+,-\bsl{Q}}^D(-\bsl{k})]^*$, $h_{-,\bsl{Q}\bsl{Q}'}^I=[h_{+,(-\bsl{Q})(-\bsl{Q}')}^I]^*$, and {\MBZ} is short for the Moir\'e Brillouin zone.
In this work, all numerical calculations with $H_0$ are done in the momentum space by using \eqnref{eq:H_0TBG_k}, \eqnref{eq:H_0D_k} and \eqnref{eq:H_0E_k}.
The numerical band structure of $H_{0}$ in the $+$ valley is shown in \figref{fig:bands_SP}(a-d) as red lines.
The definitions of various high-symmetry points in MBZ are illustrated in \figref{fig:MBZ}.

At the end of this part, we list the symmetries of $H_0$ for generic $\E$.
We have discussed TR and Moir\'e lattice translations, which are symmetries of $H_0$ for any values of $\E$.
Beside these two, $H_0$ has spin-charge $\U(2)$ symmetry in each valley, the spinless three-fold rotation symmetry $C_3$ along $z$, $C_2\TR$ symmetry (the combination of the spinless two-fold rotation $C_2$ along $z$ and the TR operation), an effective unitary anti-symmetry $C_{2x}P$, and the charge conjugate anti-symmetry $\C$.
Here anti-symmetry means that the symmmetry operation anti-commutes with the Hamiltonian, \ie, $ C_{2x}P H_0 (C_{2x}P)^{-1} = -H_0$ and $ \C H_0 \C^{-1} = -H_0$.
(See the symmetry representations in \appref{app:sym_rep}.)

\subsection{Coulumb Interaction}
In this part, we review the Coulomb interaction in the BM-type model for MATSTG following \refcite{BAB20210211TSTGI,BAB20210628TSTGII}. 

The Coulomb interaction in MATSTG is screened by the top and bottom gates, which are parallel to the MATSTG sample.
For simplicity, we assume that MATSTG lies in the middle of two gates, and then the Coulomb interaction between two electrons separated by $\bsl{r}$ has the following form 
\eq{
\label{eq:Coulomb_V}
V(\bsl{r}) = \frac{1}{\A} \sum_{\bsl{p}} e^{ -\ii \bsl{p}\cdot \bsl{r} } V(\bsl{p})\ ,
}
where  
\eq{
V(\bsl{p}) = \pi \xi^2 V_\xi \frac{ \tanh( \xi |\bsl{p}|/2) }{ \xi |\bsl{p}|/2 }\ ,
}
$\xi$ is the distance between two gates, and $V_\xi = \frac{e^2}{4\pi \epsilon \xi}$ with $e$ the elementary charge and $\epsilon$ the dielectric constant.
Throughout this work, we choose 
\eq{
\label{eq:parameter_values_int}
\xi = 100 \AA\text{ and } V_\xi = 24 \text{meV} 
}
in EUS for all numerical calculations, unless specified otherwise.
In \eqnref{eq:Coulomb_V}, we have included the screening due to the two gates.
It is clear that 
\eq{
\label{eq:Coulomb_V_sym}
V^*(\bsl{r}) = V(\bsl{r})\ \text{and}\ V(g\bsl{r})=V(\bsl{r}) \ \forall g\in \O(2)\ .
}

With the form of the Coulomb interaction (\eqnref{eq:Coulomb_V}), the Hamiltonian for the interaction reads 
\eq{
\label{eq:H_int_r}
H_{int} = \frac{1}{2} \int d^2 r d^2 r' V(\bsl{r}-\bsl{r}') :\rho(\bsl{r}): :\rho(\bsl{r}'):\ ,
}
where $\rho(\bsl{r})= \sum_{\eta,l} \psi^\dagger_{\eta,\bsl{r},l} \psi_{\eta,\bsl{r},l}$ is the electron number density operator.
The normal-ordering is defined as $:O:= O - \bra{G_0} O \ket{G_0}$ with $\ket{G_0}$ chosen such that
\eq{
\label{eq:G0_ket}
\bra{G_0} \psi^\dagger_{\eta,\bsl{r},l,\sigma,s} \psi_{\eta',\bsl{r}',l',\sigma',s'} \ket{G_0} = \frac{1}{2} \delta_{\eta\eta'} \delta(\bsl{r}-\bsl{r}') \delta_{ll'}\delta_{\sigma\sigma'} \delta_{ss'}\ .
}

The usage of the normal ordering is just to include a uniform positive charge background that makes half filling charge-neutral, as discussed in the following.
Based on the form of the interaction (\eqnref{eq:H_int_r}), $\left(-e:\rho(\bsl{r}):\right)$ should be the total charge density at $\bsl{r}$.
Since we know $-e:\rho(\bsl{r}): = -e \rho(\bsl{r}) + e \bra{G_0} \rho(\bsl{r}) \ket{G_0}$ and $-e \rho(\bsl{r})$ is the electron charge density, $e \bra{G_0} \rho(\bsl{r}) \ket{G_0}$ should be the background charge density.
Note that $e  \bra{G_0} \rho(\bsl{r}) \ket{G_0} = 12 e\  \delta(\bsl{r}=0) =\frac{e\sum_{\bsl{p}} 2\times 3 \times 2\times 2}{2 \A} $ is nothing but the charge density of a uniform positive charge background that corresponds to half filling, justifying the meaning of the normal ordering.

$H_{int}$ is invariant under TR, $C_3$, $C_2\TR$, $m_z$, Moir\'e lattice translations, $C_{2x}P$, $\C$, and $\U(2)\times\U(2)$.
(See more details in \appref{app:sym_rep}.)

\subsection{Interacting BM-type Hamiltonian and Filling}
\label{sec:BM_Total}

In this part, we review some general properties of the interacting BM-type model for MATSTG following \refcite{BAB20210211TSTGI,BAB20210628TSTGII}. 

The interacting BM-type Hamiltonian for MATSTG is the sum of the single-particle BM-type Hamiltonian and the Coulomb interaction as
\eq{
\label{eq:H_tot}
H=H_0+H_{int}\ .
}
The total Hamiltonian $H$ has $U(2)\times U(2)$, $\TR$, $C_3$, $C_2\TR$ and $T_{\bsl{R}}$ symmetries, as well as $m_z$ if combined with  the action $\E\rightarrow -\E$ on the electric field.
However, due to the opposite behaviors of $H_{int}$ and $H_0$ under $C_{2x}P$ and $\C$, $H$ does not preserve $C_{2x}P$ or $\C$, but it preserves the combination of them, \ie, $\C C_{2x}P$.
Therefore, the symmetry properties of the total Hamiltonian are
\eqa{
\label{eq:H_tot_sym}
& [\TR, H]= [C_3, H] = [C_2\TR, H]= [T_{\bsl{R}}, H] \\
&  = [\C C_{2x}P, H] = [\U(2)\times\U(2), H] = 0\\
& m_z H m_z^{-1} = \left.H\right|_{\E\rightarrow -\E}\ .
}

Based on the symmetry properties of $H$ (\eqnref{eq:H_tot_sym}), we know that we only need to study $\E\geq 0$ since the negative $\E$ are related by $m_z$.
Furthermore, we also only need to study the non-positive fillings, owing to $\C C_{2x}P$.
To see this, we first define the filling operator 
\eq{
\hat{\nu} =  \frac{1}{N} \int d^2 r :\rho(\bsl{r}):\ ,
}
where $N$ is the number of Moir\'e unit cells.
The eigenvalue $\nu$ of $\hat{\nu}$ is the filling, \ie, the averaged number of electrons per Moir\'e unit cell counted from the charge neutrality.
Owing to $[\hat{\nu}, H]=0$ derived from the charge-$\U(1)$ invariance of $H$, we can label the energy eigenstates with definite filling $\nu$.

As the filling operator anti-commutes with $\C C_{2x}P$ as $\{ \hat{\nu}, \C C_{2x}P \} = 0$ (\appref{app:sym_rep}), we only need to study $\nu\leq 0 $.
To be more specific, for any many-body energy eigenstate $\ket{\psi_{\nu,E}}$ of $H$ with filling $\nu$ and energy $E$, $\C C_{2x}P \ket{\psi_{\nu,E}}$ is an energy eigenstate with the same energy $E$ and opposite filling $-\nu$.
Then, if we obtain the set of all othornormal energy eigenstates $\{ \ket{i,\nu,E_i} \}$ with filling $\nu$, $\{ \C C_{2x}P \ket{i,\nu,E_i} \}$ is the set of all orthonormal energy eigenstates with opposite filling $-\nu$, and the two energy eiegnstates have the same energy if they are related by $C C_{2x}P$.
Therefore, we only need to diagonalize $H$ for $\nu\leq 0$, and we will adopt this simplification in later calculations.

\begin{figure*}[t]
    \centering
    \includegraphics[width=2\columnwidth]{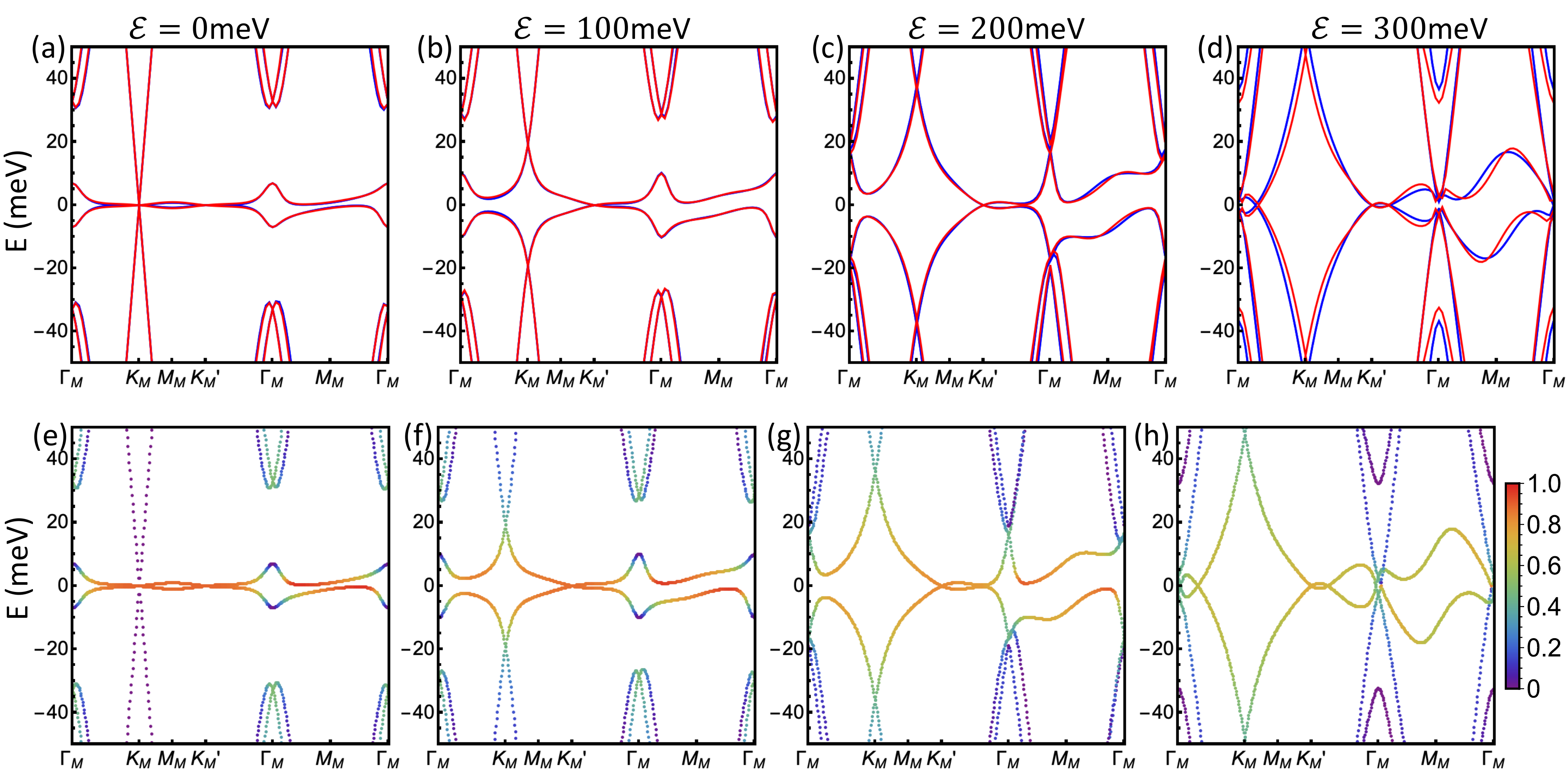}
    \caption{The single-particle band structures of MATSTG in the $+$ valley for the parameter values in \eqnref{eq:theta_value} and \tabref{tab:f-c-d_values}.
    In this figure, we use EUS.
    The momentum cutoffs of the BM-type model and the $f-c-d$ model are $2\sqrt{7}$ and $2\sqrt{3}$, respectively. (See \appref{app:fcd_SP} for details on the $f-c-d$ model.)
    In (a-d), we plot the band structure of the single-particle BM-type model \eqnref{eq:H_0} in red, and the band structure of the single-particle $f-c-d$ model \eqnref{eq:H_fcd_SP} in blue.
    $\E$ is the energy difference between the top and bottom layer generated by the displacement field.
    In (e-h), we replot the band structure of the single-particle BM-type model \eqnref{eq:H_0} in (a-d), respectively.
    The colors of the points show the (square of the absolute values of) overlaps between the Bloch states and the trial Wannier functions according to the color bar on the right of (h).
    }
    \label{fig:bands_SP}
\end{figure*}

\section{$f-c-d$ Model}
\label{sec:fcd}

In this section, we construct the heavy fermion $f-c-d$ model for MATSTG.
We start with the single-particle $f-c-d$ model, and then project the Coulomb interaction to the heavy fermion basis to obtain the interacting Hamiltonian.

\subsection{Single-Particle $f-c-d$ Model}
We start with the single-particle $f-c-d$ model.

\subsubsection{Review: $f$ and $c$ Modes}
\label{sec:f_c}

We first discuss the construction of the $f$ and $c$ modes in the TBG part of the Hamiltonian.
As shown in \eqnref{eq:H0TBG+_r}, the TBG part $H_{0,TBG}$ is just ordinary TBG with a $\sqrt{2}$ scaling of the interlayer tunneling~\cite{BAB20210211TSTGI}. 
Such rescaling can be cancelled by the same rescaling of the energy unit.
Thus, given any statement about the ordinary TBG with twist angle $\theta_{TBG}$, the same statement holds for $H_{0,TBG}$ with $\theta$ satisfying $\sin(\theta/2) = \sqrt{2} \sin(\theta_{TBG}/2)$~\cite{BAB20210211TSTGI}.
When $\theta$ is very small (\eg, around the first magic angle), the condition can be approximated by $\theta\approx \sqrt{2} \theta_{TBG}$.

According to \refcite{Song20211110MATBGHF}, in the ordinary TBG, localized heavy $f$ modes and itinerant $c$ modes can be constructed in each valley for each spin by mixing the nearly flat bands with the four lowest (two above and two below the flat bands) remote bands around $\Gamma_{\text{M}}$.
Owing to the correspondence between the TBG part $H_{0,TBG}$ of MATSTG and the ordinary TBG, we are also able to construct such $f$ and $c$ modes from $H_{0,TBG}$.
In the rest of this part, we follow \refcite{Song20211110MATBGHF} to show such construction.
The discussion in this part is the same as that in \refcite{Song20211110MATBGHF}, and thus can be viewed as a review of \refcite{Song20211110MATBGHF}. 

First, the $f$ and $c$ modes have the following expressions
\eq{
\label{eq:f_k}
f_{\eta,\bsl{k},\alpha,s}^\dagger = \sum_{\bsl{Q}\sigma}\widetilde{\psi}^\dagger_{\eta,\bsl{k},\bsl{Q},\sigma,s} \left[\widetilde{v}_{\eta,f,\alpha}(\bsl{k})\right]_{\bsl{Q}\sigma} \text{ for $\bsl{k}\in$MBZ}
}
and 
\eq{
\label{eq:c_k}
c_{\eta,\bsl{k},\beta,s}^\dagger = \sum_{\bsl{Q}\sigma}\widetilde{\psi}^\dagger_{\eta,\bsl{k},\bsl{Q},\sigma,s} \left[\widetilde{u}_{\eta,c,\beta}(\bsl{k})\right]_{\bsl{Q}\sigma} \text{ for $\left|\bsl{k}\right|\leq \Lambda_c$}\ ,
}
where $\alpha=1,2$, $\beta=1,2,3,4$, and $\Lambda_c$ is a small momentum cutoff for the $c$ modes (small compared to the length of the primitive Moir\'e reciprocal vectors).
$\widetilde{v}_{\eta,f,\alpha}(\bsl{k})$ is a smooth function of $\bsl{k}\in\dsR^2$ while keeping $f_{\eta,\bsl{k}+\bsl{G},\alpha,s}^\dagger=f_{\eta,\bsl{k},\alpha,s}^\dagger$ with $\bsl{G}$ the Moir\'e lattice vector, and $\widetilde{u}_{\eta,c,\beta}(\bsl{k})$ is a smooth function of $\bsl{k}$ for $\left|\bsl{k}\right|\leq \Lambda_c$.
Here $\widetilde{v}_{\eta,f}$ and $\widetilde{u}_{\eta,c}$ are all in the eigen-subspace of the lowest six spinless bands in valley $\eta$; $\widetilde{v}_{\eta,f}$ belongs to the subspace for the nearly flat bands  (the remote bands) far away from $\Gamma_{\text{M}}$ (at $\Gamma_{\text{M}}$).

$f$ modes are exponentially localized functions with physical symmetry representations (reps).
Specifically, we can define 
\eq{
\label{eq:f_R_k}
f^\dagger_{\eta,\bsl{R}} = \frac{1}{\sqrt{N}}\sum_{\bsl{k}\in\MBZ} e^{-\ii\bsl{k}\cdot\bsl{R}} f^\dagger_{\eta,\bsl{k}}\ ,
}
where 
\eq{
f^\dagger_{\eta,\bsl{k}} = (f^\dagger_{\eta,\bsl{k},1,\uparrow}, f^\dagger_{\eta,\bsl{k},1,\downarrow}, f^\dagger_{\eta,\bsl{k},2,\uparrow}, f^\dagger_{\eta,\bsl{k},2,\downarrow})\ .
}
The smoothness of $\widetilde{v}_{\eta,f}(\bsl{k})$ guarantees the exponential localization, and the symmetry reps in \appref{app:sym_rep} suggest that $f_{\eta,\bsl{R}}$ creates two spinful p-like orbitals localized at $\bsl{R}$.
This is why the $f$ modes are localized.

Now we construct the low-energy Hamiltonain of $H_{0,TBG}$ based on the $f$ modes and $c$ modes.
First, note that
\eqa{
\label{eq:psi_tilde_f_c_k}
\widetilde{\psi}^\dagger_{\eta,\bsl{k},\bsl{Q},\sigma, s} &= \sum_{\alpha=1,2} f^\dagger_{\eta,\bsl{k},\alpha,s} [\widetilde{v}_{\eta,f,\alpha}(\bsl{k})]_{\bsl{Q}\sigma}^* \\
& + \sum_{\beta=1}^4 c^\dagger_{\eta,\bsl{k},\beta,s} [\widetilde{u}_{\eta,c,\beta}(\bsl{k})]_{\bsl{Q}\sigma}^* \theta(\Lambda_c-|\bsl{k}|) + ...\ ,
}
where ``..." labels the high-energy modes in the subspace spanned by $\widetilde{\psi}$.
Then, we can separate out the low-energy part of $H_{0,TBG,\eta}$ (in the $f$ and $c$ basis) as 
\eq{
H_{0,TBG,\eta} = H_{0,\eta,f} + H_{0,\eta,c} + H_{0,\eta,fc} + ...\ ,
}
where $H_{0,\eta,fc}$ involves both $f$ and $c$ modes, $H_{0,\eta,f}$ only involves $f$ modes, and $ H_{0,\eta,c}$ only involves $c$ modes.
Based on the symmetry, the expressions of the low-energy terms are
\eq{
\label{eq:H_0_f}
H_{0,\eta,f} = 0\ ,
}
\eql{
\label{eq:H_0_c}
H_{0,\eta,c} = \sum^{|\bsl{k}|\leq \Lambda_c}_{\bsl{k}} c^\dagger_{\eta,\bsl{k}}  
\mat{ 
 0_{2\times 2} & v_{\star}(\eta k_x \tau_0 + \ii k_y \tau_z) \\
v_{\star}(\eta k_x \tau_0 - \ii k_y \tau_z) & M \tau_x 
} \otimes s_0
c_{\eta,\bsl{k}}\ ,
}
and
\eql{
\label{eq:H_0_fc}
H_{0,\eta,fc} = \sum^{|\bsl{k}|\leq \Lambda_c}_{\bsl{k}} f^\dagger_{\eta,\bsl{k}}  e^{- \frac{|\bsl{k}|^2\lambda^2}{2}} 
\mat{ 
\gamma \tau_0 + v_{\star}' (\eta k_x \tau_x + k_y \tau_y)   & v_{\star}'' (\eta k_x \tau_x - k_y \tau_y)
} \otimes s_0
c_{\eta,\bsl{k}}+h.c.\ ,
}
where 
\eqa{
& c_{\eta,\bsl{k}}^\dagger = (c_{\eta,\bsl{k},\Gamma_3}^\dagger, c_{\eta,\bsl{k},\Gamma_1\Gamma_2}^\dagger) \\ 
& c_{\eta,\bsl{k},\Gamma_3}^\dagger = (c_{\eta,\bsl{k},1, \uparrow}^\dagger, c_{\eta,\bsl{k},1,\downarrow}^\dagger, c_{\eta,\bsl{k},2, \uparrow}^\dagger, c_{\eta,\bsl{k},2, \downarrow}^\dagger) \\ 
& c_{\eta,\bsl{k},\Gamma_1\Gamma_2}^\dagger = (c_{\eta,\bsl{k},3, \uparrow}^\dagger, c_{\eta,\bsl{k},3,\downarrow}^\dagger, c_{\eta,\bsl{k},4, \uparrow}^\dagger, c_{\eta,\bsl{k},4, \downarrow}^\dagger)\ ,
}and $\lambda^2$ is the Wannier spread of the $f$ modes.
Here we choose $H_{0,\eta,f}=0$ because the hopping among $f$ modes is very small $(\sim 0.1 \meV)$, and we only keep terms up to $O(\bsl{k}^2)$ for $H_{0,\eta,c}$ and $H_{0,\eta,fc}$.
Owing to the zero kinetic energy of $f$ modes, they are heavy.
Here the factor $e^{- \frac{|\bsl{k}|^2\lambda^2}{2}} $ is added, since it is the coupling between a wave-packet with spread $\lambda^2$ and an itinerant electron with momentum $\bsl{k}$. 
This factor can be neglected for small $\bsl{k}$, but adding it allows us to choose a larger $\Lambda_c$ in later numerical calculations.

To determine the values of the parameters, we need to specify $\theta$.
Specifically, we choose
\eq{
\label{eq:theta_value}
\theta=1.4703^\circ\ ,
}
which is close to the first magic angle of MATSTG. 
In the rest of this work, we choose \eqnref{eq:theta_value} for all numerical calculations unless specified otherwise.
(We note that the framework discussed here is not limited to the value of $\theta$ in \eqnref{eq:theta_value}, as discussed in \appref{app:sym_rep}.)
Then, by projecting $H_{0,TBG}$ to the $f$ and $c$ modes, we can get the numerical values for the parameters in \eqnref{eq:H_0_f}-\eqref{eq:H_0_fc}, as shown in \tabref{tab:f-c-d_values}.

\begin{table}[t]
    \centering
    \begin{tabular}{|c|c|c|c|c|}
    \hline
         $M$ &  $\gamma$ & $v_\star$ &  $v_\star'$ & $v_\star''$ \\
         \hline
         -0.02678 &  0.1265 & 0.7176 & 0.2711 & 0.005768 \\
         \hhline{|=|=|=|=|=|}
        $M_1$  & $B_\gamma$ & $B_M$ & $B_{v''}$ & $\lambda$\\
         \hline
           -0.1394 & -0.09818 & -0.08583 & 0.08760 & 1.407\\
          \hline
    \end{tabular}
    \caption{Numerical values of the parameters in the single-particle $f-c-d$ model (\eqnref{eq:H_fcd_SP}) for the value of $\theta$ in \eqnref{eq:theta_value}.}
    \label{tab:f-c-d_values}
\end{table}

Before moving to other parts of the $f-c-d$ model, we note that approximate analytic expressions exist for the $f$ modes.
Explicitly, the $f$ modes have general expressions as
\eqa{
\label{eq:f_R_W}
f_{\eta,\bsl{R},\alpha,s}^\dagger  = \int d^2 r \sum_{\widetilde{l},\sigma} e^{\ii \bsl{R}\cdot \Delta K_{\widetilde{l}}} w_{\eta\alpha \widetilde{l} \sigma}(\bsl{r}-\bsl{R}) \widetilde{\psi}^\dagger_{\eta,\widetilde{l},\bsl{r},\sigma,s}\ ,
}
where
\eq{
\label{eq:Deltak}
\Delta K_t = -\bsl{q}_3 \ ,\ \Delta K_b = \bsl{q}_2\ ,
}
and
\eq{
\label{eq:W_Uf}
w_{\eta\alpha \widetilde{l} \sigma}(\bsl{r}) = \frac{1}{N\sqrt{\Omega}}\sum_{\bsl{k}}^{\MBZ}\sum_{\bsl{Q}\in\Q_{\eta,\widetilde{l}}} e^{\ii(\bsl{k}-\bsl{Q})\cdot \bsl{r}} [\widetilde{v}_{\eta,f,\alpha}(\bsl{k})]_{\bsl{Q}\sigma}
}
with $\Omega$ being the area of the Moir\'e unit cell, $\Q_{\eta,t}=\Q_{\eta}$ and $\Q_{\eta,b}=\Q_{-\eta}$.
Symmetry properties of $w_{\eta\alpha \widetilde{l} \sigma}(\bsl{r})$ are listed in \appref{app:sym_rep}.
The approximate expressions of $w_{\eta\alpha \widetilde{l} \sigma}(\bsl{r}-\bsl{R})$ are
\eqa{
\label{eq:f_ana}
& w_{+1 t A}^{\text{approx}}(\bsl{r}) = \frac{N_0}{\sqrt{2}} \frac{1}{\sqrt{\pi \lambda_1^2}} e^{-\frac{|\bsl{r}|^2}{2\lambda_1^2}} e^{-\ii \frac{\pi}{4}} \\
& w_{+1 t B}^{\text{approx}}(\bsl{r}) = \frac{N_1}{\sqrt{2}} \frac{1}{\sqrt{\pi \lambda_2^4}} e^{-\frac{|\bsl{r}|^2}{2\lambda_2^2}} (x+\ii y) e^{-\ii \frac{5\pi}{4}}
}
and other expressions can be obtained by acting with the symmetries on the basis.
With $N_0=-0.8193$, $N_1=-0.5734$, $\lambda_1=0.7502$ and $\lambda_2=0.8001$, we find the overlapping probability between the numerical $f$ modes and the analytical $f$ modes is at least $86\%$ as varying momentum, meaning that the analytical $f$ modes are good approximations.
By using the analytical expressions, we find that the low-energy bands of MATSTG are dominated by the $f$ modes, as shown in \figref{fig:bands_SP}(e-h).

\subsubsection{$d$ Modes}
\label{sec:d}

The low-energy Dirac modes are just $d^\dagger_{\eta,\bsl{p}}$ with small momentum $|\bsl{p}|<\Lambda_d$, where $\Lambda_d$ is the small momentum cutoff for the $d$ modes.
Then, the corresponding low-energy model of $d$ modes is
\eq{
\label{eq:H_0_d}
 H_{0,d,\eta} = \sum_{\bsl{p}}^{|\bsl{p}|\leq\Lambda_d} d^\dagger_{\eta,\bsl{p}}  (\eta p_x \sigma_x + p_y \sigma_y) \otimes s_0\  d_{\eta,\bsl{p}}\ ,
}
where
\eq{
d^\dagger_{\eta,\bsl{p}} = (d^\dagger_{\eta,\bsl{p},A,\uparrow},d^\dagger_{\eta,\bsl{p},A,\downarrow},d^\dagger_{\eta,\bsl{p},B,\uparrow},d^\dagger_{\eta,\bsl{p},B,\downarrow})\ .
}

\subsubsection{$f-d$ Coupling Around $\eta\KM$}
\label{sec:f_d}

The displacement field would couple $f/c$ modes to $d$ modes.
The leading-order coupling should happen between $f$ and $d$ modes around $\eta \KM$.
This is because, at $\E=0$, $d$ modes cross with the nearly-flat bands around $\eta \KM$ in valley $\eta$, and the nearly-flat bands around $\eta \KM$ are purely given by $f$ modes. (See \figref{fig:bands_SP}(a,e).)

Based on the symmetry reps in \eqnref{eq:sym_rep_f} in \secref{sec:SP_BM}, the leading-order coupling reads
\eqa{
\label{eq:H_0_fd}
H_{0,\eta,fd} & = \sum^{|\bsl{p}|\leq \Lambda_d}_{\bsl{p}} e^{-\frac{|\bsl{p}|^2\lambda^2}{2}} f^\dagger_{\eta ,\eta \KM + \bsl{p}}\ M_1 \E (\tau_0 + \eta \ii \tau_z)  s_0\  d_{\eta ,\bsl{p}}\\
& + h.c.\ ,
}
where the $\bsl{p}$-dependent terms are small (and are neglected) since the rep of $H_{0,\E}$ in the original basis is momentum independent. (See \appref{app:f_d} for details.)
Again, we add the factor $e^{- \frac{|\bsl{k}|^2\lambda^2}{2}} $ to allow a larger $\Lambda_d$ in later numerical calculations, in the same spirit of the factor for the $c$ modes~\cite{Song20211110MATBGHF}.
By projecting $H_{0,\E}$ to $f$ and $d$ at $\eta \KM$, we can get the numerical value of $M_1$ for the $\theta$ value in \eqnref{eq:theta_value}, as shown in \tabref{tab:f-c-d_values}.
Interestingly, the value of $M_1$ can also be estimated by the approximate analytical expressions of $f$ modes in \eqnref{eq:f_ana}, resulting in
\eq{
M_1 \approx \frac{N_0}{2}\sqrt{\frac{\pi \lambda_1^2}{\Omega}}= -0.1397\ ,
}
which is quite close to the numerical value, suggesting the good quality of the analytical approximation.

\subsubsection{$f-d$ and $c-d$ Couplings around $\Gamma_{\text{M}}$}
\label{sec:fd_cd}

 We did not yet include a $c-d$ coupling, since we focused on the $\eta \KM$, where $c$ does not appear at low energies.
To make our model more precise, we add the $c-d$ (as well as $f-d$) coupling around $\Gamma_{\text{M}}$.
The forms of those couplings are tedious, and we find that a more convenient way is to include them as corrections to the low-energy TBG part (\ie, \eqnref{eq:H_0_f}-\eqref{eq:H_0_fc}).
Such corrections can be obtained by using the perturbation theory, since the $f-d$ and $c-d$ couplings are small compared to the gaps between $f/c$ modes and $d$ modes around $\Gamma$, as elaborated in \appref{app:fd_cd}.
As $H_{0,\E}$ has lower symmetries than $H_{0,TBG}$, the correction would bring in terms that break the extra symmetries of $H_{0,TBG}$.
Nevertheless, we numerically find that those terms that break extra symmetries can be neglected without affecting the precision too much.
As a result, the correction due to terms that preserve the extra symmetries can be incorporated by performing the following replacement in \eqnref{eq:H_0_f}-\eqref{eq:H_0_fc} 
\eqa{
\label{eq:TBG_replacement}
\gamma\rightarrow \gamma + B_\gamma \E^2\ ,\ \ v_\star''\rightarrow v_\star'' + B_{v''} \E^2\ ,\ M\rightarrow M + B_{M} \E^2\ .
}
We can directly obtain the values of $B_\gamma$, $B_{v''}$ and $B_{M}$ for the $\theta$ value in \eqnref{eq:theta_value} from the perturbation methods, and show the results in \tabref{tab:f-c-d_values}.

\subsubsection{Single-Particle $f-c-d$ Model}
\label{sec:f-c-d_SP_sum}

Combining \secref{sec:f_c}-\ref{sec:fd_cd}, we arrive at the single-particle $f-c-d$ model as
\eq{
\label{eq:H_fcd_SP}
H_{0,\eta }^{eff} = H_{0,\eta ,f} + H_{0,\eta ,c} +  H_{0,\eta ,fc}  + H_{0,\eta ,d}+ H_{0,\eta ,fd}\ ,
}
where $H_{0,\eta ,f}$, $H_{0,\eta ,c}$ and  $H_{0,\eta ,fc}$ are \eqnref{eq:H_0_f}-\eqref{eq:H_0_fc} with the replacement in \eqnref{eq:TBG_replacement}, $H_{0,\eta ,d}$ is in \eqnref{eq:H_0_d}, and $H_{0,\eta ,fd}$ is in \eqnref{eq:H_0_fd}.
With the parameter values in \tabref{tab:f-c-d_values}, we plot the band structure of \eqnref{eq:H_fcd_SP} in valley $+$ in \figref{fig:bands_SP}(a-d).
We find that the bands of \eqnref{eq:H_fcd_SP} match those of the single-particle BM-type model $H_0$ in \eqnref{eq:H_0} very well for $0 \leq \E \leq 300$meV and for the energy window $[-50 \meV, 50 \meV]$ (in EUS).
The details on the numerical calculation can be found in \appref{app:fcd_SP}.

Before moving to the interacting part of the $f-c-d$ model, we comment on the difference between our heavy localied $f$ modes and the heavy modes mentioned in previous works~\cite{Sachdev20191202TSTGSP,Ramires2021TSTGHF} on MATSTG.

First, we emphasize that our heavy localized $f$ modes are not the heavy modes mentioned in \refcite{Sachdev20191202TSTGSP}.
\refcite{Sachdev20191202TSTGSP} directly refers to the nearly flat bands in TBG part as the ultraheavy quasi-particles: however, these cannot be localized if physical symmetry reps are required due to the nontrivial topology of the bands.
On the other hand, our $f$ modes are localized, since the Wannier obstruction has been broken by mixing the nearly-flat bands and the remote bands in the construction.

Second, although the heavy-fermion physics in MATSTG was also discussed in \refcite{Ramires2021TSTGHF}, the heavy modes in \refcite{Ramires2021TSTGHF} are different from our heavy localized $f$ modes.
In \refcite{Ramires2021TSTGHF}, the dispersionless localized modes are phenomenologically constructed by coupling the TBG nearly-flat bands to the Dirac modes at a relatively large displacement field. 
It is not clear whether their construction can be applied to small displacement fields, since at zero displacement field, the TBG nearly-flat bands are decoupled from the Dirac cones, and cannot be directly treated as localized modes due to their nontrivial topology.
Our dispersionless localized $f$ modes are constructed by combining the TBG flat bands with the remote bands around $\Gamma_{\text{M}}$, which does not rely on the displacement field.
One manifestation of such  differences is that the heavy modes in \refcite{Ramires2021TSTGHF} have in total 4 flavors per Moir\'e unit cell and couple to dispersive modes around $\text{M}_{\text{M}}$, while our heavy $f$ modes have 8 flavors per Moir\'e unit cell and couple to dispersive modes around $\Gamma_{\text{M}}$ (and around $\eta \KM$ via the displacement field).
Nevertheless, despite the difference, it is interesting to study (as future works) whether the model in \refcite{Ramires2021TSTGHF} and our $f-c-d$ model give qualitatively consistent phases after including the interaction.

\subsection{Interaction among $f$, $c$ and $d$ Modes}

We now discuss the interaction among $f$, $c$ and $d$ modes, which is derived by projecting the Coulomb interaction to the $f$, $c$ and $d$ modes.

\subsubsection{Review: Interaction Among $f$ and $c$ Modes}

Both $f$ and $c$ modes are constructed solely from the TBG part of the  model.
Therefore, the interaction among $f$ and $c$ modes should have the same form as those in \refcite{Song20211110MATBGHF}, which we will review in this part.
More details can be found in \appref{app:fcd_int}.

First, for the interaction among $f$ modes, the leading-order term is the density-density interaction, which reads
\eqa{
\label{eq:H_int_U}
H_{int,U} & =
\frac{U_1}{2}\sum_{\bsl{R}} :\rho_{f}(\bsl{R}): :\rho_{f}(\bsl{R}): \\
& \quad + \frac{U_2}{2}\sum_{ \bsl{R} , \bsl{R}' }^{|\bsl{R}-\bsl{R}'|=|\bsl{a}_{M,1}|} :\rho_{f}(\bsl{R}): :\rho_{f}(\bsl{R}'): \ ,
}
where $\rho_f(\bsl{R}) = \sum_{\eta,\alpha,s} f^\dagger_{\eta,\bsl{R},\alpha,s}f_{\eta,\bsl{R},\alpha,s}$.
The expressions of $U_1$ and $U_2$ can be found in \appref{app:fcd_int}.
In \eqnref{eq:H_int_U}, we neglect the density-density interactions of further ranges, as they are exponentially lower owing to the localized nature of the $f$ modes.

Second, the interaction among $c$ modes turns out to have the Coulomb form to the leading order as
\eq{
\label{eq:H_int_V_c}
H_{int,V,c}= \frac{1}{2} \int d^2 r d^2r' V(\bsl{r}-\bsl{r}') :\rho_c(\bsl{r}): :\rho_c(\bsl{r}'):\ ,
}
where $\rho_{c}(\bsl{r}) = \sum_{\beta} \rho_{c,\beta}(\bsl{r}) $, $\rho_{c,\beta}(\bsl{r}) = \sum_{\eta,s} c^\dagger_{\eta,\bsl{r},\beta,s}c_{\eta,\bsl{r},\beta,s}$, and 
\eq{
\label{eq:c_r}
c^\dagger_{\eta,\bsl{r},\beta,s}=\frac{1}{\sqrt{\A}}\sum_{\bsl{p}}^{|\bsl{p}|\leq \Lambda_c} e^{-\ii \bsl{p}\cdot\bsl{r}}c^\dagger_{\eta,\bsl{p},\beta,s}\ .
}

Third, the interaction between $f$ and $c$ modes has two non-negligible terms.
One term is the channel-resolved density-density interaction as
\eq{
\label{eq:H_int_W_fc}
H_{int,W,fc} =  \Omega \sum_{\bsl{R}, \beta } W_\beta :\rho_{f}(\bsl{R}): :\rho_{c,\beta}(\bsl{R}): 
}
with $W_1=W_2$ and $W_3=W_4$. The last term is 
\eqa{
\label{eq:H_int_J}
H_{int,J} & = - \frac{J \Omega}{2} \sum_{\bsl{R}} \sum_{\eta \alpha s } \sum_{\eta' \alpha' s' } (\eta \eta' + (-)^{\alpha+\alpha'}) \\
& :f^\dagger_{\eta,\bsl{R},\alpha,s}f_{\eta',\bsl{R},\alpha',s'}: :c^\dagger_{\eta',\bsl{R},\alpha'+2,s'}c_{\eta,\bsl{R},\alpha+2,s}:\ .
}
The interaction only occurs at the Moir\'e lattice positions, which is consistent with the fact that $f$ modes are localized at Moir\'e lattice positions.
The expressions of $W_\beta$ and $J$ can be found in \appref{app:fcd_int}.

\subsubsection{Interaction Among $d$ Modes}

Inherited from the total Coulomb interaction, the interaction among $d$ modes is given by the Coulomb form
\eq{
\label{eq:H_int_V_d}
H_{int,V,d}= \frac{1}{2} \int d^2 r d^2r' V(\bsl{r}-\bsl{r}') :\rho_d(\bsl{r}): :\rho_d(\bsl{r}'):\ ,
}
where $\rho_{d}(\bsl{r}) = \sum_{\eta,\sigma,s} \widetilde{d}^\dagger_{\eta,\bsl{r},\sigma,s} \widetilde{d}_{\eta,\bsl{r},\sigma,s}$ and $\widetilde{d}_{\eta,\bsl{r},\sigma,s}^\dagger = \frac{1}{\sqrt{\A}}\sum_{\bsl{p}}^{|\bsl{p}|\leq \Lambda_d} e^{-\ii \bsl{p}\cdot \bsl{r}} d^\dagger_{\eta,\bsl{p},\sigma,s}$ which becomes $d_{\eta,\bsl{r},\sigma,s}$ in the limit of $\Lambda_d\rightarrow \infty$.

\subsubsection{$f-d$ and $c-d$ Interaction}

We find that the interaction between $f$ and $d$ modes and the interaction between $c$ and $d$ modes are both in the form of density-density interaction in the leading order, as discussed in details in \appref{app:fcd_int}.
Specifically, we find that the leading-order interaction between $f$ and $d$ modes reads
\eq{
\label{eq:H_int_W_fd}
H_{int,W,fd} =  \Omega W_{fd} \sum_{\bsl{R}} :\rho_{f}(\bsl{R}): :\rho_{d}(\bsl{R}): \ ,
}
and the leading-order interaction between $c$ and $d$ modes has the Coulomb form as
\eq{
\label{eq:H_int_V_cd}
H_{int,V,cd} =  \int d^2 r d^2r' V(\bsl{r}-\bsl{r}') :\rho_c(\bsl{r}): :\rho_d(\bsl{r}'):\ .
}
The expression of $W_{fd}$ can be found in \appref{app:fcd_int}.

\subsubsection{Total Interaction}

\begin{table}[t]
    \centering
    \begin{tabular}{|c|c|c|c|c|c|c|}
    \hline
         Unit  & $U_1$ &  $U_2$ & $W_1$ &  $W_3$ & $J$ & $W_{fd}$\\
         \hline
        \eqnref{eq:unit_system} & 0.3523 & 0.02388 & 0.3409 & 0.3761 & 0.09337 &  0.3647 \\
         \hline
         EUS (meV) & 91.50 &  6.203 & 88.54 & 97.67 & 24.25 &  94.71\\
         \hline
    \end{tabular}
    \caption{Numerical values of the parameters in \eqnref{eq:H_fcd_int}.
    Values in the second line of the table is in the unit system specified in \eqnref{eq:unit_system}, while those in the third line are in EUS.
    More details can be found in \appref{app:fcd_int}.
    }
    \label{tab:fcd_int}
\end{table}

The total interaction among the $f$, $c$ and $d$ modes is the sum of 
\eqnref{eq:H_int_U},\eqref{eq:H_int_V_c},\eqref{eq:H_int_W_fc},\eqref{eq:H_int_J},\eqref{eq:H_int_V_d}, \eqref{eq:H_int_W_fd} and \eqref{eq:H_int_V_cd}, which reads
\eqa{
\label{eq:H_fcd_int}
H^{eff}_{int} & = H_{int,U} + H_{int,V,c} + H_{int,W,fc} + H_{int,J} \\
& \quad + H_{int,V,d} + H_{int,V,cd} + H_{int,W,fd}
}
We numerically evaluate the values of the interaction parameters, and the results are listed in \tabref{tab:fcd_int}.
Among the interaction strengths, we can see that the largest energy scale is $90\sim 100$meV in EUS.
We have $W_1$, $W_3$, $W_{fd}$ and $U_1$ at this scale.
Unlike \refcite{Song20211110MATBGHF}, $W_3$ is slightly larger than $U_1$ here, since the gate distance is not scaled by $1/\sqrt{2}$ for MATSTG compared to that in \refcite{Song20211110MATBGHF}. (See more details in \appref{app:fcd_int}.)
Moreover, $W_{fd}$ is also slightly larger than $U_1$ here.
Nevertheless, we would expect that the onsite repulsive interaction $U_1$ among $f$ modes is the dominant interaction channel at low energies, since $U_1$ only involves the $f$ modes (which dominate in the low energy), while $W_1$, $W_3$ and $W_{fd}$ involve the $c$ and $d$ modes with relatively higher energies.

\subsection{$f-c-d$ Model For MATSTG}

The $f-c-d$ model for MATSTG is just the sum of the single-particle part \eqnref{eq:H_fcd_SP} and the interaction \eqnref{eq:H_fcd_int} as 
\eq{
\label{eq:H_fcd}
H_{fcd} = \sum_{\eta} H_{0,\eta}^{eff} + H_{int}^{eff}\ .
}
This is the low-energy model that we propose for MATSTG with only Coulomb interaction.
The single-particle band structure (\figref{fig:bands_SP}) already shows that the $f-c-d$ model well captures the single-particle physics of MATSTG for $\E\in[-300,300]$meV and for the energy window $[-50,50]$meV in EUS.
Since the largest energy scale of the interaction is $U_1 \sim 100 meV$, the energy window corresponds to $[-U_1/2, U_1/2]$, covering the main low-energy modes affected by the interaction. 
Therefore, we expect the $f-c-d$ model \eqnref{eq:H_fcd} to work for the specified $\E$ range and energy window even at the many-body level.
We will perform Hartree-Fock calculations with the model in the following section.

\section{Numerical Hartree-Fock Calculations}
\label{sec:NHF_cal}

With our model (\eqnref{eq:H_fcd}), we perform numerical Hartree-Fock calculations for $\nu=0,-1,-2$.
We will not study the positive fillings since they are related to the negative fillings by $\C C_{2x}\P$ as discussed in \secref{sec:BM_Total}.

\begin{figure*}[t]
    \centering
    \includegraphics[width=2\columnwidth]{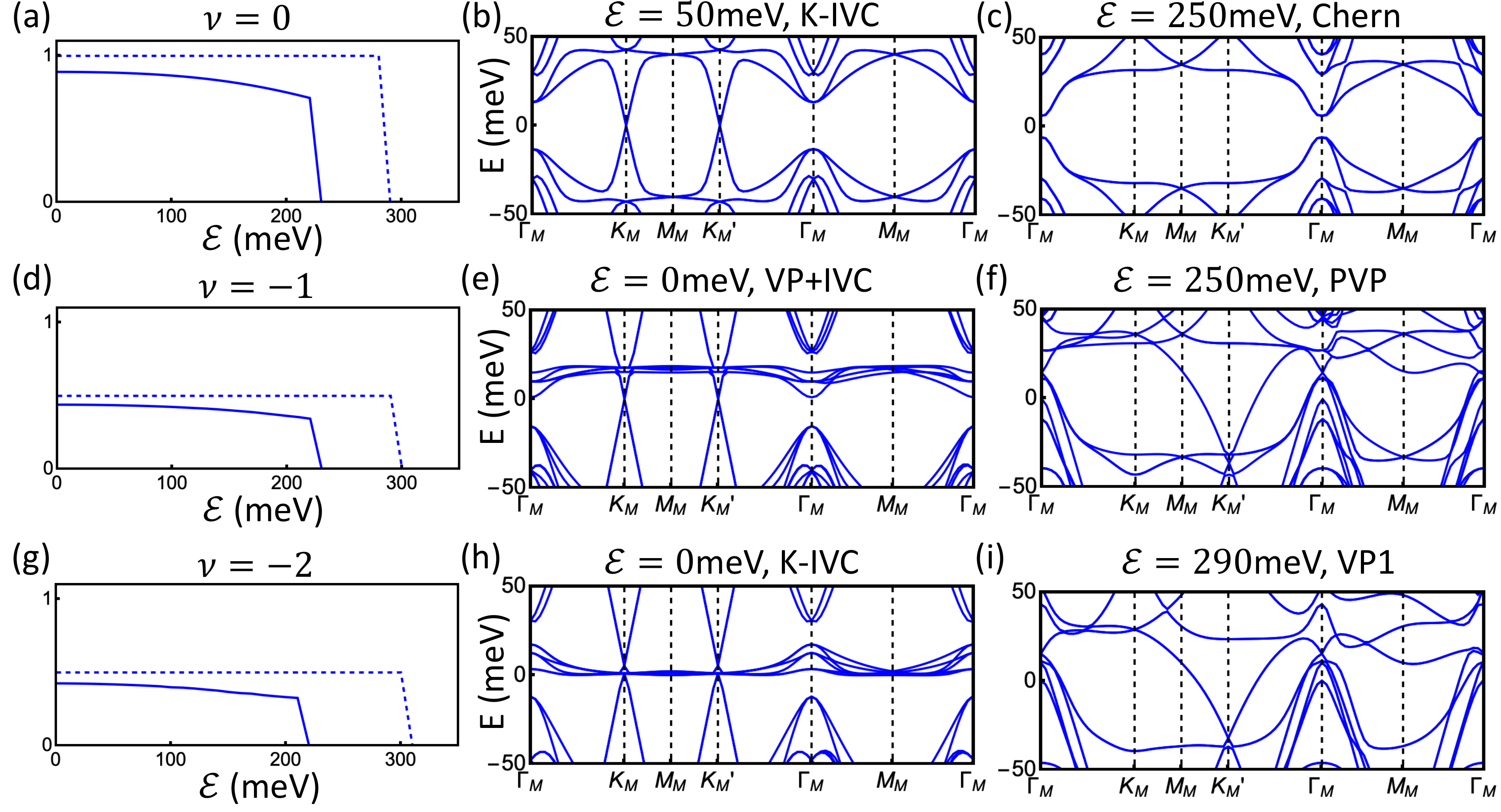}
    \caption{This figure shows the numerical Hartree-Fock results for MATSTG based on \eqnref{eq:H_fcd}, where (a-c) are for $\nu=0$, (d-f) are for $\nu=-1$ and (g-i) are for $\nu=-2$.
    (a,d,g) shows the intervalley coherence of the Hartree-Fock ground state as $\E$ varies, where the zero (nonzero) value corresponds to the absence (presence) of the intervalley coherence.
    The solid line is given by the self-consistent Hartree-Fock calculation, while the  dashed is the one-shot result.
    (b,c,e,f,h,i) are the Hartree-Fock band structures of the ground state (or one of the competing ground states) at the corresponding filling and $\E$, plotted with the density matrices given by the self-consistent Hartree-Fock calculation.
    }
    \label{fig:numerical_HF}
\end{figure*}

Similar to the TBG case \cite{Song20211110MATBGHF}, the initial states that we choose for the Hartree-Fock calculation have the following general form
\eq{
\label{eq:state_ini_HF}
\ket{\Psi_{\text{initial}}} = \prod_{\bsl{R}}\ f^\dagger_{\bsl{R}}\zeta_1 f^\dagger_{\bsl{R}}\zeta_2 \cdot\cdot\cdot f^\dagger_{\bsl{R}}\zeta_{4+\nu}
\left|\text{Fermi Sea}\right\rangle\ ,
}
where  
\eql{
f^\dagger_{\bsl{R}} = (f^\dagger_{+,\bsl{R},1 ,\uparrow}, f^\dagger_{+,\bsl{R},1 ,\downarrow}, f^\dagger_{+,\bsl{R}, 2 ,\uparrow} ,  f^\dagger_{+,\bsl{R}, 2 ,\downarrow}, f^\dagger_{-,\bsl{R},1 ,\uparrow}, f^\dagger_{-,\bsl{R},1 ,\downarrow}, f^\dagger_{-,\bsl{R}, 2 ,\uparrow}, f^\dagger_{-,\bsl{R}, 2 ,\downarrow})\ , 
}
each of $\zeta_1,...,\zeta_{4+\nu}$ has eight components,
\eg,
\eq{
\zeta_1 =\mat{ 
\left( \zeta_1 \right)_{+1 \uparrow} \\ 
\left( \zeta_1 \right)_{+1 \downarrow} \\ 
\left( \zeta_1 \right)_{+2 \uparrow} \\ 
\left( \zeta_1 \right)_{+2 \downarrow} \\ 
\left( \zeta_1 \right)_{-1 \uparrow} \\ 
\left( \zeta_1 \right)_{-1 \downarrow} \\ 
\left( \zeta_1 \right)_{-2 \uparrow} \\ 
\left( \zeta_1 \right)_{-2 \downarrow}
}\ ,
}
\eq{
f^\dagger_{\bsl{R}} \zeta_1 =\sum_{\eta,\alpha,s} f^\dagger_{\eta,\bsl{R},\alpha,s} \left( \zeta_1 \right)_{\eta\alpha s}\ ,
}
and $\ket{\text{Fermi Sea}}$ is the half-filled Fermi sea of the free $c$ and $d$ modes.
(See the choice of the initial states in \appref{app:ini_states}.)
\eqnref{eq:state_ini_HF} means that we specify different initial states by specifying different combinations of the $f$ modes, \ie, specifying 
\eq{
\label{eq:xi}
\zeta=\mat{\zeta_1 & \zeta_2 & ... & \zeta_{4+\nu}}\ .
}
We can do so because the $f$ modes and its onsite interaction dominate the low-energy physics as discussed in the last section.
By using \eqnref{eq:state_ini_HF}, we perform self-consistent Hartree-Fock calculations for $\nu=0,-1,-2$, and the results are summarized below and shown in \figref{fig:numerical_HF}. (See details in \appref{app:Hartree_Fock}.)

As shown in \figref{fig:numerical_HF}(a,d,g), for all the considered fillings, increasing the displacement field $\E$ would lead to a phase transition, at which the ground states lose intervalley coherence.

For $\nu=0$, the low-$\E$ ground states are the Kramers-intervalley-coherent (K-IVC) states, while there are four types of competing ground states at high $\E$, namely Chern states $(\ch=\pm 2)$, half-Chern states $(\ch=\pm 1)$, valley-Hall (VH) states and $C_2\TR$-invariant states, where ``competing" means that the differences in their ground-state energies are beyond our numerical resolution, VH refers to the state with nonzero valley Chern numbers but zero total Chern number, and $\ch$ stands for the Chern number.
The low-$\E$ states are metallic, while the high-$\E$ states are insulating. (See \figref{fig:numerical_HF}(b,c).) 

For $\nu=-1$, the low-$\E$ ground states are a combination of valley-polarized (VP) and intervalley-coherent (IVC) states, while there are three types of competing partially-valley-polarized (PVP) ground states at high $\E$, where PVP means that one valley has one more electron than the other valley per Moir\'e unit cell and the state has no intervalley coherence.
PVP is ``partial" because the VP state for $\nu=-1$ should have three more electrons in one valley than in the other.
Both the low-$\E$ and high-$\E$ states are metallic. (See \figref{fig:numerical_HF}(e,f).) 

For $\nu=-2$, the low-$\E$ ground states are K-IVC states, while there are four types of competing ground states at high $\E$---two types of VP states and two types of valley unpolarized states. 
Both the low-$\E$ and high-$\E$ states are metallic. (See \figref{fig:numerical_HF}(h,i).) 

All these self-consistent Hartree-Fock results obtained from our $f-c-d$ model (\eqnref{eq:H_fcd}) are generally consistent with previous numerical results in \refcite{Sachdev20210603TSTGInt,BAB20210628TSTGII,Vishwanath20211122AMTG}, verifying the validity of our $f-c-d$ model.
Moreover, our calculation finds some high-$\E$ ground states (like the half-Chern states for $\nu=0$) that are missed in \refcite{Sachdev20210603TSTGInt,BAB20210628TSTGII,Vishwanath20211122AMTG}, meaning that our calculation actually refines the previous results~\cite{Sachdev20210603TSTGInt,BAB20210628TSTGII,Vishwanath20211122AMTG}.

In particular, we find that the phase transitions characterized by the loss of intervalley coherence (\figref{fig:numerical_HF}(a,d,g)) can be qualitatively captured by the one-shot Hartree-Fock calculation, where ``one-shot" means only performing the first step of the iteration, which is numerically simple to do compared with the full self-consistent calculation and can even be done analytically as discussed in the next section. 
Furthermore, we find that the competing energies of the high-$\E$ ground states can be precisely captured in the one-shot Hartree-Fock calculation.
Therefore, our choice of the initial states in \eqnref{eq:state_ini_HF} are considerably close to the final Hartree-Fock ground states given by the self-consistent Hartree-Fock calculation, verifying the fact that the $f$ modes and their onsite repulsive interaction dominate the low-energy physics.

We note that our Hartree-Fock calculation is done only for the translationally-invariant initial states listed in \appref{app:ini_states}. It is possible that the true ground state is beyond our chosen initial states in \appref{app:ini_states} (\eg, beyond the translationally-invairant subspace). We leave a complete Hartree-Fock study as a future work.

\section{Analytical Understanding}
\label{sec:ana}

In this section, we provide an analytical understanding for the key numerical results in \secref{sec:NHF_cal}.
As discussed at the end of \secref{sec:NHF_cal}, the one-shot Hartree-Fock calculation (i) can qualitatively capture the phase transition between states with and without intervalley-coherence (\figref{fig:numerical_HF}(a,d,h)) and (ii) can precisely capture the competing energies of the several found high-$\E$ states.
Therefore, we will use the analytical one-shot Hartree-Fock Hamiltonian of the $f-c-d$ model (\eqnref{eq:H_fcd}) derived from the expression of the initial states (\eqnref{eq:state_ini_HF}) to answer two questions: (i) why the states without intervalley coherence are favored at high $\E$, and (ii) why those high-$\E$ ground states have nearly the same energies.

Let us start with the first question: why the states without intervalley coherence are favored at high $\E$.
Since we care about high $\E$, let us consider the limit where $\E$ is infinitely large.
The validity of this assumption will be discussed right beneath \propref{prop:high_field_states}. 
The low-energy itinerant modes are mainly around $\Gamma_{\text{M}}$ and $\pm\KM$.
In the following, we will look at $\pm\KM$ first and then look at $\Gamma_{\text{M}}$.

We want to minimize the total energy of all the occupied states at $\KM$ and $-\KM$, which is labelled by $E_{\pm \KM}$.
To do so, let us define $\zeta_\eta$.
We know $\zeta_{l}$ (with $l = 1,2,...,4+\nu$) in \eqnref{eq:state_ini_HF} has eight components as $(\zeta_{l})_{\eta\alpha s}$, where $\eta$, $\alpha$ and $s$ are indices of the $f$ modes.
We define $\zeta_\eta$ as a $4\times (4+\nu)$ matrix such that $(\zeta_\eta)_{\alpha s,l} = (\zeta_{l})_{\eta\alpha s}$, which means that 
\eq{
\label{eq:xi_eta}
\zeta=\mat{\zeta_+ \\ \zeta_-}\ .
}
Then, as elaborated in \appref{app:simple_rule_high_E}, in the high-$\E$ limit, the one-shot Hartree-Fock Hamiltonian at $\eta\KM$ of MBZ to the first order of $|U_1/\E|$ (up to unitary transformation and total energy shift) reads
\eqa{
\label{eq:Gamma_approx_E}
& \mat{ 
\epsilon_0 \mathds{1}_{4\times 4} & & \\
 & \epsilon_1 \mathds{1}_{4\times 4} & \\
 & & \nu(U_1+6U_2)\mathds{1}_{4\times 4}
} \\
& - U_1 
\left(\begin{array}{c|c}
\mat{ |\chi_{0,1}|^2 &   \\   & |\chi_{1,1}|^2} \otimes (\zeta_\eta \zeta_\eta^\dagger - \frac{1}{2})  &  \\
\hline
   &  \zeta_{-\eta} \zeta_{-\eta}^\dagger - \frac{1}{2}
\end{array}\right) \ ,
}
where 
\eq{
\mat{ 
\nu (U_1 + 6 U_2) & \sqrt{2} M_1 \E \\ 
\sqrt{2} M_1 \E & W_{fd} \nu
} \chi_{\gamma} = \epsilon_\gamma \chi_{\gamma} \ ,
}
$\gamma = 0,1$, $\chi_\gamma$ is real, and 
\eqa{
\epsilon_\gamma & = \frac{\nu (U_1 + 6 U_2 + W_{fd} )}{2} \\
& \quad + (-)^\gamma \sqrt{\left[ \frac{\nu (U_1+ 6 U_2 - W_{fd})}{2} \right]^2 + 2 M_1^2 \E^2 }\ .
}
(See \appref{app:simple_rule_high_E} for details.)
Since the chemical potential can be estimated as $\mu \approx \nu(U_1+6U_2)$ ($f$ modes give the filling) as discussed in \appref{app:simple_rule_high_E} (also in \refcite{Song20211110MATBGHF}), the occupied states of the approximated Hamiltonian in \eqnref{eq:Gamma_approx_E} are all eigenstates of
\eq{
[\epsilon_1-\nu(U_1+6U_2)] \mathds{1}_{4\times 4} - U_1 |\chi_{1,1}|^2  (\zeta_\eta \zeta_\eta^\dagger - \frac{1}{2})
}
and all negative-energy eigensstates of 
\eq{
- U_1 (\zeta_{\eta} \zeta_{\eta}^\dagger - \frac{1}{2})
}
for both $\eta=\pm$, where we have subtracted the chemical potential.
The total energy of these occupied states give $E_{\pm \KM}$ to the first order of $|U_1/\E|$.

Now let us minimize $E_{\pm \KM}$.
To express $E_{\pm \KM}$, we use $\lambda_{i}$ $(i=1,2,...,8)$ to label the eigenvalues of 
\eq{
\mat{
\zeta_{+} \zeta_{+}^\dagger & \\
 & \zeta_{-} \zeta_{-}^\dagger
}\ .
}
Then, we can choose $\lambda_1\geq \lambda_2\geq ... \geq \lambda_n \geq 1/2 \geq \lambda_{n+1} \geq ... \geq \lambda_8$ without loss of generality, resulting in
\eqa{
& E_{\pm \KM}  = 8 [\epsilon_1-\nu(U_1+6U_2)] - U_1 |\chi_{1,1}|^2  \nu \\
&\ \ \ - U_1 \sum_{i=1}^n (\lambda_i -\frac{1}{2}) + O(U_1^2/\E)\ ,
\label{eq:energy}
}
where we have used
\eq{
\sum_{\eta} \Tr[\zeta_\eta \zeta_\eta^\dagger] = \Tr[\zeta \zeta^\dagger ] = 4+\nu\ .
}

To proceed, we note that $\lambda_i\in[0,1]$ and $\sum_{i=1}^8 \lambda_i = 4 + \nu$.
Then, we know
\eq{
E_{\pm \KM} \geq  8 [\epsilon_1-\nu(U_1+6U_2)] - U_1 |\chi_{1,1}|^2  \nu - U_1 \frac{4+\nu}{2} + O(U_1^2/\E) \ .
}
As elabrated in \appref{app:simple_rule_high_E}, it turns out that the equality happens if and only if 
\eq{
\label{eq:mat_sim_xi_pm}
\mat{
\zeta_{+} \zeta_{+}^\dagger & \\
 & \zeta_{-} \zeta_{-}^\dagger
} 
\cong
\text{diag}(\underbrace{1,1,...,1}_{4+\nu},\underbrace{0,0,...,0}_{4-\nu})\ ,
}
which is equivalent to $\zeta_{+} \zeta_{-}^\dagger = 0$ (\ie, zero intervalley coherence).
Here $\cong$ means being equal up to any unitary transformations.
Therefore, we know $E_{\pm \KM}$ is minimized if and only if the intervalley coherence of the state vanishes.

Now we turn to the $\Gamma_{\text{M}}$ point.
As discussed in \appref{app:simple_rule_high_E} (and also in \refcite{Song20211110MATBGHF}), the main origin of the symmetry breaking is the $J$ interaction term, which appears in the diagonal block of the one-shot Hartree-Fock Hamiltonian for $c^\dagger_{\Gamma_1\Gamma_2}$, expressed as $\nu W_3 + h_{\Gamma_{1}\Gamma_{2}}$.
In our case, $h_{\Gamma_{1}\Gamma_{2}}$ reads
\eqa{
\label{eq:hGamma1Gamma2}
& h_{\Gamma_{1}\Gamma_{2}} = \widetilde{M} \eta_0\sigma_x s_0 \\
& -  \frac{J}{2} \left( \eta_z \sigma_0 s_0 \zeta \zeta^\dagger \eta_z\sigma_0 s_0 + \eta_0 \sigma_z s_0 \zeta \zeta^\dagger \eta_0\sigma_z s_0  - \mathds{1}_{8\times 8} \right)\ ,
}
where $\widetilde{M} = M + B_M \E^2$.
Since we consider the high-$\E$ limit, we have $|\widetilde{M}|\gg J$.
Then, the energy difference between different states given by $h_{\Gamma_{1}\Gamma_{2}}$ should be of order $J$, which is generally much smaller than the energy difference at $\pm\KM$ which is of the order $U_1$.
Therefore, we should only focus on the states with lowest $E_{\pm \KM}$, $\ie$ states with zero intervalley coherence.
In other words, the discussion at $\pm\KM$ already suggests that only states without intervalley coherence should be favored at large $\E$.

To further pick out the high-$\E$ ground states among all states without IVC, let us minimize the energy at $\Gamma_{\text{M}}$.
Since we are considering the high-$\E$ limit, we have $|\widetilde{M}|\gg |\nu(U_1+6 U_2-W_3)|$.
Then, by minimizing the total energy of all the occupied states of $\nu W_3 + h_{\Gamma_{1}\Gamma_{2}}$ (\ie, states of $h_{\Gamma_{1}\Gamma_{2}}$ that are energetically lower than $\nu(U_1+6 U_2-W_3)$)  while keeping the intervalley coherence zero, we find that the energetically favored states are (and only are) the states whose $\zeta \zeta^\dagger$ (up to $U(2)\times U(2)$) are also spin-diagonal with $4+\nu$ diagonal blocks (labelled by valley and spin) being $(\sigma_0\pm\sigma_z)/2$ and $4-\nu$ diagonal blocks being zero. 
(See details in \appref{app:simple_rule_high_E}.)
Eventually, we arrive at the following rule for the high-$\E$ ground states, which resolves the first question raised at the beginning of this section.
\begin{proposition}
\label{prop:high_field_states}
For $\nu=0,-1,-2$, at the one-shot Hartree-Fock level, a state is energetically favored at high $\E$ if and only if its $\zeta \zeta^\dagger$, up to $U(2)\times U(2)$, is spin-valley diagonal with $4+\nu$ diagonal blocks (labelled by valley and spin) being $(\sigma_0\pm\sigma_z)/2$ and $4-\nu$ diagonal blocks being zero.
\end{proposition}

Now let us discuss the validity of the derivation that leads to \propref{prop:high_field_states}.
We know that the derivation is done in the limit that $\E$ is infinitely large, which seems to contradict the fact that the $f-c-d$ model is valid within $\E=300$meV (EUS), since $|\sqrt{2} M_1 \E| \sim U_1$ for $\E=300$meV.
However, we show in \appref{app:simple_rule_300_nu0} that the derivation should still be valid for $\nu=0$ at $\E=300\meV$, since the quantities required to be small in the derivation are still small for $\nu=0$ at $\E=300\meV$.
Although the derivation is not entirely reasonable for $\nu=-1,-2$, we find that \propref{prop:high_field_states} is consistent with the self-consistent Hartree-Fock calculation for $\nu=-1,-2$.
Specifically, we numerate all initial states that satisfy \propref{prop:high_field_states} for $\nu=0,-1,-2$, and we find that they all become high-$\E$ ground states in the self-consistent Hartree-Fock calculation discussed in \secref{sec:NHF_cal}.

Before proceeding to the second question raised at the beginning of this section, let us provide an understanding of the appearance of the phase transition with gradually increasing $\E$.
In the earlier part of this section, we have shown that the Hartree-Fock Hamiltonian at $\pm\KM$ should favor states without intervalley coherence at high $\E$; on the other hand, \refcite{Song20211110MATBGHF} suggests the TBG part around $\Gamma_{\text{M}}$ should favor states with nonzero intervalley coherence.
Then, the transition should be a result of the competition between $\Gamma_{\text{M}}$ and $\pm \KM$.
To make it concrete, let us consider $\nu=0$ and treat $\E$ perturbatively, to consider the case where $\E$ is gradually increased.
We focus on the competence between K-IVC and Chern states.
By using second order perturbation, we derive the effective energies for the two states at $\pm \KM$ as
\eqa{
& E^{\text{K-IVC},\nu =0}_{\pm\KM} = -4 \sum_{\bsl{p}}^{|\bsl{p}|<\Lambda_d} |\bsl{p}| \\
& E^{\text{Ch},\nu =0}_{\pm\KM} = -4 \sum_{\bsl{p}}^{|\bsl{p}|<\Lambda_d} \sqrt{|\bsl{p}|^2+  \frac{16 M_1^4 \E^4}{U_1^2}} 
}
and at $\Gamma_{\text{M}}$ as 
\eqa{
 &E^{\text{K-IVC},\nu =0}_{\Gamma_{\text{M}}} = -\sum_{\bsl{k}}^{|\bsl{k}|<\Lambda_c} \left[\sqrt{U_1^2 + 16 (|v'_{\star}\bsl{p}|-|\widetilde{\gamma}|)^2}\right.\\
& \quad \left.+\sqrt{U_1^2 + 16 (|v'_{\star}\bsl{p}|+|\widetilde{\gamma}|)^2}\right] \\
& E^{\text{Ch},\nu =0}_{\Gamma_{\text{M}}}   = -\sum_{\bsl{k}}^{|\bsl{k}|<\Lambda_c} \sum_{z=\pm}\left[\sqrt{U_1^2 + 16  |v'_{\star}\bsl{p}|^2 }+ z\sqrt{U_1^2 + 16  \widetilde{\gamma}^2 } \right]\ ,
}
where $\E$ is treated perturbatively.
(See \appref{app:E_perturbatively} for details.)
Then, the total effective energies are
\eqa{
& E^{\text{K-IVC},\nu =0}_{eff} = E^{\text{K-IVC},\nu =0}_{\Gamma_{\text{M}}} + E^{\text{K-IVC},\nu =0}_{\pm\KM} \\
& E^{\text{Ch},\nu =0}_{eff} = E^{\text{Ch},\nu =0}_{\Gamma_{\text{M}}} + E^{\text{Ch},\nu =0}_{\pm\KM}\ .
}
As elaborated in \appref{app:E_perturbatively}, at $\E=0$, we have $E^{\text{K-IVC},\nu =0}_{eff}< E^{\text{Ch},\nu =0}_{eff} $ since $E^{\text{K-IVC},\nu =0}_{\pm\KM}= E^{\text{Ch},\nu =0}_{\pm\KM}$ and $E^{\text{K-IVC},\nu =0}_{\Gamma_{\text{M}}}<E^{\text{Ch},\nu =0}_{\Gamma_{\text{M}}}$.
Moreover, at $\E=\E_c$ ($\approx 294.816$meV in EUS) that satisfies $\gamma+B_\gamma \E_c^2=0$, we have $E^{\text{K-IVC},\nu =0}_{eff} > E^{\text{Ch},\nu =0}_{eff}$ since $E^{\text{K-IVC},\nu =0}_{\pm\KM} > E^{\text{Ch},\nu =0}_{\pm\KM} $ and $E^{\text{K-IVC},\nu =0}_{\Gamma_{\text{M}}} = E^{\text{Ch},\nu =0}_{\Gamma_{\text{M}}}$, demonstrating the existence of the transition (as increasing $\E$ from $\E=0$ to $\E=\E_c$).

Combining the low-$\E$ with the high-$\E$ picture, we arrive at the following picture. 
At low $\E$, $\Gamma_{\text{M}}$ dominates and favors nonzero intervalley coherence.
At high $\E$, $\pm\KM$ dominate and favor zero intervalley coherence, and the secondary $\Gamma_{\text{M}}$ effect picks out specific states among all states without inter-valley coherence.

Now we turn to the second question: why the numerically found high-$\E$ low-energy states have competing energies.
We answer this question by showing that those high-$\E$ states have exactly the same Hartree-Fock energies at the one-shot level.
At the one-shot level, we find (\appref{app:one-shot_sym_energy}) that the Hartree-Fock Hamiltonian is block diagonalzied in spin and valley for all the high-$\E$ ground states for $\nu=0,-1,-2$.
Interestingly, the one-shot Hartree-Fock Hamiltnoians for different types of states are related by performing spinless version of $C_{2}\TR$, noted as $C_2\overline{\TR}$, or spinless TR symmetries on certain blocks.
Taking VH and Chern states at $\nu=0$ as an example, we have
\eq{
H^{A,OS}= H^{A,OS}_{+,\uparrow}+H^{A,OS}_{+,\downarrow}+H^{A,OS}_{-,\uparrow}+H^{A,OS}_{-,\downarrow} - E_0^{OS}\ ,
}
where $A=$VH and Chern, and $OS$ is short for one-shot, $E_0^{OS}$ depends on the initial states only through the filling, and $H^{\Chern,OS}$ is related to $H^{\VH,OS}$ as
\eqa{
H^{\Chern,OS} & = H^{\VH,OS}_{+,\uparrow}+H^{\VH,OS}_{+,\downarrow}+ C_2\overline{\TR} H^{\VH,OS}_{-,\uparrow}(C_2\overline{\TR})^{-1} \\
& \quad + C_2\overline{\TR} H^{\VH,OS}_{-,\downarrow} (C_2\overline{\TR})^{-1}\ .
}
Therefore, the one-shot Hartree-Fock energies are exactly the same for the high-$\E$ ground states with the same filling.
(See \appref{app:one-shot_sym_energy} for more details.)

Before concluding the paper, we compare and contrast our analytic discussion to those in \refcite{Sachdev20210603TSTGInt,Vishwanath20211122AMTG,Efetov20220414TSTGExpAndTheory}
Instead of choosing the $f-c-d$ basis in our work, \refcite{Sachdev20210603TSTGInt,Vishwanath20211122AMTG,Efetov20220414TSTGExpAndTheory} chose the TBG nearly-flat bands and the Dirac cones as basis for the analytical discussions.
As a result, \refcite{Sachdev20210603TSTGInt,Vishwanath20211122AMTG,Efetov20220414TSTGExpAndTheory} did not give a general simple analytic rule for high-$\E$ states as  \propref{prop:high_field_states} or simple symmetry argument for competing energies as ours, indicating the great simplification brought by our $f-c-d$ model. 
Furthermore, \refcite{Sachdev20210603TSTGInt,Vishwanath20211122AMTG,Efetov20220414TSTGExpAndTheory} do not provide an understanding of the appearance of the transition; the simple picture of the heavy fermion model explains the transition based on the competition between the energies at $\Gamma_{\text{M}}$ and $\pm\KM$ points.

\section{Conclusion and Discussion}
\label{sec:conclusion}

In conclusion, we construct an effective heavy fermionic $f-c-d$ model for MATSTG with localized heavy $f$ modes and itinerant $c$ and $d$ modes.
Our $f-c-d$ model can reproduce the previously-obtained single-particle band structure of MATSTG in the energy window $[-50 \meV, 50 \meV]$ and for displacement field $\E\in[0,300]$meV in EUS.
Our $f-c-d$ model can also reproduce and refine the previous numerical Hartree-Fock results for $\nu=0,-1,-2$.
Remarkably, based on our $f-c-d$ model at $\nu=0,-1,-2$, we propose a simple analytical rule for the high-$\E$ ground states, which explains the general loss of intervalley coherence observed in numerical results, and we find analytical symmetry arguments that explain the completing energies of the nearly-degenerate high-$\E$ ground states.

For experiments, we predict that at charge neutrality and high displacement fields, Chern gaps for $\ch=\pm 1,\pm 2$ can be observed by scanning tunneling microscope in the presence of an out-of-plane magnetic field.
In particular, we predict that $\ch=\pm 2$ gaps should be most pronounced, since our arguments in \appref{app:stablize_Chern} show that the orbital effect of the magnetic field can lower the energy of the Chern states.
More specifically, the projection of the orbital effect of the magnetic field to the two TR-realted Chern states is proportional to $\sigma_z$, which always lowers the energy of one Chern state regardless of the sign of the coefficient.
We leave a more detailed study of such prediction for the future.
Our work both generalizes and puts on firmer footing through analytical reasoning the applicability and the importance of the topological heavy fermion model in naturally explaining the emergence of Coulomb interaction-driven correlated phases in Moir\'e multilayer graphene systems.

\section{Acknowledgement}

J.Y. thanks Yang-Zhi Chou, Seth Davis, Biao Lian, Jay D. Sau, Zhi-Da Song, and Fang Xie for helpful discussions.
This work is supported by the Laboratory of Physical Sciences at the University of Maryland.
J.Y. and B. A. B. were supported by DOE Grant No. DE-SC0016239. 
Additional support was provided by Gordon and Betty Moore Foundation through
the EPiQS Initiative, Grant GBMF11070 and Grant No. GBMF8685 towards the Princeton theory program.

\appendix
\begin{widetext}

\tableofcontents

\section{More Details on the Basis of the Hamiltonian}
\label{app:sym_rep}

In this section, we provide more details on the basis of the Hamiltonian and the furnished symmetry reps.

Let us use $\pm\K$ to label the two graphene valleys with $\K=\frac{4\pi}{3  a_G}(1,0)$ in EUS.
(Recall that EUS is the unit system in which $\AA$ is the length unit and $\meV$ is the energy unit, as discussed at the beginning of \secref{sec:BM_TSTG}.)
Given a single graphene, its electron basis near $\pm \K$ reads
\eq{
c^\dagger_{\pm \K + \bsl{p}, \sigma, s} = \frac{1}{\sqrt{N_{G}}}\sum_{\bsl{R}_G} e^{\ii (\pm \K + \bsl{p})\cdot (\bsl{R}_G + \bsl{\tau}_\sigma) }c^\dagger_{\bsl{R}_G + \bsl{\tau}_\sigma, s}\ ,
}
where $N_G$ is the number of lattice points for graphene, $\bsl{R}_G$ labels the Bravais lattice points of graphene, $\bsl{\tau}_\sigma$ labels the vector for the sublattice, and $c^\dagger_{\bsl{R}_G + \bsl{\tau}_\sigma, s}$ creates an electron with $p_z$ orbital and spin $s$ at $\bsl{R}_G + \bsl{\tau}_\sigma$.
We note that $c$ here is for the electron basis of the original graphene following the notation in \refcite{Song20211110MATBGHF}, not to be confused with the $c$ modes in \eqnref{eq:c_k}.

Now we rotate the graphene by a generic angle $\theta$ counter-clockwisely about the out-of-plane axis (denoted as $C_\theta$) and shift the graphene along the out-of-plane axis by $d_z$, we have 
\eq{
T_{d_z}C_{\theta} c^\dagger_{\bsl{R}_G + \bsl{\tau}_\sigma} C_{\theta}^{-1} T_{d_z}^{-1}= c^\dagger_{d_z, C_{\theta}(\bsl{R}_G + \bsl{\tau}_\sigma)} e^{-\ii\frac{s_z}{2}\theta}
}
and
\eq{
T_{d_z}C_{\theta} c^\dagger_{\pm \K + \bsl{p}, \sigma} C_{\theta}^{-1} T_{d_z}^{-1} = \frac{1}{\sqrt{N_{G}}}\sum_{C_\theta \bsl{R}_G} e^{\ii (\pm C_\theta\K + C_\theta\bsl{p})\cdot (C_\theta\bsl{R}_G + C_\theta\bsl{\tau}_\sigma) } c^\dagger_{d_z, C_{\theta}\bsl{R}_G + C_\theta \bsl{\tau}_\sigma} e^{-\ii\frac{s_z}{2}\theta}\ ,
}
where we define
\eqa{
& c^\dagger_{\bsl{R}_G + \bsl{\tau}_\sigma} = (c^\dagger_{\bsl{R}_G + \bsl{\tau}_\sigma, \uparrow} , c^\dagger_{\bsl{R}_G + \bsl{\tau}_\sigma, \downarrow}) \\
& c^\dagger_{\pm \K + \bsl{p}, \sigma} = (c^\dagger_{\pm \K + \bsl{p}, \sigma, \uparrow} , c^\dagger_{\pm \K + \bsl{p}, \sigma, \downarrow}) \ .
}
We can then define 
\eqa{
& c^\dagger_{d_z,\theta, \pm C_\theta\K + \bsl{p}, \sigma,s} = \frac{1}{\sqrt{N_{G}}}\sum_{C_\theta \bsl{R}_G} e^{\ii (\pm C_\theta\K + \bsl{p})\cdot (C_\theta\bsl{R}_G + C_\theta\bsl{\tau}_\sigma) } c^\dagger_{d_z, C_{\theta}\bsl{R}_G + C_\theta \bsl{\tau}_\sigma,s} \\
&
c^\dagger_{d_z,\theta, \pm C_\theta\K + \bsl{p}}  = (c^\dagger_{d_z,\theta, \pm C_\theta\K + \bsl{p}, A,\uparrow}  , c^\dagger_{d_z,\theta, \pm C_\theta\K + \bsl{p}, A,\downarrow}, c^\dagger_{d_z,\theta, \pm C_\theta\K + \bsl{p}, B,\uparrow}  , c^\dagger_{d_z,\theta, \pm C_\theta\K + \bsl{p}, B ,\downarrow} )
}
which gives 
\eq{
\label{eq:psi_theta}
c^\dagger_{d_z,\theta, \pm C_\theta\K + C_\theta\bsl{p}} = T_{d_z}C_{\theta} c^\dagger_{\pm \K + \bsl{p}} C_{\theta}^{-1} T_{d_z}^{-1} \sigma_0 e^{\ii\frac{s_z}{2}\theta}\ .
}
Based on \eqnref{eq:psi_theta}, we clearly see that the symmetry reps of $C_3$, $C_2$ and $\TR$ symmetries are the same for $c^\dagger_{d_z,\theta, \pm C_\theta\K + \bsl{p}, \sigma, s}$ and $c^\dagger_{0,0, \pm \K + \bsl{p}, \sigma ,s}$, since $C_3$, $C_2$ and $\TR$ commutes with $C_{\theta}$ and $T_{d_z}$.
The symmetry reps of $m_z$ are also the same, except that $d_z$ is flipped by $m_z$ in $c^\dagger_{d_z, \theta, \pm C_\theta\K + \bsl{p}, \sigma,s }$.
Specifically, we have
\eqa{
\label{eq:sym_reps_varphi}
& C_3 c^\dagger_{d_z,\theta, \pm C_\theta\K + \bsl{p}} C_3^{-1} =  c^\dagger_{d_z,\theta, \pm C_\theta\K + C_3\bsl{p}} e^{\pm \ii \frac{2\pi}{3} \sigma_z} s_0 \\
& C_2 c^\dagger_{d_z,\theta, \pm C_\theta\K + \bsl{p}} C_2^{-1} =  c^\dagger_{d_z,\theta, \mp C_\theta\K - \bsl{p}} \sigma_x s_0 \\
& \TR c^\dagger_{d_z,\theta, \pm C_\theta\K + \bsl{p}} \TR^{-1} =  c^\dagger_{d_z,\theta, \mp C_\theta\K - \bsl{p}} \sigma_0 \ii s_y \\
& m_z c^\dagger_{d_z,\theta, \pm C_\theta\K + \bsl{p}} m_z^{-1} =  c^\dagger_{-d_z,\theta, \pm C_\theta\K + \bsl{p}} (-\sigma_0 s_0)\ .
}
(Recall that $C_3$, $C_2$ and $m_z$ are defined to be spinless operators.)
The lattice translations for $c^\dagger_{d_z,\theta, \pm C_\theta\K + \bsl{p}, \sigma}$ now becomes 
\eq{
T_{C_\theta\bsl{R}_G} c^\dagger_{d_z,\theta, \pm C_\theta\K + \bsl{p}} T_{C_\theta\bsl{R}_G}^{-1} = c^\dagger_{d_z,\theta, \pm C_\theta\K + \bsl{p}} e^{-\ii (\pm C_\theta\K + \bsl{p}).C_\theta\bsl{R}_G}\ .
}

Now we take the continuum limit, \ie, treating the graphene lattice as a continuous media. 
Then, $C_\theta\bsl{R}_G \rightarrow \bsl{r}$ with $\bsl{r}$ taking continuous values in $\dsR^2$, $\pm C_\theta\K$ and $\sigma$ become internal degrees of freedom, and $\bsl{p}$ now also takes values in $\dsR^2$.
Specifically, we have
\eq{
c^\dagger_{d_z,\theta, \pm C_\theta\K + \bsl{p}, \sigma, s} \rightarrow c^\dagger_{d_z,\theta, \pm C_\theta\K , \bsl{p}, \sigma, s} \ .
}
Symmetric reps of $c^\dagger_{d_z,\theta, \pm C_\theta\K , \bsl{p}, \sigma, s}$ and $c^\dagger_{d_z,\theta, \pm C_\theta\K + \bsl{p}, \sigma, s}$ are exactly the same as \eqnref{eq:sym_reps_varphi} for $C_3$, $C_2$, $\TR$ and $m_z$.
The translation operation of $c^\dagger_{d_z,\theta, \pm C_\theta\K , \bsl{p}, \sigma, s}$ becomes continuous as
\eq{
T_{\bsl{r}} c^\dagger_{d_z,\theta, \pm C_\theta\K , \bsl{p}, \sigma, s} T_{\bsl{r}}^{-1} = c^\dagger_{d_z,\theta, \pm C_\theta\K , \bsl{p}, \sigma, s} e^{-\ii (\pm C_\theta\K + \bsl{p})\cdot \bsl{r}}\ .
}
$\psi^\dagger_{\eta,\bsl{r},l,\sigma,s}$ in \eqnref{eq:H0+_r} is defined as
\eq{
\psi^\dagger_{\eta,\bsl{r},l,\sigma,s} = c^\dagger_{d_{z,l},\theta_l, \eta C_{\theta_l}\K , \bsl{r}, \sigma, s}\ ,
}
where $\theta_l$ is the twist angle for the $l$th layer, $d_{z,l}$ is the position of the $l$th layer along the out-of-plane axis, and 
\eq{
c^\dagger_{d_z,\theta, \pm C_\theta\K , \bsl{r}, \sigma, s} = \frac{1}{\sqrt{\mathcal{A}}}\sum_{\bsl{p}} e^{-\ii \bsl{r}\cdot\bsl{p}} c^\dagger_{d_z,\theta, \pm C_\theta\K , \bsl{p}, \sigma, s}\ .
}
Then, we have 
\eq{
T_{\bsl{r}_0} \psi^\dagger_{\eta,\bsl{r},l,\sigma,s} T_{\bsl{r}_0}^{-1} = \psi^\dagger_{\eta,\bsl{r}+\bsl{r}_0,l,\sigma,s} e^{-\ii \eta (C_{\theta_l}\K)\cdot \bsl{r}_0}\ .
}
Eventually, based on \eqnref{eq:psi_tilde_r}, we know that $\bsl{K}_{t/b}$ in \eqnref{eq:psi_tilder_r_translation} are determined by $\bsl{K}_{t} = C_{-\theta/2} \K$ and $\bsl{K}_{b} = C_{\theta/2}\K$, since we choose $\theta_1=\theta_3 = -\theta/2$ and $\theta_2 = \theta/2$.

We define
\eq{
\psi^\dagger_{\eta,\bsl{r},l} = (\psi^\dagger_{\eta,\bsl{r},l,A,\uparrow},\psi^\dagger_{\eta,\bsl{r},l,A,\downarrow},\psi^\dagger_{\eta,\bsl{r},l,B,\uparrow},\psi^\dagger_{\eta,\bsl{r},l,B,\downarrow})\ ,
}
and define $\widetilde{\psi}_{\eta,\bsl{r},\widetilde{l}}^\dagger$ and $d_{\eta,\bsl{r}}^\dagger$ by \eqnref{eq:psi_tilde_r}-\eqref{eq:d_r}.
Then, $C_3$ is represented as
\eqa{
& C_3 \widetilde{\psi}_{\eta,\bsl{r},\widetilde{l}}^\dagger C_3^{-1} = \widetilde{\psi}_{\eta,C_3\bsl{r},\widetilde{l}}^\dagger\ e^{\eta\ii \sigma_z \frac{2\pi}{3}} s_0 \\
& C_3 d_{\eta,\bsl{r}}^\dagger C_3^{-1} = d_{\eta,C_3\bsl{r}}^\dagger\ e^{\eta\ii \sigma_z \frac{2\pi}{3}} s_0\ ;
}
$C_2\TR$ is represented as
\eqa{
& C_2\TR \widetilde{\psi}_{\eta,\bsl{r},\widetilde{l}}^\dagger (C_2\TR)^{-1} = \widetilde{\psi}_{\eta,-\bsl{r},\widetilde{l}}^\dagger\ \sigma_x \ii s_y \\
& C_2\TR d_{\eta,\bsl{r}}^\dagger (C_2\TR)^{-1} = d_{\eta,-\bsl{r}}^\dagger\ \sigma_x \ii s_y\ ;
}
we can define an effective $C_{2x}$ as
\eqa{
& C_{2x} \widetilde{\psi}_{\eta,\bsl{r},\widetilde{l}}^\dagger (C_{2x})^{-1} = \sum_{\widetilde{l}'}\widetilde{\psi}_{\eta,C_{2x}\bsl{r},\widetilde{l}'}^\dagger  (-\sigma_x) s_0 \mat{ 0 & 1 \\ 1 & 0}_{\widetilde{l}'\widetilde{l}} \\
& C_{2x} d_{\eta,\bsl{r}}^\dagger (C_{2x})^{-1} = d_{-\eta,C_{2x}\bsl{r}}^\dagger\ \sigma_x s_0\ ,
}
we can also define an effective $P$ as
\eqa{
& P \widetilde{\psi}_{\eta,\bsl{r},\widetilde{l}}^\dagger P^{-1} = \sum_{\widetilde{l}'}\widetilde{\psi}_{\eta,-\bsl{r},\widetilde{l}'}^\dagger \eta \mat{ 0 & -1 \\ 1 & 0}_{\widetilde{l}'\widetilde{l}} \\
& P d_{\eta,\bsl{r}}^\dagger P^{-1} = \eta d_{-\eta,-\bsl{r}}^\dagger\ ,
}
then $C_{2x}P$ is represented as
\eqa{
& C_{2x}P \widetilde{\psi}_{\eta,\bsl{r},\widetilde{l}}^\dagger (C_{2x}P)^{-1} = \widetilde{\psi}_{\eta,-C_{2x}\bsl{r},\widetilde{l}}^\dagger\ (-1)^{\widetilde{l}} \eta (-\sigma_x) s_0 \\
& C_{2x}P d_{\eta,\bsl{r}}^\dagger (C_{2x}P)^{-1} = d_{\eta,-C_{2x}\bsl{r}}^\dagger\ \eta \sigma_x s_0
}
with $(-1)^t=-(-1)^b=1$ and $C_{2x}\bsl{r}=(x,-y)^T$; the rep of $T_{\bsl{R}}$ is in \eqnref{eq:psi_tilder_r_translation}; $\TR$ is represented as
\eqa{
& \TR \widetilde{\psi}_{\eta,\bsl{r},\widetilde{l}}^\dagger \TR^{-1} = \widetilde{\psi}_{-\eta,\bsl{r},\widetilde{l}}^\dagger\ \sigma_0 \ii s_y \\
& \TR d_{\eta,\bsl{r}}^\dagger \TR^{-1} = d_{-\eta,\bsl{r}}^\dagger\ \sigma_0 \ii s_y \ ;
}
$\C$ is represented as
\eqa{
& \C \widetilde{\psi}_{\eta,\bsl{r},\widetilde{l}}^\dagger \C^{-1} = \widetilde{\psi}_{\eta,\bsl{r},\widetilde{l}}^T \\
& \C d_{\eta,\bsl{r}}^\dagger \C^{-1} = d_{\eta,\bsl{r}}^T
}
with $\C^2=1$.
The symmetry reps in the momentum space can be naturally obtained by using \eqnref{eq:BM_basis_k_Q}. 

The symmetry properties of $:\rho(\bsl{r}):$ are 
\eq{
\left\{
\begin{array}{l}
\TR :\rho(\bsl{r}): \TR^{-1} = :\rho(\bsl{r}):\\
C_3 :\rho(\bsl{r}): C_3^{-1} = :\rho(C_3\bsl{r}):\\
C_2\TR :\rho(\bsl{r}): (C_2\TR)^{-1} =: \rho(-\bsl{r}):\\
m_z :\rho(\bsl{r}): m_z^{-1} = :\rho(\bsl{r}):\\
T_{\bsl{R}} :\rho(\bsl{r}): T_{\bsl{R}}^{-1} = :\rho(\bsl{r}+\bsl{R}):\\
C_{2x}P :\rho(\bsl{r}): (C_{2x}P)^{-1} = :\rho(-C_{2x}\bsl{r}):\\
\C :\rho(\bsl{r}): (\C)^{-1} = -:\rho(\bsl{r}):
\end{array}
\right.
}
Combined with the fact that $\rho(\bsl{r})$ is invariant under the spin-charge $\U(2)$ in each valley, we have
\eqa{
\label{eq:H_int_sym}
& [\TR, H_{int}]= [C_3, H_{int}] = [C_2\TR, H_{int}] = [m_z,H_{int}]  \\
& = [T_{\bsl{R}}, H_{int}]= [C_{2x}P, H_{int}] = [\C, H_{int}] \\
& = [\U(2)\times\U(2), H_{int}] = 0\ .
}

The symmetry reps of $f$ and $c$ are particularly important for deriving the low-energy effective model.
The relevant high-symmetry points in MBZ for $H_{0,TBG}$ are $\Gamma_{\text{M}}$, $\KM$, and $\text{M}_{\text{M}}$ (shown in \figref{fig:MBZ}).
Based on the origin of the $f$ modes, we know that the symmetry reps of $f$ should carry the symmetry reps of the nearly flat bands at $\KM$ and $\text{M}_{\text{M}}$, and carry one 2D irreducible rep (irrep) of the remote bands at $\Gamma_{\text{M}}$.
According to \eqnref{eq:f_k}, the symmetry rep of $f$ is determined by the form of $\widetilde{v}_{\eta,f,\alpha}(\bsl{k})$.
Then, we require $\widetilde{v}_{\eta,f}$ to guarantee the following the rep of $f$
\eqa{
\label{eq:sym_rep_f}
& C_2 \TR f_{+,\bsl{k}}^\dagger (C_2 \TR)^{-1} = f_{+,\bsl{k}}^\dagger \tau_x \ii s_y \\
& C_3 f_{+,\bsl{k}}^\dagger (C_3)^{-1} = f_{+,C_3 \bsl{k}}^\dagger e^{\ii\tau_z \frac{2\pi}{3}} s_0 \\
& C_{2x} f_{+,\bsl{k}}^\dagger C_{2x}^{-1} = f_{+,C_{2x}\bsl{k}}^\dagger \tau_x  s_0 \\
& P f_{+,\bsl{k}}^\dagger P^{-1} = f_{+,-\bsl{k}}^\dagger \ii \tau_z   s_0 \\
& T_{\bsl{R}} f_{+,\bsl{k}}^\dagger T_{\bsl{R}}^{-1} = f_{+,\bsl{k}}^\dagger e^{ -\ii (\bsl{K}_{b}+\bsl{q}_2+\bsl{k})\cdot \bsl{R}}\\
&  f_{-,\bsl{k}}^\dagger = \TR  f_{+,-\bsl{k}}^\dagger \TR^{-1} \tau_0 (-\ii s_y)\ .
}
The spinlesss parts of the reps here are the same as those of $f$ in \refcite{Song20211110MATBGHF}, except the extra $ e^{ -\ii (\bsl{K}_{b}+\bsl{q}_2)\cdot \bsl{R} }$ factor in the rep of translation, which will be discussed carefully below.
Furthermore, to guarantee the exponential decay of the Wannier functions of $f$ modes, we have to require $\widetilde{v}_{\eta,f}$ to be smooth while keeping
\eqa{
\label{eq:Uf_G-shift}
[\widetilde{v}_{\eta,f,\alpha}(\bsl{k}+\bsl{G})]_{\bsl{Q}\sigma} = [\widetilde{v}_{\eta,f,\alpha}(\bsl{k})]_{\bsl{Q}-\bsl{G}\sigma} \ .
}
The existence of such smooth $\widetilde{v}_{\eta,f}$ is numerically verified in \refcite{Song20211110MATBGHF}.
Note that the 2D irrep carried by $f_{+,\Gamma_{\text{M}}}$ is just the spinless $\Gamma_3$ if we only consider $D_3$ (spanned by $C_3$ and $C_{2x}$)~\cite{BCServer}.

Now we show that the extra $ e^{ -\ii (\bsl{K}_{b}+\bsl{q}_2)\cdot \bsl{R} }$  factor of $f$ modes under Moir\'e lattice translations (shown in \eqnref{eq:sym_rep_f}) can be safely neglected for any values of angle, similar to \refcite{Song20211110MATBGHF}.

First, we show under certain special values of the angles, we can make $\bsl{K}_b+\bsl{q}_2$ a Moir\'e reciprocal lattice vector, and thus $ e^{ -\ii (\bsl{K}_{b}+\bsl{q}_2)\cdot \bsl{R} }$ becomes 1.
Combined with \eqnref{eq:q}, \eqnref{eq:K_tb} and \eqnref{eq:bM} , we have 
\eqa{
 & \bsl{K}_b+\bsl{q}_2  \in \bsl{b}_{M,1} \dsZ + \bsl{b}_{M,2} \dsZ \\
 & \Leftrightarrow  \frac{1}{2}(\cot(\theta/2)-\sqrt{3}, 0 )  \in  \left\{ \left. \left(\frac{\sqrt{3}}{2} n_1 + \sqrt{3} n_2 , -\frac{3}{2} n_1\right) \right| n_1,n_2\in\dsZ \right\} \\
 & \Leftrightarrow  \frac{1}{2} \cot(\theta/2)-  \frac{1}{2}  \sqrt{3}  \in    \sqrt{3} \dsZ \ .
}
Therefore, we can choose $\theta$ to satisfy 
\eq{
\label{eq:theta_condition}
\frac{1}{2}\cot\left( \frac{\theta}{2} \right) - \frac{\sqrt{3}}{2} =0 \mod \sqrt{3}
} 
such that $ e^{ -\ii (\bsl{K}_{b}+\bsl{q}_2)\cdot \bsl{R} }$ becomes 1.

Second, even if $\theta$ does not satisfy \eqnref{eq:theta_condition}, we can define an operation as 
\eq{
 Y_{\bsl{R}} f_{\eta,\bsl{k}}^\dagger Y_{\bsl{R}}^{-1} = f_{\eta,\bsl{k}}^\dagger e^{ \ii \eta (\bsl{K}_{b}+\bsl{q}_2)\cdot \bsl{R}}\ ,
}
where $Y_{\bsl{R}}$ belongs to the valley $U(1)$, which is obeyed by the system.
Then, we can redefine $Y_{\bsl{R}} T_{\bsl{R}}$ as the new lattice translation, which does not has the $e^{ -\ii (\bsl{K}_{b}+\bsl{q}_2)\cdot \bsl{R} }$ factor in \eqnref{eq:sym_rep_f}.
This is what was done in \refcite{Song20211110MATBGHF}.

For the convenience of the derivation in this study, we simply choose $\theta$ to have the value in \eqnref{eq:theta_value}, which approximately satisfies \eqnref{eq:theta_condition}.
Nevertheless, \eqnref{eq:theta_value} is not required for omitting the $e^{ -\ii (\bsl{K}_{b}+\bsl{q}_2)\cdot \bsl{R} }$ factor of the lattice translation in \eqnref{eq:sym_rep_f}.

At $\Gamma_{\text{M}}$, the remote bands have one remaining 2D irrep (also corresponding to $\Gamma_3$ of $D_3$) of the remote bands at $\Gamma_{\text{M}}$, and the near-flat bands have two 1D irreps (corresponding to $\Gamma_1$ and $\Gamma_2$ of $D_3$).
They should be carried by the $c$ modes.
As a result, the reps furnished by the $c$ modes are
\eqa{
\label{eq:sym_rep_c}
& C_2 \TR c_{+,\bsl{k}}^\dagger (C_2 \TR)^{-1} = c_{+,\bsl{k}}^\dagger \mat{ \tau_x & \\  & \tau_x } \ii s_y \\
& C_3 c_{+,\bsl{k}}^\dagger (C_3)^{-1} = c_{+,C_3 \bsl{k}}^\dagger  \mat{ e^{\ii\tau_z \frac{2\pi}{3}} & \\  &  \tau_0 }  s_0 \\
& C_{2x} c_{+,\bsl{k}}^\dagger C_{2x}^{-1} = c_{+,C_{2x}\bsl{k}}^\dagger \mat{ \tau_x & \\  & \tau_x }  s_0 \\
& P c_{+,\bsl{k}}^\dagger P^{-1} = c_{+,-\bsl{k}}^\dagger \mat{ - \ii \tau_z & \\  & - \ii \tau_z }   s_0 \\
& T_{\bsl{R}} c_{+,\bsl{k}}^\dagger T_{\bsl{R}}^{-1} = c_{+,\bsl{k}}^\dagger e^{ -\ii (\bsl{K}_{b}+\bsl{q}_2+\bsl{k})\cdot \bsl{R}}\\
& c_{-,\bsl{k}}^\dagger = \TR  c_{+,-\bsl{k}}^\dagger \TR^{-1} \mathds{1}_4 (-\ii s_y)\ ,
}
where $c_{\eta,\bsl{k}}^\dagger = (c_{\eta,\bsl{k},\Gamma_3}^\dagger, c_{\eta,\bsl{k},\Gamma_1\Gamma_2}^\dagger)$, $c_{\eta,\bsl{k},\Gamma_3}^\dagger = (c_{\eta,\bsl{k},1}^\dagger,c_{\eta,\bsl{k},2}^\dagger)$, and $c_{\eta,\bsl{k},\Gamma_1\Gamma_2}^\dagger = (c_{\eta,\bsl{k},3}^\dagger,c_{\eta,\bsl{k},4}^\dagger)$.
Note that $\tau_{0,x,y,z}$ carries the index $\alpha$ for $f_{\eta,\bsl{k}}^\dagger$ and carries the index $\beta$ for $c_{\eta,\bsl{k},\Gamma_3}^\dagger$ and $c_{\eta,\bsl{k},\Gamma_1\Gamma_2}^\dagger$.
According to \eqnref{eq:c_k}, \eqnref{eq:sym_rep_c} is guaranteed by choosing a special $\widetilde{u}_{\eta,c,\beta}(\bsl{k})$ with $|\bsl{k}|<\Lambda_c$.
Furthermore, in order to guarantee the resultant effective Hamiltonian to have a smooth matrix rep, we need to require $\widetilde{u}_{\eta,c,\beta}(\bsl{k})$ to be smooth.
Such required $\widetilde{u}_{\eta,c,\beta}(\bsl{k})$ always exists for $|\bsl{k}|<\Lambda_c$.
The reason is that $\widetilde{u}_{\eta,c,\beta}(\bsl{k})$ is effectively defined on an open manifold instead of a torus, as we do not impose any relation between $\widetilde{u}_{\eta,c,\beta}(\bsl{k})$ and $\widetilde{u}_{\eta,c,\beta}(\bsl{k}+\bsl{G})$ if both $\bsl{k}$ and $\bsl{k}+\bsl{G}$ have magnitudes smaller than $\Lambda_c$.

We would like the compare the lattice translations of the $f$, $c$ and $d$ modes after considering \eqnref{eq:theta_condition}, which read
\eqa{
\label{eq:f_c_d_translation}
&  T_{\bsl{R}} f_{\eta,\bsl{k}}^\dagger T_{\bsl{R}}^{-1} = f_{\eta,\bsl{k}}^\dagger e^{ -\ii  \bsl{k}\cdot \bsl{R}} \\
& T_{\bsl{R}} c_{\eta,\bsl{k}}^\dagger T_{\bsl{R}}^{-1} = c_{\eta,\bsl{k}}^\dagger e^{ -\ii  \bsl{k} \cdot \bsl{R}}\\
& T_{\bsl{R}} d_{\eta,\bsl{p}}^\dagger T_{\bsl{R}}^{-1} = d_{\eta,\bsl{p}}^\dagger e^{ -\ii (\eta \bsl{K}_{t}+\bsl{p})\cdot \bsl{R}} = d_{\eta,\bsl{p}}^\dagger e^{ -\ii (\eta \bsl{K}_{b}+ \eta \bsl{q}_2+(\bsl{p}+\eta \KM))\cdot \bsl{R}} = d_{\eta,\bsl{p}}^\dagger e^{ -\ii (\bsl{p}+\eta \KM)\cdot \bsl{R}}\ ,
}
where we used \eqnref{eq:sym_rep_f}, \eqnref{eq:sym_rep_c}, \eqnref{eq:q}, \eqnref{eq:psi_tilder_r_translation}-\eqref{eq:psi_tilde_d_p}. and \figref{fig:MBZ}.
According to \eqnref{eq:f_c_d_translation}, $c_{\eta,\bsl{k}}^\dagger$ and $f_{\eta,\bsl{k}}^\dagger$ transforms in the same way under the Moir\'e lattice translations.
It means that $c_{\eta,\bsl{k}}^\dagger$ is around the $\GM$ point of the $f_{\eta,\bsl{k}}^\dagger$ modes for small $\bsl{k}$.
On the other hand, according to \eqnref{eq:f_c_d_translation}, $d_{\eta,\bsl{k}-\eta \KM}^\dagger$ and $f_{\eta,\bsl{k}}^\dagger$ transforms in the same way under the Moir\'e lattice translations, \ie, $\bsl{k}$ in $d_{\eta,\bsl{k}-\eta \KM}^\dagger$ is the same as $\bsl{k}$ in $f_{\eta,\bsl{k}}^\dagger$.
Thus, $d_{\eta,\bsl{p}}^\dagger$ with small $\bsl{p}$ are around the $\eta \KM$ point of the $f$ modes.

Symmetry properties of $w_{\eta\alpha \widetilde{l} \sigma}(\bsl{r})$ (defined in \eqnref{eq:W_Uf}) are listed below.
\eqa{
\label{eq:W_in_f_sym}
& w_{-\alpha \widetilde{l} \sigma}(\bsl{r}) = w_{+\alpha \widetilde{l} \sigma}^*(\bsl{r}) \\
& w_{+1 t A}(C_3^{-1}\bsl{r}) = w_{+1t A}(\bsl{r}) \\
& w_{+1 t B}(C_3^{-1}\bsl{r}) = w_{+1t B}(\bsl{r}) e^{-\ii 2\pi/3} \\
& w_{+1 t \sigma}(\bsl{r}) = -\ii w_{+ 1 t \sigma}^*(C_{2x}^{-1}\bsl{r}) \\
& w_{+1 b \sigma}(\bsl{r}) = - w_{+ 1 t \sigma}^*(-C_{2x}^{-1}\bsl{r}) \\
& w_{+2 \widetilde{l} \sigma}(\bsl{r}) = w_{+ 1 \widetilde{l} \overline{\sigma}}^*(-\bsl{r}) \ ,
}
where $\overline{\sigma}=A/B$ for $\sigma=B/A$.

\section{More Details on the Single-Particle $f-c-d$ Model}
\label{app:fcd_SP}

\subsection{More Details on the $f-d$ Coupling Around $\eta\KM$}
\label{app:f_d}

In this part, we will present more details on the $f-d$ Coupling Around $\eta\KM$.

In general, the $f-d$ coupling reads
\eqa{
H_{0,\eta,fd} = \sum_{\bsl{k}\in\MBZ}\sum_{\bsl{p}}^{|\bsl{p}|<\Lambda_d} f_{\eta,\bsl{k}}^\dagger \widetilde{h}_{\eta,fd}(\bsl{k},\bsl{p})\otimes s_0 d_{\eta,\bsl{p}} + h.c.\ ,
}
where we have used $\U(2)\times \U(2)$ to rule out the inter-valley coupling and the spin-orbit coupling, and $\widetilde{h}_{\eta,fd}(\bsl{k}+\bsl{G},\bsl{p})=\widetilde{h}_{\eta,fd}(\bsl{k},\bsl{p})$ for any Moir\'e reciprocal lattice vector $\bsl{G}$.
$H_{0,\eta,fd}$ preserves the Moir\'e lattice translation $T_{\bsl{R}}$, $C_{2}\TR$, $C_{3}$,  $C_{2x}P$, TR, and the combination of $m_z$ and $\E\rightarrow-\E$.
Then, according to \eqnref{eq:sym_rep_f} and \eqnref{eq:psi_tilder_r_translation}, $T_{\bsl{R}}$ gives
\eqa{
\label{eq:H_0_eta_fd_intermidiate}
 & T_{\bsl{R}} H_{0,\eta,fd} T_{\bsl{R}} ^{-1} = H_{0,\eta,fd} \ \forall \bsl{R} \\
 & \Leftrightarrow \widetilde{h}_{\eta,fd}(\bsl{k},\bsl{p}) e^{-\ii (\eta \bsl{q}_2 + \eta \bsl{q}_3 + \bsl{k} - \bsl{p})\cdot \bsl{R}} = \widetilde{h}_{\eta,fd}(\bsl{k},\bsl{p}) \ \forall \bsl{R} \\
 & \Leftrightarrow \widetilde{h}_{\eta,fd}(\bsl{k},\bsl{p}) = \sum_{\bsl{G}} \delta_{\bsl{p},\bsl{k}-\eta \KM + \bsl{G}} \widetilde{h}_{\eta,fd}(\bsl{k},\bsl{k}-\eta \KM + \bsl{G}) \\
 & \Leftrightarrow H_{0,\eta,fd} = \sum_{\bsl{k}\in\MBZ}\sum_{\bsl{p}}^{|\bsl{p}|<\Lambda_d} f_{\eta,\bsl{k}}^\dagger \sum_{\bsl{G}} \delta_{\bsl{p},\bsl{k}-\eta \KM + \bsl{G}} \widetilde{h}_{\eta,fd}(\bsl{k},\bsl{k}-\eta \KM + \bsl{G})\otimes s_0 d_{\eta,\bsl{p}} + h.c.\\
 & \Leftrightarrow H_{0,\eta,fd} = \sum_{\bsl{p}'}\sum_{\bsl{p}}^{|\bsl{p}|<\Lambda_d} f_{\eta,\bsl{p}'}^\dagger \delta_{\bsl{p},\bsl{p}'-\eta \KM} \widetilde{h}_{\eta,fd}(\bsl{p}',\bsl{p}'-\eta \KM)\otimes s_0 d_{\eta,\bsl{p}} + h.c.\\
 & \Leftrightarrow H_{0,\eta,fd} = \sum_{\bsl{p}}^{|\bsl{p}|<\Lambda_d} f_{\eta,\bsl{p}+\eta\KM}^\dagger h_{\eta,fd}(\bsl{p})\otimes s_0 d_{\eta,\bsl{p}} + h.c.\ ,
}
where $ h_{\eta,fd}(\bsl{p}) =  \widetilde{h}_{\eta,fd}(\bsl{p}+\eta \KM,\bsl{p})$.
Since we are considering the coupling around $\eta \KM$, we only consider $\bsl{p}$ to the first order. 
Then, $C_{2}\TR$ and $C_{3}$ give
\eqa{
& \left\{
\begin{array}{l}
 C_{2}\TR:\ \tau_x h_{+,fd}^*(\bsl{p}) \tau_x =  h_{+,fd}(\bsl{p})\\ 
 C_{3}:\ e^{\ii \tau_z 2\pi/3} h_{+,fd}(\bsl{p}) e^{-\ii \tau_z 2\pi/3} =  h_{+,fd}(C_3\bsl{p})
\end{array}
\right. \\
& \Leftrightarrow 
h_{+,fd}(\bsl{p}) = \widetilde{M}_1 \tau_0 + \ii \widetilde{M}_1'\tau_z + v_1 (p_x + \ii p_y) (\tau_x-\ii \tau_y) + v_1' (p_x - \ii p_y) (\tau_x + \ii \tau_y) + O(p^2)\ ,
}
where $\widetilde{M}_1,\widetilde{M}_1'$ are real and $v_1,v_1'$ are complex.
Furthermore, $C_{2x}P$ gives
\eq{
\tau_y h_{+,fd}(\bsl{p}) \tau_x = - h_{+,fd}(C_{2y}\bsl{p}) \Leftrightarrow \widetilde{M}_1 = \widetilde{M}_1' \ \&\ v_1 = v_1'^* = |v_1| e^{\ii \frac{\pi}{4}}\ .
}
Therefore, we have
\eq{
\label{eq:h_+_fd_intermidiate}
h_{+,fd}(\bsl{p}) = \widetilde{M}_1 (\tau_0 + \ii \tau_z) + v_{fd,1} \left[ e^{\ii \pi/4} (p_x + \ii p_y) (\tau_x-\ii \tau_y) + e^{-\ii \pi/4} (p_x - \ii p_y) (\tau_x + \ii \tau_y) \right] + O(p^2)\ ,
}
where the combination of $m_z$ and $\E\rightarrow-\E$ requires that $\widetilde{M}_1$ and $v_{fd,1}$ are odd in $\E$.
To further simplify $h_{+,fd}(\bsl{p})$, we project $H_{0,\E,+}$ in \eqnref{eq:H_0E_k} to the $f$ and $d$ basis at $\KM$.
Explicitly, we have 
\eqa{
\label{eq:H_0_E_+_proj_KM}
H_{0,\E,+} & = \sum_{\bsl{k}\in\MBZ} \sum_{\bsl{Q}\in \Q_+} \sum_{\sigma,s} \frac{\E}{2}  \widetilde{\psi}^\dagger_{+,\bsl{k},\bsl{Q},\sigma,s} d_{+,\bsl{k},\bsl{Q},\sigma,s} +h.c. \\
& = \sum_{\bsl{k}\in\MBZ} \sum_{\bsl{Q}\in \Q_+} \sum_{\sigma,s} \frac{\E}{2}  \sum_{\alpha} f^\dagger_{+,\bsl{k},\alpha,s} [\widetilde{v}_{+,f,\alpha}(\bsl{k})]_{\bsl{Q}\sigma}^* d_{+,\bsl{k},\bsl{Q},\sigma,s} +h.c. + ...\\
& = \sum_{\bsl{p}}^{|\bsl{p}|<\Lambda_d} \sum_{\sigma,s} \frac{\E}{2}  \sum_{\alpha} f^\dagger_{+,\bsl{p}+\KM,\alpha,s} \sum_{\bsl{Q}\in \Q_+} [\widetilde{v}_{+,f,\alpha}(\bsl{p}+\KM)]_{\bsl{Q}\sigma}^* d_{+,\bsl{p}+\KM-\bsl{Q},\sigma,s} +h.c. + ...\\
& = \sum_{\bsl{p}}^{|\bsl{p}|<\Lambda_d} \sum_{\sigma,s}  \sum_{\alpha} f^\dagger_{+,\bsl{p}+\KM,\alpha,s} \frac{\E}{2}  [\widetilde{v}_{+,f,\alpha}(\bsl{p}+\KM)]_{\bsl{Q}=\KM,\sigma}^* d_{+,\bsl{p},\sigma,s} +h.c. + ...\ ,
}
where we have used \eqnref{eq:psi_tilde_f_c_k} for the second equality, and $...$ contains high-energy modes.
By comparing \eqnref{eq:H_0_E_+_proj_KM} to \eqnref{eq:H_0_eta_fd_intermidiate}, we arrive at
\eq{
\label{eq:H_0_E_+_proj_KM_1}
\left[h_{+,fd}(\bsl{p}) \right]_{\alpha\sigma}= \frac{\E}{2} [\widetilde{v}_{+,f,\alpha}(\bsl{p}+\KM)]_{\bsl{Q}=\KM,\sigma}^* \approx \frac{\E}{2} [\widetilde{v}_{+,f,\alpha}(\KM)]_{\bsl{Q}=\KM,\sigma}^* \ ,
}
where we neglect the momentum dependence of $\widetilde{v}_{\eta,f,\alpha}(\bsl{k})$ in the last step since \refcite{Song20211110MATBGHF} shows the momentum dependence of $\widetilde{v}_{\eta,f,\alpha}(\bsl{k})$  should be small as the $f$ modes have small Wannier spread and have Wannier center at 1a position.
Owing to \eqnref{eq:H_0_E_+_proj_KM_1} and \eqnref{eq:h_+_fd_intermidiate}, we get 
\eqa{
& \widetilde{M}_1 = M_1 \E \\
& v_{fd,1} = 0 
}
with the value of $M_1$ in \tabref{tab:f-c-d_values}.
Therefore, combined with the $\TR$ symmetry and the extra exponential decay factor, we have \eqnref{eq:H_0_fd} as the leading-order term of the $f-d$ coupling around $\eta \KM$.

\subsection{More details on $f-d$ and $c-d$ Couplings around $\Gamma_{\text{M}}$}
\label{app:fd_cd}

In this part, we provide more details on the how we project out the $f-d$ and $c-d$ couplings around $\GM$. 
We will focus on the $+$ valley, since the Hamiltonian at the $-$ valley can be obtained by the TR symmtry.

According to \eqnref{eq:H_0D_k}, the $d$ modes with lowest energies at $\GM$ in the $+$ valley are $d^\dagger_{+,0,\bsl{q}_1}$, $d^\dagger_{+,0,\bsl{q}_2}$ and $d^\dagger_{+,0,\bsl{q}_3}$, which gives energies $\pm 1$ owing to \eqnref{eq:q}.
Then, we consider the following $f-d$ and $c-d$ couplings around $\Gamma$ 
\eq{
\label{eq:fc-d_Gamma}
\mat{  c^\dagger_{+,\bsl{k},\Gamma_3} & c^\dagger_{+,\bsl{k},\Gamma_1\Gamma2} & f^\dagger_{+,\bsl{k}}} 
\mat{ h_{+,c-d,\Gamma} \\ h_{+,f-d,\Gamma}  } \otimes s_0
\mat{ d_{+,\bsl{k},\bsl{q}_1} \\ d_{+,\bsl{k},\bsl{q}_2} \\ d_{+,\bsl{k},\bsl{q}_3} }\ ,
}
where $f$ and $c$ are defined in \eqnref{eq:f_k} and \eqnref{eq:c_k}, respectively, $h_{+,f-d,\Gamma}$ is a $2\times 6$ matrix, and $h_{+,c-d,\Gamma}$ is a $4\times 6$ matrix.
Here we neglect the momentum dependence of the $f-d$ and $c-d$ coupling away from $\GM$ since the matrix rep of \eqnref{eq:H_0E_k} is momentum independent.
To obtain the forms of $h_{+,f-d,\Gamma}$ and $h_{+,c-d,\Gamma}$, we use \eqnref{eq:psi_tilde_f_c_k} to project \eqnref{eq:H_0E_k} to the low-energy modes around $\GM$:
\eqa{
\label{eq:H_0_E_+_proj_GM}
H_{0,\E,+} & = \sum_{\bsl{k}\in\MBZ} \sum_{\bsl{Q}\in \Q_+} \sum_{\sigma,s} \frac{\E}{2}  \widetilde{\psi}^\dagger_{+,\bsl{k},\bsl{Q},\sigma,s} d_{+,\bsl{k},\bsl{Q},\sigma,s} +h.c. \\
& = \sum_{\bsl{k}}^{|\bsl{k}|<\Lambda_c} \sum_{\bsl{Q}\in \Q_+} \sum_{\sigma,s} \frac{\E}{2}  \left(\sum_{\alpha} f^\dagger_{+,\bsl{k},\alpha,s} [\widetilde{v}_{\eta,f,\alpha}(\bsl{k})]_{\bsl{Q}\sigma}^* d_{+,\bsl{k},\bsl{Q},\sigma,s} + \sum_{\beta} c^\dagger_{+,\bsl{k},\beta,s} [\widetilde{u}_{\eta,f,\beta}(\bsl{k})]_{\bsl{Q}\sigma}^* d_{+,\bsl{k},\bsl{Q},\sigma,s}\right) +h.c. + ...\\
& = \sum_{\bsl{k}}^{|\bsl{k}|<\Lambda_c} \sum_{\bsl{Q} = \bsl{q}_1, \bsl{q}_3 ,\bsl{q}_3} \sum_{\sigma,s}   \left(\sum_{\alpha} f^\dagger_{+,\bsl{k},\alpha,s} \frac{\E}{2} [\widetilde{v}_{\eta,f,\alpha}(\bsl{k})]_{\bsl{Q}\sigma}^* d_{+,\bsl{k},\bsl{Q},\sigma,s} + \sum_{\beta} c^\dagger_{+,\bsl{k},\beta,s} \frac{\E}{2} [\widetilde{u}_{\eta,f,\beta}(\bsl{k})]_{\bsl{Q}\sigma}^* d_{+,\bsl{k},\bsl{Q},\sigma,s}\right) +h.c. + ...\\
& = \sum_{\bsl{k}}^{|\bsl{k}|<\Lambda_c} \sum_{\bsl{Q} = \bsl{q}_1, \bsl{q}_3 ,\bsl{q}_3} \sum_{\sigma,s}   \left(\sum_{\alpha} f^\dagger_{+,\bsl{k},\alpha,s} \frac{\E}{2} [\widetilde{v}_{\eta,f,\alpha}(0)]_{\bsl{Q}\sigma}^* d_{+,\bsl{k},\bsl{Q},\sigma,s} + \sum_{\beta} c^\dagger_{+,\bsl{k},\beta,s} \frac{\E}{2} [\widetilde{u}_{\eta,f,\beta}(0)]_{\bsl{Q}\sigma}^* d_{+,\bsl{k},\bsl{Q},\sigma,s}\right) +h.c. + ...\ ,
}
where we neglect the momentum dependence of $\widetilde{v}_{\eta,f,\alpha}(\bsl{k})$ and $\widetilde{u}_{\eta,f,\beta}(\bsl{k})$ again, and $...$ contains high-energy modes.
By comparing \eqnref{eq:H_0_E_+_proj_GM} to \eqnref{eq:fc-d_Gamma}, we can get the forms of $h_{+,f-d,\Gamma}$ and $h_{+,c-d,\Gamma}$, which read 
\eq{
h_{+,f-d,\Gamma} = \frac{\E}{2} \mat{ [\widetilde{v}_{+,f,1}(0)]_{\bsl{q}_1,A}& [\widetilde{v}_{+,f,1}(0)]_{\bsl{q}_1,B} & [\widetilde{v}_{+,f,1}(0)]_{\bsl{q}_2,A}& [\widetilde{v}_{+,f,1}(0)]_{\bsl{q}_2,B} & [\widetilde{v}_{+,f,1}(0)]_{\bsl{q}_3,A}& [\widetilde{v}_{+,f,1}(0)]_{\bsl{q}_3,B} \\
[\widetilde{v}_{+,f,2}(0)]_{\bsl{q}_1,A}& [\widetilde{v}_{+,f,2}(0)]_{\bsl{q}_1,B} & [\widetilde{v}_{+,f,2}(0)]_{\bsl{q}_2,A}& [\widetilde{v}_{+,f,2}(0)]_{\bsl{q}_2,B} & [\widetilde{v}_{+,f,2}(0)]_{\bsl{q}_3,A}& [\widetilde{v}_{+,f,2}(0)]_{\bsl{q}_3,B} 
}^*
}
and 
\eq{
h_{+,c-d,\Gamma} = \frac{\E}{2} \mat{ [\widetilde{u}_{+,c,1}(0)]_{\bsl{q}_1,A}& [\widetilde{u}_{+,c,1}(0)]_{\bsl{q}_1,B} & [\widetilde{u}_{+,c,1}(0)]_{\bsl{q}_2,A}& [\widetilde{u}_{+,c,1}(0)]_{\bsl{q}_2,B} & [\widetilde{u}_{+,c,1}(0)]_{\bsl{q}_3,A}& [\widetilde{u}_{+,c,1}(0)]_{\bsl{q}_3,B} \\
[\widetilde{u}_{+,c,2}(0)]_{\bsl{q}_1,A}& [\widetilde{u}_{+,c,2}(0)]_{\bsl{q}_1,B} & [\widetilde{u}_{+,c,2}(0)]_{\bsl{q}_2,A}& [\widetilde{u}_{+,c,2}(0)]_{\bsl{q}_2,B} & [\widetilde{u}_{+,c,2}(0)]_{\bsl{q}_3,A}& [\widetilde{u}_{+,c,2}(0)]_{\bsl{q}_3,B} \\
[\widetilde{u}_{+,c,3}(0)]_{\bsl{q}_1,A}& [\widetilde{u}_{+,c,3}(0)]_{\bsl{q}_1,B} & [\widetilde{u}_{+,c,3}(0)]_{\bsl{q}_2,A}& [\widetilde{u}_{+,c,3}(0)]_{\bsl{q}_2,B} & [\widetilde{u}_{+,c,3}(0)]_{\bsl{q}_3,A}& [\widetilde{u}_{+,c,3}(0)]_{\bsl{q}_3,B} \\
[\widetilde{u}_{+,c,4}(0)]_{\bsl{q}_1,A}& [\widetilde{u}_{+,c,4}(0)]_{\bsl{q}_1,B} & [\widetilde{u}_{+,c,4}(0)]_{\bsl{q}_2,A}& [\widetilde{u}_{+,c,4}(0)]_{\bsl{q}_2,B} & [\widetilde{u}_{+,c,4}(0)]_{\bsl{q}_3,A}& [\widetilde{u}_{+,c,4}(0)]_{\bsl{q}_3,B} 
}^*\ .
}

On the other hand, according to \eqnref{eq:H_0_f}, \eqnref{eq:H_0_c} and \eqnref{eq:H_0_fc}, the $f$ and $c$ block without the $f-d$ and $c-d$ corrections (\ie, the $f$ and $c$ block that comes from $H_{0,TBG}$ in \eqnref{eq:H_0TBG_k}) reads 
\eq{
 \mat{   c^\dagger_{+,\bsl{k},\Gamma_3} & c^\dagger_{+,\bsl{k},\Gamma_1\Gamma2} & f^\dagger_{+,\bsl{k}}}  h_{+,0}(\bsl{k}) \otimes s_0\mat{  c_{+,\bsl{k},\Gamma_3} \\ c_{+,\bsl{k},\Gamma_1\Gamma2} \\ f_{+,\bsl{k}} }
}
with 
\eq{
  h_{+,0}(\bsl{k}) = \mat{ 
  0_{2\times 2} & v_{\star}( k_x \tau_0 + \ii k_y \tau_z) & \gamma \tau_0 + v_{\star}' ( k_x \tau_x + k_y \tau_y) \\
 v_{\star}( k_x \tau_0 - \ii k_y \tau_z) & M \tau_x & v_{\star}'' ( k_x \tau_x - k_y \tau_y)  \\
  \gamma \tau_0 + v_{\star}' ( k_x \tau_x + k_y \tau_y)   & v_{\star}'' ( k_x \tau_x - k_y \tau_y) & 0_{2\times 2} \\
}\ .
}
Based on \eqnref{eq:H_0D_k}, the Hamiltonian for the low-energy $d$ modes around $\Gamma$ reads
\eq{
\mat{ d^\dagger_{+,\bsl{k},\bsl{q}_1} & d^\dagger_{\bsl{k},\bsl{q}_2} & d^\dagger_{+,\bsl{k},\bsl{q}_3} } h_{+,1}(\bsl{k}) \otimes s_0\mat{ d_{+,\bsl{k},\bsl{q}_1} \\ d_{+,\bsl{k},\bsl{q}_2} \\ d_{+,\bsl{k},\bsl{q}_3} }
}
with
\eq{
h_{+,1}(\bsl{k}) = 
\mat{
 (\bsl{k}-\bsl{q}_1)\cdot\bsl{\sigma} & 0_{2\times 2} & 0_{2\times 2} \\
 0_{2\times 2} & (\bsl{k}-\bsl{q}_2)\cdot\bsl{\sigma} & 0_{2\times 2} \\
 0_{2\times 2} & 0_{2\times 2} & (\bsl{k}-\bsl{q}_3)\cdot\bsl{\sigma} 
}\ .
}

Now we show it is reasonable for us to treat $h_{+,f-d,\Gamma}$ and $h_{+,c-d,\Gamma}$ as perturbations, and then we will project out the $d$ modes around $\Gamma$.
Since $h_{+,f-d,\Gamma}$ and $h_{+,c-d,\Gamma}$ depend linearly on $\E$, we choose $\E=300\meV$ (EUS) and find that the absolute values of the matrix elements of $h_{+,f-d,\Gamma}$ and $h_{+,c-d,\Gamma}$ are no larger than $0.22$, while the gaps between the levels of $h_{+,0}(0)$ and the levels of $h_{+,1}(0)$ are no smaller than $0.87$.
Therefore, we can treat $h_{+,f-d,\Gamma}$ and $h_{+,c-d,\Gamma}$ as perturbations.
Then, according to the second-order perturbation theory, we may project out the $d$ modes.
Explicitly, we have
\eq{
\mat{ h_{+,0}(\bsl{k}) & \mat{ h_{+,c-d,\Gamma} \\ h_{+,f-d,\Gamma} } \\
\mat{  h_{+,c-d,\Gamma} \\ h_{+,f-d,\Gamma} }^\dagger  & h_{+,1}(\bsl{k})} \otimes s_0
}
as the effective Hamiltonian around $\GM$, and by treating $h_{+,f-d,\Gamma}$ and $h_{+,c-d,\Gamma}$ as perturbations, $h_{+,0}(\bsl{k})$ gains the following correction according to the second-order perturbation theory~\cite{Winkler2003SOC}:
\eq{
\label{eq:f-d_c-d_correction}
\mat{ h_{+,c-d,\Gamma} \\ h_{+,f-d,\Gamma} } \frac{1}{h_{+,0}(\bsl{k}) -  h_{+,1}(\bsl{k})} \mat{  h_{+,c-d,\Gamma} \\ h_{+,f-d,\Gamma} }^\dagger\ .
}
To the first order of $\bsl{k}$, \eqnref{eq:f-d_c-d_correction} brings in many different corrections.
By comparing the resultant band structure of the $f-c-d$ model to that of the BM model, we find the most important three corrections are those in \eqnref{eq:TBG_replacement}.
Specifically, the correction to $M$ accounts for the increasing gap of the nearly flat bands at $\GM$ when increasing $\E$, the correction to $\gamma$ accounts for the decreasing gap of the remote bands at $\GM$ when increasing $\E$, and the correction to $v_\star''$ accounts for the change of the band structure along $\GM-\MM$ due to $\E$.
The values of $B$ parameters of \eqnref{eq:TBG_replacement}, which are shown in \tabref{tab:f-c-d_values} are also directly given by \eqnref{eq:f-d_c-d_correction}.

\subsection{More details on the band structure calculation}

In this part, we present more details on how the numerical calculations for \figref{fig:bands_SP} are carried out.
Owing to the exponentially-decayed factor in $H_{0,\eta ,fd}$ and $H_{0,\eta ,fc}$, we are allowed to extend $\Lambda_c$ and $\Lambda_d$ to outside MBZ~\cite{Song20211110MATBGHF}.
Then, we can reexpress $c^\dagger_{\eta,\bsl{p}}$ and $d^\dagger_{\eta,\bsl{p}}$ in \eqnref{eq:H_fcd_SP} as $c^\dagger_{\eta,\bsl{k}-\bsl{G}}$ and $d^\dagger_{\eta,\bsl{k}-\bsl{Q}}$, respectively, where $\bsl{k}\in\MBZ$, $\bsl{G}$ is the reciprocal lattice vector, and $\bsl{Q}\in\Q_\eta$.
In this case, the original definitions of $\Lambda_c$ and $\Lambda_d$ require $|\bsl{k}-\bsl{G}|\leq\Lambda_c$ and $|\bsl{k}-\bsl{Q}|\leq\Lambda_d$.
For the convenience of numerical calculation, we alter the definitions, and instead require $|\bsl{G}|\leq\Lambda_c$ and $|\bsl{Q}|\leq\Lambda_d$, while keeping $\bsl{k}$ running over the entire MBZ.
As a result, the terms in the single-particle $f-c-d$ model in \eqnref{eq:H_fcd_SP} become
\eq{
H_{0,\eta,f}=0\ ,
}
\eq{
H_{0,\eta ,c}=\sum_{\bsl{k}\in \MBZ}\sum_{|\bsl{G}|\leq \Lambda_c} c^\dagger_{\eta,\bsl{k}-\bsl{G}}  
\mat{ 
 0_{2\times 2}  & v_\star(\eta (k_x-G_x) \tau_0 + \ii (k_y-G_y) \tau_z) \\
v_\star (\eta (k_x-G_x) \tau_0 - \ii (k_y-G_y) \tau_z) & (M+B_M \E^2) \tau_x 
} s_0
c_{\eta,\bsl{k}-\bsl{G}}\ ,
}
\eqa{
H_{0,\eta ,fc}  = & \sum_{\bsl{k}\in \MBZ}\sum_{|\bsl{G}|\leq \Lambda_c} 
e^{- \frac{|\bsl{k}-\bsl{G}|^2\lambda^2}{2}} 
\left[f^\dagger_{\eta,\bsl{k}}    
[(\gamma+B_\gamma \E^2) \tau_0 +  v_\star'  (\bsl{k}-\bsl{G})\cdot(\eta  \tau_x , \tau_y)]    s_0  c_{\eta,\bsl{k}-\bsl{G},\Gamma_3}\right. \\
& \left.+f^\dagger_{\eta,\bsl{k}}  (v_\star''+B_{v''} \E^2) (\bsl{k}-\bsl{G})\cdot(\eta  \tau_x , -\tau_y)
 s_0  c_{\eta,\bsl{k}-\bsl{G},\Gamma_1\Gamma_2}\right]
+h.c.
}
with $c_{\eta,\bsl{k}-\bsl{G},\Gamma_3}^\dagger = (c_{\eta,\bsl{k}-\bsl{G},1}^\dagger,c_{\eta,\bsl{k}-\bsl{G},2}^\dagger)$ and $c_{\eta,\bsl{k}-\bsl{G},\Gamma_1\Gamma_2}^\dagger = (c_{\eta,\bsl{k}-\bsl{G},3}^\dagger,c_{\eta,\bsl{k}-\bsl{G},4}^\dagger)$,
\eq{
H_{0,\eta,d}=
\sum_{\bsl{k}\in\MBZ}\sum_{\bsl{Q}\in\Q_\eta}^{|\bsl{Q}|\leq \Lambda_d} d^\dagger_{\eta,\bsl{k}-\bsl{Q}}  (\eta (k_x-Q_x) \sigma_x + (k_y-Q_y) \sigma_y)  s_0\  d_{\eta,\bsl{k}-\bsl{Q}}\ ,
}
and
\eq{
H_{0,\eta ,fd}= \sum_{\bsl{k}\in\MBZ}\sum_{\bsl{Q}\in\Q_\eta}^{|\bsl{Q}|\leq\Lambda_d} e^{- \frac{|\bsl{k}-\bsl{Q}|^2\lambda^2}{2}} f^\dagger_{\eta ,\bsl{k}}\ M_1 \E (\tau_0 + \eta \ii \tau_z)  s_0\  d_{\eta ,\bsl{k}-\bsl{Q}} + h.c.\ .
}
\figref{fig:bands_SP} is plotted by choosing $\Lambda_c=\Lambda_d=2\sqrt{3}$, \eqnref{eq:theta_value} and \tabref{tab:f-c-d_values}.

\section{More details on the interaction among $f$, $c$ and $d$ modes}
\label{app:fcd_int}

In this section, we provide more details on the interaction among $f$, $c$ and $d$ modes, which is derived by projecting the gate-screened Coulomb interaction \eqnref{eq:H_int_r} to the $f$-$c$-$d$ basis.
Throughout this section, we assume $\Lambda_c$ and $\Lambda_d$ to be  small, \ie, $\Lambda_c,\Lambda_d\ll 1$.
Owing to the assumption that $\Lambda_c$ is small, we, just for the convenience, formally define 
\eqa{
\label{eq:Uc_G-shift}
[\widetilde{u}_{\eta,c,\beta}(\bsl{k}+\bsl{G})]_{\bsl{Q}\sigma} = [\widetilde{u}_{\eta,c,\beta}(\bsl{k})]_{\bsl{Q}-\bsl{G}\sigma} \text{ for $\bsl{k}< \Lambda_c$, only when $\Lambda_c\ll 1$}\ ,
}
where $\widetilde{u}_{\eta,c,\beta}$ is defined in \eqnref{eq:c_k}, and $\bsl{G}$ is any Moir\'e reciprocal lattice vector.
Note that if $\Lambda_c$ becomes large such that there exists $\bsl{k}$ and $\bsl{k}+\bsl{G}$ with  $|\bsl{k}|, |\bsl{k}+\bsl{G}|<\Lambda_c$, then  \eqnref{eq:Uc_G-shift} cannot be used anymore, since we want $c^\dagger_{\bsl{k}}$ to be independent from $c^\dagger_{\bsl{k}+\bsl{G}}$.

As the density operator $\rho(\bsl{r})$ can be split into two parts as
\eq{
\rho(\bsl{r}) = \widetilde{\rho}(\bsl{r}) + \rho_D (\bsl{r})
}
with 
\eq{ 
\widetilde{\rho}(\bsl{r}) = \sum_{\eta,\widetilde{l}} \widetilde{\psi}_{\eta,\bsl{r},\widetilde{l}}^\dagger \widetilde{\psi}_{\eta,\bsl{r},\widetilde{l}}\ \text{ for the TBG block and}\ \rho_D (\bsl{r}) = \sum_{\eta}d^\dagger_{\eta,\bsl{r}} d_{\eta,\bsl{r}}\ ,
}
the interaction \eqnref{eq:H_int_r} can be split into three parts as
\eq{
H_{int} = H_{int}^{TBG} + H_{int}^{TBG-D} + H_{int}^{D}\ ,
}
where
\eqa{
\label{eq:H_int_three_parts}
& H_{int}^{TBG} = \frac{1}{2} \int d^2 r d^2 r' V(\bsl{r}-\bsl{r}') :\widetilde{\rho}(\bsl{r}): :\widetilde{\rho}(\bsl{r}'):\\
& H_{int}^{TBG-D} = \int d^2 r d^2 r' V(\bsl{r}-\bsl{r}') :\widetilde{\rho}(\bsl{r}): :\rho_D(\bsl{r}'): \\
& H_{int}^{D} = \frac{1}{2} \int d^2 r d^2 r' V(\bsl{r}-\bsl{r}') :\rho_D(\bsl{r}): :\rho_D(\bsl{r}'): \ .
}

Before discussing each part in \eqnref{eq:H_int_three_parts}, we derive the following expressions for the convenience of the latter evaluation of the the normal ordering,
Based on the choice of $\ket{G_0}$ in \eqnref{eq:G0_ket}, we have
\eq{
\bra{G_0}\psi^\dagger_{\eta,\bsl{p},l,\sigma,s}\psi_{\eta',\bsl{p}',l',\sigma',s'}\ket{G_0} = \frac{1}{2}\delta_{\eta\eta'}\delta_{\bsl{p}\bsl{p}'}\delta_{ll'}\delta_{\sigma\sigma'}\delta_{ss'}\ .
}
Then, combined with \eqnref{eq:psi_tilde_r}, \eqnref{eq:d_r}, \eqnref{eq:psi_tilde_d_p} and \eqnref{eq:BM_basis_k_Q}, we have
\eqa{
\label{eq:G0ave_psi_tilde_d}
& \bra{G_0}\widetilde{\psi}^\dagger_{\eta,\bsl{k},\bsl{Q},\sigma,s}\widetilde{\psi}_{\eta',\bsl{k}',\bsl{Q}',\sigma',s'}\ket{G_0} = \frac{1}{2}\delta_{\eta\eta'}\delta_{\bsl{k}\bsl{k}'}\delta_{\bsl{Q}\bsl{Q}'}\delta_{\sigma\sigma'}\delta_{ss'}\\
& \bra{G_0}\widetilde{\psi}^\dagger_{\eta,\bsl{k},\bsl{Q},\sigma,s}d_{\eta',\bsl{p},\sigma',s'}\ket{G_0} =0 \\
& \bra{G_0}d_{\eta,\bsl{p},\sigma,s}^\dagger d_{\eta',\bsl{p}',\sigma',s'}\ket{G_0} = \frac{1}{2}\delta_{\eta\eta'}\delta_{\bsl{p}\bsl{p}'}\delta_{\sigma\sigma'}\delta_{ss'}\ .
}
Then, combined with \eqnref{eq:f_k} and \eqnref{eq:c_k}, we have
\eqa{
\label{eq:G0ave_f_c_d}
& \bra{G_0}f^\dagger_{\eta,\bsl{k},\alpha,s}f_{\eta',\bsl{k}',\alpha',s'}\ket{G_0} = \frac{1}{2} \delta_{\eta\eta'} \delta_{\bsl{k}\bsl{k}'} \delta_{\alpha\alpha'}\delta{ss'} \\
& \bra{G_0}f^\dagger_{\eta,\bsl{k},\alpha,s}c_{\eta',\bsl{k}',\beta',s'}\ket{G_0} = 0 \\
& \bra{G_0}c^\dagger_{\eta,\bsl{k},\beta,s}c_{\eta',\bsl{k}',\beta',s'}\ket{G_0} = \frac{1}{2} \delta_{\eta\eta'} \delta_{\bsl{k}\bsl{k}'} \delta_{\beta\beta'}\delta{ss'}\\
& \bra{G_0}f^\dagger_{\eta,\bsl{k},\alpha,s}d_{\eta',\bsl{p}',\sigma',s'}\ket{G_0} = 0 \\
& \bra{G_0}c^\dagger_{\eta,\bsl{k},\beta,s}d_{\eta',\bsl{p}',\sigma',s'}\ket{G_0} = 0 \ .
}

In the following, we will discuss how we project the three parts in \eqnref{eq:H_int_three_parts} onto the $f$, $c$ and $d$ modes.
All the numerical evaluations throughout this section are done with the parameter values in \tabref{tab:f-c-d_values} and \eqnref{eq:parameter_values_int}.

\subsection{Review on $H_{int}^{TBG}$}

We discuss $H_{int}^{TBG}$ in \eqnref{eq:H_int_three_parts} first.
Since $H_{int}^{TBG}$ only depends on $\widetilde{\psi}$, its projection onto the $f$ and $c$ modes should have the same form as the interaction terms in the ordinary MATBG as discussed \refcite{Song20211110MATBGHF}.
This subsection is a review of the interaction in \refcite{Song20211110MATBGHF}, except that we use the parameter values for MATSTG.

To do the projection, we first need to figure out the projection of $\widetilde{\psi}_{\eta,\bsl{r},\widetilde{l},\sigma,s}^\dagger$ to $f^\dagger$ and $c^\dagger$.
Combining \eqnref{eq:psi_tilde_f_c_k} with 
\eq{
\widetilde{\psi}_{\eta,\bsl{r},\widetilde{l}}^\dagger 
= \frac{1}{\sqrt{\A}}\sum_{\bsl{p}} e^{-\ii \bsl{p}\cdot\bsl{r}} \widetilde{\psi}_{\eta,\bsl{p},\widetilde{l}}^\dagger 
= \frac{1}{\sqrt{\A}}\sum_{\bsl{k}}^{\MBZ} \sum_{\bsl{Q}\in\Q_{\eta,\widetilde{l}}} e^{-\ii (\bsl{k}-\bsl{Q})\cdot\bsl{r}} \widetilde{\psi}_{\eta,\bsl{k}-\bsl{Q},\widetilde{l}^\eta_{\bsl{Q}}}^\dagger
= \frac{1}{\sqrt{\A}}\sum_{\bsl{k}}^{\MBZ} \sum_{\bsl{Q}\in\Q_{\eta,\widetilde{l}}} e^{-\ii (\bsl{k}-\bsl{Q})\cdot\bsl{r}} \widetilde{\psi}_{\eta,\bsl{k},\bsl{Q}}^\dagger\ ,
}
we have
\eqa{
\widetilde{\psi}_{\eta,\bsl{r},\widetilde{l},\sigma,s}^\dagger 
& = \frac{1}{\sqrt{\A}}\sum_{\bsl{k}}^{\MBZ} \sum_{\bsl{Q}\in\Q_{\eta,\widetilde{l}}} e^{-\ii (\bsl{k}-\bsl{Q})\cdot\bsl{r}} \sum_{\alpha=1,2} f^\dagger_{\eta,\bsl{k},\alpha,s} [\widetilde{v}_{\eta,f,\alpha}(\bsl{k})]_{\bsl{Q}\sigma}^* \\
& \quad + \frac{1}{\sqrt{\A}}\sum_{\bsl{k}}^{|\bsl{k}|<\Lambda_c} \sum_{\bsl{Q}\in\Q_{\eta,\widetilde{l}}} e^{-\ii (\bsl{k}-\bsl{Q})\cdot\bsl{r}} \sum_{\beta=1,...,4} c^\dagger_{\eta,\bsl{k},\beta,s} [\widetilde{u}_{\eta,c,\beta}(\bsl{k})]_{\bsl{Q}\sigma}^*  + ...\ ,
}
where ``$...$" means the high-energy modes.
Combined with \eqnref{eq:f_R_k} and \eqnref{eq:c_r}, we get 
\eqa{
\widetilde{\psi}_{\eta,\bsl{r},\widetilde{l},\sigma,s}^\dagger 
& = \frac{1}{\sqrt{N \A}} \sum_{\alpha=1,2}\sum_{\bsl{R}} f^\dagger_{\eta,\bsl{R},\alpha,s} \sum_{\bsl{k}}^{\MBZ} \sum_{\bsl{Q}\in\Q_{\eta,\widetilde{l}}} e^{-\ii (\bsl{k}-\bsl{Q})\cdot(\bsl{r}-\bsl{R})} e^{\ii \bsl{Q}\cdot\bsl{R}}  [\widetilde{v}_{\eta,f,\alpha}(\bsl{k})]_{\bsl{Q}\sigma}^* \\
& \quad + \sum_{\beta=1,...,4} c^\dagger_{\eta,\bsl{r},\beta,s} \sum_{\bsl{Q}\in\Q_{\eta,\widetilde{l}}} e^{\ii  \bsl{Q}\cdot\bsl{r}}  [\widetilde{u}_{\eta,c,\beta}(\bsl{k})]_{\bsl{Q}\sigma}^*  + ...\ ,
}
where $f^\dagger_{\eta,\bsl{R},\alpha,s}$ is defined in \eqnref{eq:f_R_k}, and $c^\dagger_{\eta,\bsl{r},\beta,s}$ is defined in \eqnref{eq:c_r}.
By defining 
\eq{
\label{eq:g_in_c}
g_{\eta\beta\widetilde{l}\sigma}(\bsl{r}) = \sum_{\bsl{Q}\in\Q_{\eta,\widetilde{l}}} e^{\ii  \bsl{Q}\cdot\bsl{r}}  [\widetilde{u}_{\eta,c,\beta}(0)]_{\bsl{Q}\sigma}
}
and using $\Delta K_{\widetilde{l}}$ defined in \eqnref{eq:Deltak}, we eventually get
\eqa{
\label{eq:psi_tilde_f_c_r}
\widetilde{\psi}_{\eta,\bsl{r},\widetilde{l},\sigma,s}^\dagger 
\approx \sum_{\alpha=1,2}\sum_{\bsl{R}} f^\dagger_{\eta,\bsl{R},\alpha,s} e^{-\ii \eta \Delta K_{\widetilde{l}} \cdot \bsl{R}} w_{\eta\alpha\widetilde{l}\sigma}(\bsl{r}-\bsl{R})^* + \sum_{\beta=1,...,4} c^\dagger_{\eta,\bsl{r},\beta,s} g_{\eta\beta\widetilde{l}\sigma}^*(\bsl{r}) \ ,
}
where we use \eqnref{eq:W_Uf} and
\eq{
\label{eq:DeltaK_R}
e^{-\ii \eta \Delta K_{\widetilde{l}} \cdot \bsl{R}} = e^{\ii \eta (-)^{\widetilde{l}} \bsl{q}_1\cdot \bsl{R}} = e^{\ii \bsl{Q}\cdot \bsl{R}} \text{ for }\bsl{Q}\in \Q_{\eta,\widetilde{l}}\ ,
} 
and $(-)^{t}=-(-)^{b}=1$.
With \eqnref{eq:psi_tilde_f_c_r}, we can derive the projection of $\widetilde{\rho}(\bsl{r})$ to $f$ and $c$ modes, resulting in
\eqa{
\widetilde{\rho}(\bsl{r}) & = \sum_{\eta,\widetilde{l},\sigma,s} \widetilde{\psi}_{\eta,\bsl{r},\widetilde{l},\sigma,s}^\dagger \widetilde{\psi}_{\eta,\bsl{r},\widetilde{l},\sigma,s} \\
& \approx \sum_{\eta, \widetilde{l}, \sigma, s}\left[\sum_{\alpha=1,2}\sum_{\bsl{R}} f^\dagger_{\eta,\bsl{R},\alpha,s} e^{-\ii \eta \Delta K_{\widetilde{l}} \cdot \bsl{R}} w_{\eta\alpha\widetilde{l}\sigma}^*(\bsl{r}-\bsl{R}) + \sum_{\beta=1,...,4} c^\dagger_{\eta,\bsl{r},\beta,s} g_{\eta\beta\widetilde{l}\sigma}^*(\bsl{r})\right]\\
& \quad\times\left[\sum_{\alpha'=1,2}\sum_{\bsl{R}'} f_{\eta,\bsl{R}',\alpha',s} e^{\ii \eta \Delta K_{\widetilde{l}} \cdot \bsl{R}'} w_{\eta\alpha'\widetilde{l}\sigma}(\bsl{r}-\bsl{R}') + \sum_{\beta'=1,...,4} c_{\eta,\bsl{r},\beta',s} g_{\eta\beta'\widetilde{l}\sigma}(\bsl{r})\right]\\
& = \sum_{\eta, \widetilde{l}, \sigma, s}\left\{\sum_{\alpha=1,2}\sum_{\bsl{R}}\sum_{\alpha'=1,2}\sum_{\bsl{R}'} f^\dagger_{\eta,\bsl{R},\alpha,s}f_{\eta,\bsl{R}',\alpha',s} e^{-\ii \eta \Delta K_{\widetilde{l}} \cdot \bsl{R} + \ii \eta \Delta K_{\widetilde{l}} \cdot \bsl{R}'} w_{\eta\alpha\widetilde{l}\sigma}^*(\bsl{r}-\bsl{R}) w_{\eta\alpha'\widetilde{l}\sigma}(\bsl{r}-\bsl{R}')\right. \\
& \qquad +\left[ \sum_{\beta=1,...,4} \sum_{\alpha=1,2}\sum_{\bsl{R}} c^\dagger_{\eta,\bsl{r},\beta,s} g_{\eta\beta\widetilde{l}\sigma}^*(\bsl{r})f_{\eta,\bsl{R},\alpha,s} e^{\ii \eta \Delta K_{\widetilde{l}} \cdot \bsl{R}} w_{\eta\alpha\widetilde{l}\sigma}(\bsl{r}-\bsl{R}) + h.c.\right]\\
&\qquad \left.
+ \sum_{\beta=1,...,4} \sum_{\beta'=1,...,4} c^\dagger_{\eta,\bsl{r},\beta,s} c_{\eta,\bsl{r},\beta',s} g_{\eta\beta\widetilde{l}\sigma}^*(\bsl{r}) g_{\eta\beta'\widetilde{l}\sigma}(\bsl{r})
\right\} \ .
}
At the single particle-level, we mentioned that it is legitimate to neglect the hopping among $f$ modes due to the extreme localization of the Wannier functions, meaning that we can adopt the following approximation
\eq{
\label{eq:W_W_onsite}
w_{\eta\alpha\widetilde{l}\sigma}(\bsl{r}-\bsl{R}) w_{\eta'\alpha'\widetilde{l}'\sigma'}^*(\bsl{r}-\bsl{R}') \approx 0 \text{ if } \bsl{R}\neq \bsl{R}'\ .
}
With this approximation, we have 
\eqa{
\widetilde{\rho}(\bsl{r}) & \approx \sum_{\eta, \widetilde{l}, \sigma, s}\left\{\sum_{\alpha=1,2}\sum_{\bsl{R}}\sum_{\alpha'=1,2} f^\dagger_{\eta,\bsl{R},\alpha,s}f_{\eta,\bsl{R},\alpha',s}  w_{\eta\alpha\widetilde{l}\sigma}^*(\bsl{r}-\bsl{R}) w_{\eta\alpha'\widetilde{l}\sigma}(\bsl{r}-\bsl{R})\right. \\
& \qquad +\left[ \sum_{\beta=1,...,4} \sum_{\alpha=1,2}\sum_{\bsl{R}} c^\dagger_{\eta,\bsl{r},\beta,s} g_{\eta\beta\widetilde{l}\sigma}^*(\bsl{r})f_{\eta,\bsl{R},\alpha,s} e^{\ii \eta \Delta K_{\widetilde{l}} \cdot \bsl{R}} w_{\eta\alpha\widetilde{l}\sigma}(\bsl{r}-\bsl{R}) + h.c.\right]\\
&\qquad \left.
+ \sum_{\beta=1,...,4} \sum_{\beta'=1,...,4} c^\dagger_{\eta,\bsl{r},\beta,s} c_{\eta,\bsl{r},\beta',s} g_{\eta\beta\widetilde{l}\sigma}^*(\bsl{r}) g_{\eta\beta'\widetilde{l}\sigma}(\bsl{r})
\right\} \ .
}
Furthermore, according to the symmetry properties of the Wannier functions \eqnref{eq:W_in_f_sym}, we have 
\eqa{
\label{eq:W_in_f_relations}
& w_{+2\widetilde{l}\sigma}(\bsl{r}) = w^*_{+1\widetilde{l}\sigma}(-\bsl{r})=[(-)^{\widetilde{l}}\ii w_{+1\underline{\widetilde{l}}\overline{\sigma}}(\bsl{r})]^* = (-)^{\underline{\widetilde{l}}}\ii w^*_{+1\underline{\widetilde{l}}\overline{\sigma}}(\bsl{r}) \\ 
& w_{- \alpha \widetilde{l}\sigma}(\bsl{r})=w_{+\alpha \widetilde{l}\sigma}^*(\bsl{r})
}
with $\underline{\widetilde{l}}=t/b$ for $\widetilde{l}=b/t$ and $\overline{\sigma}=A/B$ for $\sigma=B/A$, just as \refcite{Song20211110MATBGHF}.
Then, we have
\eq{
\sum_{\widetilde{l}\sigma} w_{+1\widetilde{l}\sigma}(\bsl{r}-\bsl{R})  w_{+2\widetilde{l}\sigma}^*(\bsl{r}-\bsl{R}) =-\ii \sum_{\widetilde{l}\sigma} w_{+1\widetilde{l}\sigma}(\bsl{r}-\bsl{R}) (-)^{\underline{\widetilde{l}}} w_{+1\underline{\widetilde{l}}\overline{\sigma}}(\bsl{r}-\bsl{R}) = 0\ ,
}
resulting in  
\eq{
\sum_{\widetilde{l}\sigma} w_{\eta \alpha\widetilde{l}\sigma}(\bsl{r}-\bsl{R}) w_{\eta   \alpha' \widetilde{l}\sigma}^*(\bsl{r}-\bsl{R}) = 0 \text{ for } \alpha\neq \alpha' \ .
}
This expression brings simplification to the projection of $\widetilde{\rho}(\bsl{r})$ as
\eqa{
\widetilde{\rho}(\bsl{r}) & \approx \sum_{\eta, \widetilde{l}, \sigma, s}\left\{\sum_{\alpha=1,2}\sum_{\bsl{R}} f^\dagger_{\eta,\bsl{R},\alpha,s}f_{\eta,\bsl{R},\alpha,s}  w_{\eta\alpha\widetilde{l}\sigma}^*(\bsl{r}-\bsl{R}) w_{\eta\alpha\widetilde{l}\sigma}(\bsl{r}-\bsl{R})\right. \\
& \qquad +\left[ \sum_{\beta=1,...,4} \sum_{\alpha=1,2}\sum_{\bsl{R}} c^\dagger_{\eta,\bsl{r},\beta,s} g_{\eta\beta\widetilde{l}\sigma}^*(\bsl{r})f_{\eta,\bsl{R},\alpha,s} e^{\ii \eta \Delta K_{\widetilde{l}} \cdot \bsl{R}} w_{\eta\alpha\widetilde{l}\sigma}(\bsl{r}-\bsl{R}) + h.c.\right]\\
&\qquad \left.
+ \sum_{\beta=1,...,4} \sum_{\beta'=1,...,4} c^\dagger_{\eta,\bsl{r},\beta,s} c_{\eta,\bsl{r},\beta',s} g_{\eta\beta\widetilde{l}\sigma}^*(\bsl{r}) g_{\eta\beta'\widetilde{l}\sigma}(\bsl{r})
\right\} \ .
}
Furthermore, \eqnref{eq:W_in_f_relations} shows that $\sum_{\widetilde{l}\sigma}\left| w_{\eta\alpha\widetilde{l}\sigma}(\bsl{r}) \right|^2$ is independent of $\eta$ and $\alpha$, and then we can define 
\eq{
\label{eq:n_f}
n_f(\bsl{r}) = \sum_{\widetilde{l}\sigma}\left| w_{\eta\alpha\widetilde{l}\sigma}(\bsl{r}) \right|^2\ .
}
Then, we have
\eqa{
\widetilde{\rho}(\bsl{r}) & \approx \sum_{\bsl{R}} \rho_f(\bsl{R}) n_f(\bsl{r}-\bsl{R}) +\sum_{\eta,s}\left[ \sum_{\beta=1,...,4} \sum_{\alpha=1,2}\sum_{\bsl{R}} c^\dagger_{\eta,\bsl{r},\beta,s} f_{\eta,\bsl{R},\alpha,s}  \sum_{\widetilde{l}\sigma} g_{\eta\beta\widetilde{l}\sigma}^*(\bsl{r}) e^{\ii \eta \Delta K_{\widetilde{l}} \cdot \bsl{R}} w_{\eta\alpha\widetilde{l}\sigma}(\bsl{r}-\bsl{R}) + h.c.\right]\\
&\quad 
+ \sum_{\eta,s} \sum_{\beta=1,...,4} \sum_{\beta'=1,...,4} c^\dagger_{\eta,\bsl{r},\beta,s} c_{\eta,\bsl{r},\beta',s}  \sum_{\widetilde{l}\sigma} g_{\eta\beta\widetilde{l}\sigma}^*(\bsl{r}) g_{\eta\beta'\widetilde{l}\sigma}(\bsl{r})
 \ ,
}
where $\rho_f(\bsl{R})$ is defined under \eqnref{eq:H_int_U}.
By further defining
\eqa{
\label{eq:rho_ff_cc_fc_cf}
& \rho_{ff}(\bsl{r})= \sum_{\bsl{R}} \rho_f(\bsl{R}) n_f(\bsl{r}-\bsl{R}) \\
& \rho_{cc}(\bsl{r})= \sum_{\eta,s} \sum_{\beta=1,...,4} \sum_{\beta'=1,...,4} c^\dagger_{\eta,\bsl{r},\beta,s} c_{\eta,\bsl{r},\beta',s}  \sum_{\widetilde{l}\sigma} g_{\eta\beta\widetilde{l}\sigma}^*(\bsl{r}) g_{\eta\beta'\widetilde{l}\sigma}(\bsl{r}) \\
& \rho_{cf}(\bsl{r})= \sum_{\eta,s} \sum_{\beta=1,...,4} \sum_{\alpha=1,2}\sum_{\bsl{R}} c^\dagger_{\eta,\bsl{r},\beta,s} f_{\eta,\bsl{R},\alpha,s} \sum_{\widetilde{l}\sigma} g_{\eta\beta\widetilde{l}\sigma}^*(\bsl{r}) e^{\ii \eta \Delta K_{\widetilde{l}} \cdot \bsl{R}} w_{\eta\alpha\widetilde{l}\sigma}(\bsl{r}-\bsl{R}) \\
& \rho_{fc}(\bsl{r})= \rho_{cf}^\dagger(\bsl{r})\ ,
}
we eventually arrive at
\eq{
\label{eq:rho_tilde_proj}
\widetilde{\rho}(\bsl{r}) \approx \rho_{ff}(\bsl{r}) + \rho_{cc}(\bsl{r}) + \rho_{cf}(\bsl{r}) +\rho_{fc}(\bsl{r}) \ .
}

With \eqnref{eq:rho_tilde_proj}, the expanded $H_{int}^{TBG}$ becomes
\eqa{
\label{eq:H_int_TBG_rho}
H_{int}^{TBG} & \approx \frac{1}{2} \int d^2 r d^2 r' V(\bsl{r}-\bsl{r}') :\rho_{ff}(\bsl{r}): :\rho_{ff}(\bsl{r}'): \\
& \quad + \frac{1}{2} \int d^2 r d^2 r' V(\bsl{r}-\bsl{r}') :\rho_{cc}(\bsl{r}): :\rho_{cc}(\bsl{r}'): \\
& \quad +  \int d^2 r d^2 r' V(\bsl{r}-\bsl{r}') :\rho_{ff}(\bsl{r}): :\rho_{cc}(\bsl{r}'): \\
& \quad + \left[\frac{1}{2} \int d^2 r d^2 r' V(\bsl{r}-\bsl{r}') :\rho_{cf}(\bsl{r}): :\rho_{cf}(\bsl{r}'): +h.c. \right]\\
& \quad + \left[\frac{1}{2} \int d^2 r d^2 r' V(\bsl{r}-\bsl{r}') :\rho_{ff}(\bsl{r}): :\rho_{cf}(\bsl{r}'): +h.c. \right]\\
& \quad + \left[\frac{1}{2} \int d^2 r d^2 r' V(\bsl{r}-\bsl{r}') :\rho_{ff}(\bsl{r}): :\rho_{fc}(\bsl{r}'): +h.c. \right] \\
& \quad + \left[\frac{1}{2} \int d^2 r d^2 r' V(\bsl{r}-\bsl{r}') :\rho_{cc}(\bsl{r}): :\rho_{cf}(\bsl{r}'): +h.c. \right]\\
& \quad + \left[\frac{1}{2} \int d^2 r d^2 r' V(\bsl{r}-\bsl{r}') :\rho_{cc}(\bsl{r}): :\rho_{fc}(\bsl{r}'): +h.c. \right]\\
& \quad + \frac{1}{2} \int d^2 r d^2 r' V(\bsl{r}-\bsl{r}') \left[:\rho_{cf}(\bsl{r}): :\rho_{fc}(\bsl{r}'): + :\rho_{fc}(\bsl{r}): :\rho_{cf}(\bsl{r}'):\right]\ .
}
In the following, we will discuss each term in \eqnref{eq:H_int_TBG_rho}, as were discussed in \refcite{Song20211110MATBGHF}.

\subsubsection{$\frac{1}{2} \int d^2 r d^2 r' V(\bsl{r}-\bsl{r}') :\rho_{ff}(\bsl{r}): :\rho_{ff}(\bsl{r}'): $}

To simplify this term, we first evaluate the Fourier transformation of $n_f(\bsl{r})$ as
\eqa{
\label{eq:nf_p}
n_f(\bsl{p})  = \int d^2 r\ n_f(\bsl{r}) e^{\ii \bsl{p}\cdot \bsl{r} }  = \int d^2 r \sum_{\widetilde{l}\sigma} |w_{+1\widetilde{l}\sigma}(\bsl{r})|^2 e^{\ii \bsl{p}\cdot\bsl{r}}  = \frac{1}{N} \sum_{\bsl{k}\in \MBZ} U^\dagger_{+,f,1}(\bsl{k}+\bsl{p}) U_{+,f,1}(\bsl{k})\ ,
}
where \eqnref{eq:W_Uf} is used.
Based on \eqnref{eq:W_in_f_sym}, we can derive useful symmetry properties of $n_f(\bsl{r})$ and $n_f(\bsl{p})$ as
\eqa{ 
\label{eq:n_f_sym}
\left\{
\begin{array}{l}
    n_f(\bsl{r}) =  n_f^*(-\bsl{r}) \\
    n_f(\bsl{r}) =  n_f(C_3\bsl{r}) \\
    n_f(\bsl{r}) =  n_f(C_{2x}\bsl{r}) \\
    n_f(\bsl{r}) =  n_f(-\bsl{r}) 
\end{array}
\right.
\text{ and }
\left\{
\begin{array}{l}
    n_f(\bsl{p}) =  n_f^*(\bsl{p}) \\
    n_f(\bsl{p}) =  n_f(C_3\bsl{p}) \\
    n_f(\bsl{p}) =  n_f(C_{2x}\bsl{p}) \\
    n_f(\bsl{p}) =  n_f(-\bsl{p}) 
\end{array}
\right.\ .
}

With the definition of $n_{f}(\bsl{p})$, we have 
\eqa{
& \frac{1}{2} \int d^2 r d^2 r' V(\bsl{r}-\bsl{r}') :\rho_{ff}(\bsl{r}): :\rho_{ff}(\bsl{r}'):  \\
& = \frac{1}{2}\sum_{\bsl{R},\bsl{R}'} \int d^2 r d^2 r' V(\bsl{r}-\bsl{r}') n_f(\bsl{r}-\bsl{R}) n_f(\bsl{r}'-\bsl{R}') :\rho_{f}(\bsl{R}): :\rho_{f}(\bsl{R}'): \\
& =  \frac{1}{2}\sum_{\bsl{R},\bsl{R}'} :\rho_{f}(\bsl{R}): :\rho_{f}(\bsl{R}'): U(\bsl{R}-\bsl{R}')\ ,
}
where 
\eqa{
U(\bsl{R}-\bsl{R}') & =\int d^2 r d^2 r' V(\bsl{r}-\bsl{r}') n_f(\bsl{r}-\bsl{R}) n_f(\bsl{r}'-\bsl{R}') \\
& =\int d^2 r d^2 r' V(\bsl{r}-\bsl{r}') n_f(\bsl{r}) n_f(\bsl{r}'-\bsl{R}'+\bsl{R}) \\
& =\int d^2 r d^2 r' \frac{1}{\A^3}\sum_{\bsl{p}_1} n_f(\bsl{p}_1) e^{-\ii \bsl{p}_1 \cdot \bsl{r} } \sum_{\bsl{p}} V(\bsl{p}) e^{-\ii \bsl{p} \cdot (\bsl{r}-\bsl{r}') } \sum_{\bsl{p}_2} n_f(\bsl{p}_2) e^{-\ii \bsl{p}_2 \cdot (\bsl{r}-\bsl{R}'+\bsl{R}) } \\
& = \frac{1}{\A} \sum_{\bsl{p}} n_f^*(\bsl{p}) V(\bsl{p}) n_f(\bsl{p}) e^{-\ii \bsl{p}\cdot(\bsl{R}-\bsl{R}')}\ .
}

Numerically, we find
\eq{
U(0)= 91.50 \meV\ ,\ U(\bsl{a}_{1})= 5.387 \meV\ ,\ U(\bsl{a}_{1}-\bsl{a}_{2})= 0.5079 \meV 
}
in EUS, which shows that $U(\bsl{R})$ decays very as $|\bsl{R}|$ increases.
Therefore, we only keep the terms up to the nearest-neighboring terms and get 
\eq{
\frac{1}{2} \int d^2 r d^2 r' V(\bsl{r}-\bsl{r}') :\rho_{ff}(\bsl{r}): :\rho_{ff}(\bsl{r}'): = H_{int,U}\ ,
}
where $H_{int,U}$ is defined in \eqnref{eq:H_int_U}.
In \eqnref{eq:H_int_U}, we have 
\eq{
U_1=U(0)
}
and 
\eq{
U_2=\frac{1}{6}\sum_{\bsl{R}\neq 0} U(\bsl{R}) \ .
}
Since $U(\bsl{R})$ decays very fast as $|\bsl{R}|$ increases, $U_2$ is dominated by the $|\bsl{R}|=|\bsl{a}_1|$ contribution.
The reason for choosing an expression of $U_2$ that is not equal to $U(\bsl{a}_{1})$ is that such choice can reduce the error in calculating the interaction-induced chemical potential shift, as discussed in \refcite{Song20211110MATBGHF}.
The numerical values of $U_1$ and $U_2$ are in \tabref{tab:fcd_int}.

\subsubsection{$\frac{1}{2} \int d^2 r d^2 r' V(\bsl{r}-\bsl{r}') :\rho_{cc}(\bsl{r}): :\rho_{cc}(\bsl{r}'): $}

To simplify this term, we first derive the expression of $\sum_{\widetilde{l}\sigma} g_{\eta\beta\widetilde{l}\sigma}^*(\bsl{r}) g_{\eta\beta'\widetilde{l}\sigma}(\bsl{r})$ as
\eqa{
\sum_{\widetilde{l}\sigma} g_{\eta\beta\widetilde{l}\sigma}(\bsl{r}) g_{\eta\beta'\widetilde{l}\sigma}^*(\bsl{r}) & = \sum_{\widetilde{l}\sigma}  \sum_{\bsl{Q}\in\Q_{\eta,\widetilde{l}}} e^{-\ii  \bsl{Q}\cdot\bsl{r}}  [\widetilde{u}_{\eta,c,\beta}(0)]_{\bsl{Q}\sigma} \sum_{\bsl{Q}'\in\Q_{\eta,\widetilde{l}}} e^{\ii  \bsl{Q}'\cdot\bsl{r}}  [\widetilde{u}_{\eta,c,\beta'}(0)]_{\bsl{Q}'\sigma}^* \\
& = \sum_{\widetilde{l}\sigma}  \sum_{\bsl{Q}\in\Q_{\eta,\widetilde{l}}} \sum_{\bsl{G}} e^{-\ii  \bsl{G}\cdot\bsl{r}}  [\widetilde{u}_{\eta,c,\beta'}(0)]_{\bsl{Q}-\bsl{G}\sigma}^* [\widetilde{u}_{\eta,c,\beta}(0)]_{\bsl{Q}\sigma} \\
& =  \sum_{\bsl{Q}\in\Q } \sum_{\sigma} \sum_{\bsl{G}} e^{-\ii  \bsl{G}\cdot\bsl{r}}  [\widetilde{u}_{\eta,c,\beta'}(\bsl{G})]_{\bsl{Q}\sigma}^* [\widetilde{u}_{\eta,c,\beta}(0)]_{\bsl{Q}\sigma} \\
& =  \sum_{\bsl{G}} e^{-\ii  \bsl{G}\cdot\bsl{r}}  \widetilde{u}_{\eta,c,\beta'}^\dagger(\bsl{G}) \widetilde{u}_{\eta,c,\beta}(0) \\
}
where \eqnref{eq:Uc_G-shift} and \eqnref{eq:g_in_c} are used.
Then, we have
\eq{
\label{eq:g_star_g}
\sum_{\widetilde{l}\sigma} g_{\eta\beta\widetilde{l}\sigma}^*(\bsl{r}) g_{\eta\beta'\widetilde{l}\sigma}(\bsl{r}) = \sum_{\bsl{G}} e^{-\ii  \bsl{G}\cdot\bsl{r}}   \widetilde{u}_{\eta,c,\beta}^\dagger(\bsl{G})\widetilde{u}_{\eta,c,\beta'}(0)\ .
}
Then,
\eqa{
& \frac{1}{2} \int d^2 r d^2 r' V(\bsl{r}-\bsl{r}') :\rho_{cc}(\bsl{r}): :\rho_{cc}(\bsl{r}'): \\
& = \frac{1}{2} \int d^2 r d^2 r' V(\bsl{r}-\bsl{r}')  \sum_{\eta,s} \sum_{\beta} \sum_{\beta'} :c^\dagger_{\eta,\bsl{r},\beta,s} c_{\eta,\bsl{r},\beta',s} : \sum_{\widetilde{l}\sigma}  g_{\eta\beta\widetilde{l}\sigma}^*(\bsl{r}) g_{\eta\beta'\widetilde{l}\sigma}(\bsl{r}) \\
& \quad \times  \sum_{\eta_1,s_1} \sum_{\beta_1} \sum_{\beta_1'} :c^\dagger_{\eta_1,\bsl{r}',\beta_1,s} c_{\eta_1,\bsl{r}',\beta'_1,s_1}:  \sum_{\widetilde{l}_1\sigma_1} g_{\eta_1\beta_1\widetilde{l}_1\sigma_1}^*(\bsl{r}') g_{\eta_1\beta_1'\widetilde{l}_1\sigma_1}(\bsl{r}')\\
&  = \frac{1}{2} \int d^2 r d^2 r' V(\bsl{r}-\bsl{r}')  \sum_{\eta,s} \sum_{\beta,\beta'} \sum_{\eta_1,s_1} \sum_{\beta_1,\beta_1'} :c^\dagger_{\eta,\bsl{r},\beta,s} c_{\eta,\bsl{r},\beta',s} : :c^\dagger_{\eta_1,\bsl{r}',\beta_1,s_1} c_{\eta_1,\bsl{r}',\beta'_1,s_1}: \\
& \quad \times \sum_{\bsl{G}} e^{-\ii  \bsl{G}\cdot\bsl{r}}  \widetilde{u}_{\eta,c,\beta}^\dagger(\bsl{G}) \widetilde{u}_{\eta,c,\beta'}(0) \sum_{\bsl{G}'} e^{-\ii  \bsl{G}'\cdot\bsl{r}'}  \widetilde{u}_{\eta,c,\beta_1}^\dagger(\bsl{G}') \widetilde{u}_{\eta,c,\beta_1'}(0) \\
&  = \frac{1}{2} \frac{1}{\A} \sum_{\bsl{k}}^{\MBZ}\sum_{\bsl{G},\bsl{G}',\bsl{G}''} V(\bsl{k}+\bsl{G}'') \sum_{\eta,s} \sum_{\beta,\beta'} \sum_{\eta_1,s_1} \sum_{\beta_1,\beta_1'} \int d^2 r:c^\dagger_{\eta,\bsl{r},\beta,s} c_{\eta,\bsl{r},\beta',s} : e^{-\ii  (\bsl{k}+\bsl{G}''+\bsl{G})\cdot\bsl{r}}  \\
& \quad \times \int  d^2 r'  :c^\dagger_{\eta_1,\bsl{r}',\beta_1,s} c_{\eta_1,\bsl{r}',\beta'_1,s_1}: e^{-\ii  (-\bsl{k}-\bsl{G}''+\bsl{G}')\cdot\bsl{r}'}  \widetilde{u}_{\eta,c,\beta}^\dagger(\bsl{G}) \widetilde{u}_{\eta,c,\beta'}(0)  U_{\eta_1,c,\beta_1}^\dagger(\bsl{G}') U_{\eta_1,c,\beta_1'}(0)\ .
}
Owing to small $\Lambda_c\ll |\bsl{q}_1| = 1$, we have 
\eq{
\label{eq:small_Lambda_c_1}
\int d^2 r:c^\dagger_{\eta,\bsl{r},\beta,s} c_{\eta,\bsl{r},\beta',s} : e^{-\ii  (\bsl{k}+\bsl{G}''+\bsl{G})\cdot\bsl{r}} = 0 \text{ if }\bsl{G}''+\bsl{G} \neq 0\ .
}
Then, we have
\eqa{
& \frac{1}{2} \int d^2 r d^2 r' V(\bsl{r}-\bsl{r}') :\rho_{cc}(\bsl{r}): :\rho_{cc}(\bsl{r}'): \\
&  = \frac{1}{2} \frac{1}{\A} \sum_{\bsl{k}}^{\MBZ}\sum_{\bsl{G}} V(\bsl{k}+\bsl{G}) \sum_{\eta,s} \sum_{\beta,\beta'} \sum_{\eta_1,s_1} \sum_{\beta_1,\beta_1'} \int d^2 r:c^\dagger_{\eta,\bsl{r},\beta,s} c_{\eta,\bsl{r},\beta',s} : e^{-\ii  \bsl{k} \cdot\bsl{r}}  \\
& \quad \times \int  d^2 r'  :c^\dagger_{\eta_1,\bsl{r}',\beta_1,s_1} c_{\eta_1,\bsl{r}',\beta'_1,s_1}: e^{\ii \bsl{k} \cdot\bsl{r}'}  \widetilde{u}_{\eta,c,\beta}^\dagger(-\bsl{G}) \widetilde{u}_{\eta,c,\beta'}(0)  U_{\eta_1,c,\beta_1}^\dagger(\bsl{G}) U_{\eta_1,c,\beta_1'}(0)\ .
}
Furthermore, we numerically find that 
\eq{
\sum_{\bsl{G}} V(\bsl{k}+\bsl{G}) \widetilde{u}_{\eta,c,\beta}^\dagger(-\bsl{G}) U_{\eta_1,c,\beta'}(0)  \widetilde{u}_{\eta,c,\beta_1}^\dagger(\bsl{G}) U_{\eta_1,c,\beta_1'}(0) \approx V(\bsl{k})\delta_{\beta\beta'}\delta_{\beta_1\beta_1'} 
}
with only $8\%$ error.
Then, we have 
\eqa{
& \frac{1}{2} \int d^2 r d^2 r' V(\bsl{r}-\bsl{r}') :\rho_{cc}(\bsl{r}): :\rho_{cc}(\bsl{r}'): \\
& \approx \frac{1}{2} \int d^2 r \int  d^2 r' \frac{1}{\A} \sum_{\bsl{k}}^{\MBZ}V(\bsl{k}) e^{-\ii \bsl{k} \cdot(\bsl{r}-\bsl{r}')}\sum_{\eta,s} \sum_{\beta} \sum_{\eta_1,s_1} \sum_{\beta_1} :c^\dagger_{\eta,\bsl{r},\beta,s} c_{\eta,\bsl{r},\beta,s} :  :c^\dagger_{\eta_1,\bsl{r}',\beta_1,s} c_{\eta_1,\bsl{r}',\beta_1,s_1}:  \ .
}
Again owing to small $\Lambda_c$, we  can extend the summation of $\bsl{k}$ from $\MBZ$ to $\dsR^2$, leading to 
\eqa{
& \frac{1}{2} \int d^2 r d^2 r' V(\bsl{r}-\bsl{r}') :\rho_{cc}(\bsl{r}): :\rho_{cc}(\bsl{r}'): \\
&  \approx \frac{1}{2} \int d^2 r \int  d^2 r' \frac{1}{\A} \sum_{\bsl{p}}V(\bsl{p}) e^{-\ii \bsl{p} \cdot(\bsl{r}-\bsl{r}')}\sum_{\eta,s} \sum_{\beta} \sum_{\eta_1,s_1} \sum_{\beta_1} :c^\dagger_{\eta,\bsl{r},\beta,s} c_{\eta,\bsl{r},\beta,s} :  :c^\dagger_{\eta_1,\bsl{r}',\beta_1,s} c_{\eta_1,\bsl{r}',\beta_1,s_1}:  \ ,
}
resulting in 
\eqa{
\frac{1}{2} \int d^2 r d^2 r' V(\bsl{r}-\bsl{r}') :\rho_{cc}(\bsl{r}): :\rho_{cc}(\bsl{r}'):  \approx H_{int,V,c}\ ,
}
where $H_{int,V,c}$ is defined in \eqnref{eq:H_int_V_c}.

\subsubsection{$\int d^2 r d^2 r' V(\bsl{r}-\bsl{r}') :\rho_{ff}(\bsl{r}): :\rho_{cc}(\bsl{r}'):$}

First, by using \eqnref{eq:g_star_g} and \eqnref{eq:n_f_sym}, we have 
\eqa{
& \int d^2 r d^2 r' V(\bsl{r}-\bsl{r}') :\rho_{ff}(\bsl{r}): :\rho_{cc}(\bsl{r}'): = \frac{1}{\A}\sum_{\bsl{R}} :\rho_f(\bsl{R}): \sum_{\eta,s}\sum_{\beta\beta'} \sum_{\bsl{k}}^{\MBZ}e^{-\ii\bsl{k}\cdot\bsl{R}}  \\
& \quad \times \sum_{\bsl{G},\bsl{G}'}\int d^2 r' e^{\ii(\bsl{k}+\bsl{G}'-\bsl{G})\cdot \bsl{r}'} \widetilde{u}_{\eta,c,\beta}^\dagger(\bsl{G})  \widetilde{u}_{\eta,c,\beta'}(0)  V(\bsl{k}+\bsl{G}') n_f(\bsl{k}+\bsl{G}') :c^\dagger_{\eta,\bsl{r}',\beta,s} c_{\eta,\bsl{r}',\beta',s}:\ .
}
Then, by using \eqnref{eq:small_Lambda_c_1} derived from the small $\Lambda_c$, we get 
\eqa{
& \int d^2 r d^2 r' V(\bsl{r}-\bsl{r}') :\rho_{ff}(\bsl{r}): :\rho_{cc}(\bsl{r}'): = \frac{1}{\A}\sum_{\bsl{R}} :\rho_f(\bsl{R}): \sum_{\eta,s}\sum_{\beta\beta'} \sum_{\bsl{k}}^{\MBZ}e^{-\ii\bsl{k}\cdot\bsl{R}}  \\
& \quad \times \int d^2 r' e^{\ii \bsl{k} \cdot \bsl{r}'}:c^\dagger_{\eta,\bsl{r}',\beta,s} c_{\eta,\bsl{r}',\beta',s}: \sum_{\bsl{G}} \widetilde{u}_{\eta,c,\beta}^\dagger(\bsl{G})  \widetilde{u}_{\eta,c,\beta'}(0)  V(\bsl{k}+\bsl{G}) n_f(\bsl{k}+\bsl{G}) \ .
}
Again owing to small $\Lambda_c$, we can choose $\bsl{k}=0$ in $V(\bsl{k}+\bsl{G}) n_f(\bsl{k}+\bsl{G})$ as a good approximation, resulting in 
\eqa{
& \int d^2 r d^2 r' V(\bsl{r}-\bsl{r}') :\rho_{ff}(\bsl{r}): :\rho_{cc}(\bsl{r}'): \\
& \approx \frac{1}{\A}\sum_{\bsl{R}} :\rho_f(\bsl{R}): \sum_{\eta,s}\sum_{\beta\beta'} \sum_{\bsl{k}}^{\MBZ}e^{-\ii\bsl{k}\cdot\bsl{R}} \int d^2 r' e^{\ii \bsl{k} \cdot \bsl{r}'}:c^\dagger_{\eta,\bsl{r}',\beta,s} c_{\eta,\bsl{r}',\beta',s}: [X_{\eta}]_{\beta\beta'} \ ,
}
where
\eq{
[X_{\eta}]_{\beta\beta'}=\sum_{\bsl{G}} \widetilde{u}_{\eta,c,\beta}^\dagger(\bsl{G})  \widetilde{u}_{\eta,c,\beta'}(0)  V(\bsl{G}) n_f(\bsl{G})\ .
}
Based on with \eqnref{eq:sym_rep_c} and \eqnref{eq:n_f_sym}, we have
\eqa{
& X_{+}=\mat{ e^{-\ii \frac{2\pi}{3}  \sigma_z} & \\ & \sigma_0} X_{+} \mat{ e^{\ii \frac{2\pi}{3}  \sigma_z} & \\ & \sigma_0} =\mat{\sigma_x & \\ & \sigma_x} X_{+} \mat{\sigma_x & \\ & \sigma_x} \\
& =  \mat{\sigma_x & \\ & \sigma_x} X_{\eta}^* \mat{\sigma_x & \\ & \sigma_x} =\mat{\sigma_z & \\ & \sigma_z} X_{+} \mat{\sigma_z & \\ & \sigma_z} = X_-^*
}
resulting in 
\eq{
X_\eta = \Omega \mat{ W_1 \sigma_0 & \\ & W_3 \sigma_0}
}
with $W_1,W_3\in\dsR$.
Then, combined with the fact that small $\Lambda_c$ allows us to extend the summation of $\bsl{k}$ to $\dsR^2$, we arrive at 
\eqa{
& \int d^2 r d^2 r' V(\bsl{r}-\bsl{r}') :\rho_{ff}(\bsl{r}): :\rho_{cc}(\bsl{r}'): \\
&\approx
\frac{1}{\A}\sum_{\bsl{R}} :\rho_f(\bsl{R}): \sum_{\eta,s}\sum_{\beta\beta'} \sum_{\bsl{k}}^{\MBZ}e^{-\ii\bsl{k}\cdot\bsl{R}} \int d^2 r' e^{\ii \bsl{k} \cdot \bsl{r}'}:c^\dagger_{\eta,\bsl{r}',\beta,s} c_{\eta,\bsl{r}',\beta',s}:  \Omega W_\beta \delta_{\beta\beta'} \\
& =
\Omega \sum_{\bsl{R}} :\rho_f(\bsl{R}): \sum_{\eta,s}\sum_{\beta} \frac{1}{\A}\sum_{\bsl{p}} e^{-\ii\bsl{p}\cdot\bsl{R}} \int d^2 r' e^{\ii \bsl{p} \cdot \bsl{r}'}:c^\dagger_{\eta,\bsl{r}',\beta,s} c_{\eta,\bsl{r}',\beta,s}:   W_\beta  \\
& = H_{int,W,fc}\ ,
}
where $H_{int,W,fc}$ is defined in \eqnref{eq:H_int_W_fc}.

\subsubsection{$\frac{1}{2} \int d^2 r d^2 r' V(\bsl{r}-\bsl{r}') \left[:\rho_{cf}(\bsl{r}): :\rho_{fc}(\bsl{r}'): + :\rho_{fc}(\bsl{r}): :\rho_{cf}(\bsl{r}'):\right]$}

To simplify this term, first note that
\eqa{
\label{eq:g_W_in_cf}
& \sum_{\widetilde{l}\sigma} g^*_{\eta\beta\widetilde{l}\sigma}(\bsl{r}) e^{\eta \ii \Delta K_{\widetilde{l}}\cdot \bsl{R}} w_{\eta\alpha\widetilde{l}\sigma}(\bsl{r}-\bsl{R}) \\
& = \sum_{\widetilde{l}\sigma} \sum_{\bsl{Q}\in\Q_{\eta,\widetilde{l}}} e^{-\ii  \bsl{Q}\cdot\bsl{r}}  [\widetilde{u}_{\eta,c,\beta}(0)]_{\bsl{Q}\sigma}^* e^{\eta \ii \Delta K_{\widetilde{l}}\cdot \bsl{R}} \frac{1}{N\sqrt{\Omega}}\sum_{\bsl{k}}^{\MBZ}\sum_{\bsl{Q}'\in\Q_{\eta,\widetilde{l}}} e^{\ii(\bsl{k}-\bsl{Q}')\cdot \bsl{r}} [\widetilde{v}_{\eta,f,\alpha}(\bsl{k})]_{\bsl{Q}'\sigma} \\
& = \frac{1}{N\sqrt{\Omega}} \sum_{\widetilde{l}\sigma} \sum_{\bsl{k}}^{\MBZ} e^{\ii \bsl{k} \cdot (\bsl{r}-\bsl{R})} \sum_{\bsl{Q},\bsl{Q}'\in\Q_{\eta,\widetilde{l}}} e^{\ii  (\bsl{Q}-\bsl{Q}')\cdot\bsl{r}}  [\widetilde{u}_{\eta,c,\beta}(0)]_{\bsl{Q}\sigma}^* [\widetilde{v}_{\eta,f,\alpha}(\bsl{k})]_{\bsl{Q}'\sigma} \\
& = \frac{1}{N\sqrt{\Omega}}\sum_{\bsl{k}}^{\MBZ} \sum_{\bsl{G}} e^{\ii (\bsl{k}-\bsl{G}) \cdot (\bsl{r}-\bsl{R})} \sum_{\bsl{Q},\sigma} [\widetilde{u}_{\eta,c,\beta}(0)]_{\bsl{Q}\sigma}^* [\widetilde{v}_{\eta,f,\alpha}(\bsl{k})]_{\bsl{Q}+\bsl{G}\sigma} \\
& = \frac{1}{N\sqrt{\Omega}}\sum_{\bsl{p}} e^{\ii \bsl{p} \cdot (\bsl{r}-\bsl{R})}  \widetilde{u}_{\eta,c,\beta}(0)^\dagger  \widetilde{v}_{\eta,f,\alpha}(\bsl{p}) \ ,
}
where we use \eqnref{eq:W_Uf}, \eqnref{eq:g_in_c}, \eqnref{eq:Uf_G-shift} and \eqnref{eq:DeltaK_R}.
With \eqnref{eq:g_W_in_cf} and \eqnref{eq:W_W_onsite}, we have
\eqa{
& \frac{1}{2} \int d^2 r d^2 r' V(\bsl{r}-\bsl{r}')  :\rho_{cf}(\bsl{r}): :\rho_{fc}(\bsl{r}'): \\
& \approx \frac{1}{2}\frac{1}{\A^2 N} \int d^2 r d^2 r' \sum_{\eta,s}\sum_{\beta\alpha\bsl{R}} :c^\dagger_{\eta,\bsl{r},\beta,s} f_{\eta,\bsl{R},\beta,s}: \sum_{\eta',s'}\sum_{\beta'\alpha'} :f^\dagger_{\eta',\bsl{R},\alpha',s'} c_{\eta',\bsl{r}',\beta',s'}: \\
& \quad \times \sum_{\bsl{p},\bsl{p}_1,\bsl{p}_2}e^{\ii \bsl{p}_1\cdot\bsl{r}-\ii \bsl{r}'\cdot \bsl{p}_2 - \ii \bsl{p}_1\cdot \bsl{R} + \ii \bsl{p}_2\cdot \bsl{R}} V(\bsl{p}) \widetilde{u}^\dagger_{\eta,c,\beta}(0) \widetilde{v}_{\eta,f,\alpha}(\bsl{p}_1+\bsl{p})\widetilde{v}_{\eta',f,\alpha'}^\dagger(\bsl{p}_2+\bsl{p}) \widetilde{u}_{\eta',c,\beta'}(0)\ .
}
Here $\bsl{p}_1$ and $\bsl{p}_2$ are carried by the $c$ modes, and thus we can set them to be zero in $\widetilde{u}^\dagger_{\eta,c,\beta}(0) \widetilde{v}_{\eta,f,\alpha}(\bsl{p}_1+\bsl{p})\widetilde{v}_{\eta',f,\alpha'}^\dagger(\bsl{p}_2+\bsl{p}) \widetilde{u}_{\eta',c,\beta'}(0)$ as a good approximation, resulting in 
\eqa{
& \frac{1}{2} \int d^2 r d^2 r' V(\bsl{r}-\bsl{r}')  :\rho_{cf}(\bsl{r}): :\rho_{fc}(\bsl{r}'): \\
& \approx \frac{1}{2}\frac{1}{\A^2 N} \int d^2 r d^2 r' \sum_{\eta,s}\sum_{\beta\alpha\bsl{R}} :c^\dagger_{\eta,\bsl{r},\beta,s} f_{\eta,\bsl{R},\beta,s}: \sum_{\eta',s'}\sum_{\beta'\alpha'} :f^\dagger_{\eta',\bsl{R},\alpha',s'} c_{\eta',\bsl{r}',\beta',s'}: \\
& \quad \times \sum_{\bsl{p},\bsl{p}_1,\bsl{p}_2}e^{\ii \bsl{p}_1\cdot\bsl{r}-\ii \bsl{r}'\cdot \bsl{p}_2 - \ii \bsl{p}_1\cdot \bsl{R} + \ii \bsl{p}_2\cdot \bsl{R}} V(\bsl{p}) \widetilde{u}^\dagger_{\eta,c,\beta}(0) \widetilde{v}_{\eta,f,\alpha}(\bsl{p})\widetilde{v}_{\eta',f,\alpha'}^\dagger(\bsl{p}) \widetilde{u}_{\eta',c,\beta'}(0) \\
& = \frac{\Omega}{2} \sum_{\eta,s}\sum_{\alpha\beta\bsl{R}} :c^\dagger_{\eta,\bsl{R},\beta,s}f_{\eta,\bsl{R},\alpha,s}: \sum_{\eta',s'}\sum_{\alpha'\beta'}:f^\dagger_{\eta,'\bsl{R},\alpha',s'} c_{\eta',\bsl{R},\beta',s'}: J^*_{\eta\alpha\beta,\eta'\alpha'\beta'}\ ,
}
where
\eq{
\label{eq:J}
J_{\eta\alpha\beta,\eta'\alpha'\beta'} = \frac{1}{\A}\sum_{\bsl{p}}V(\bsl{p})  \widetilde{v}^\dagger_{\eta,f,\alpha}(\bsl{p})\widetilde{u}_{\eta,c,\beta}(0) \widetilde{u}_{\eta',c,\beta'}(0)\widetilde{v}^\dagger_{\eta',f,\alpha'}(\bsl{p})
}
which satisfies 
\eq{
J_{\eta\alpha\beta,\eta'\alpha'\beta'} = J_{\eta'\alpha'\beta',\eta\alpha\beta}^*\ .
}
Similarly, we have
\eqa{
& \frac{1}{2} \int d^2 r d^2 r' V(\bsl{r}-\bsl{r}')  :\rho_{fc}(\bsl{r}): :\rho_{cf}(\bsl{r}'): \\
& = \frac{1}{2} \int d^2 r d^2 r' V(\bsl{r}-\bsl{r}')  :\rho_{fc}(\bsl{r}'): :\rho_{cf}(\bsl{r}): \\
& \approx \frac{\Omega}{2} \sum_{\bsl{R}} \sum_{\eta,s,\alpha,\beta}\sum_{\eta',s',\alpha'\beta'}  :f^\dagger_{\eta',\bsl{R},\alpha',s'} c_{\eta',\bsl{R},\beta',s'}: :c^\dagger_{\eta,\bsl{R},\beta,s}f_{\eta,\bsl{R},\alpha,s}: J^*_{\eta\alpha\beta,\eta'\alpha'\beta'}\ .
}
As a result, we have
\eqa{
& \frac{1}{2} \int d^2 r d^2 r' V(\bsl{r}-\bsl{r}') \left[:\rho_{cf}(\bsl{r}): :\rho_{fc}(\bsl{r}'): + :\rho_{fc}(\bsl{r}): :\rho_{cf}(\bsl{r}'):\right] \\
& \approx -\Omega \sum_{\bsl{R}} \sum_{\eta,s,\alpha,\beta}\sum_{\eta',s',\alpha'\beta'} J_{\eta\alpha\beta,\eta'\alpha'\beta'} :f^\dagger_{\eta,\bsl{R},\alpha,s} f_{\eta',\bsl{R},\alpha',s'}: :c^\dagger_{\eta',\bsl{R},\beta',s'} c_{\eta,\bsl{R},\beta,s} : + const.\ .
}

Numerically, we find the biggest components of $J_{\eta\alpha\beta,\eta'\alpha'\beta'}$ are equal to
$J_{\eta13,\eta13}$, $J_{\eta24,\eta24}$, $J_{\eta24,-\eta13}$, and $J_{\eta13,-\eta24}$, whose magnitudes are $24.25\meV$ in EUS.
The next biggest components of $J_{\eta\alpha\beta,\eta'\alpha'\beta'}$ have magnitudes being $6.478\meV$ in EUS, which are roughly a quarter of those of the biggest components.
Therefore, we only keep the biggest components of $J_{\eta\alpha\beta,\eta'\alpha'\beta'}$.
Furthermore, based on \eqnref{eq:sym_rep_f} and \eqnref{eq:sym_rep_c}, we find~\cite{Song20211110MATBGHF} that 
\eq{
J_{\eta13,\eta13}=J_{\eta24,\eta24}=-J_{\eta24,-\eta13}=-J_{\eta13,-\eta24}
}
which is independent of $\eta$.
Then, we define 
\eq{
J=J_{+13,+13}\ ,
}
leading to
\eq{
\frac{1}{2} \int d^2 r d^2 r' V(\bsl{r}-\bsl{r}') \left[:\rho_{cf}(\bsl{r}): :\rho_{fc}(\bsl{r}'): + :\rho_{fc}(\bsl{r}): :\rho_{cf}(\bsl{r}'):\right] \approx H_{int,J}+const.\ ,
}
where $H_{int,J}$ is defined in \eqnref{eq:H_int_J}.

\subsubsection{$\frac{1}{2} \int d^2 r d^2 r' V(\bsl{r}-\bsl{r}') :\rho_{cf}(\bsl{r}): :\rho_{cf}(\bsl{r}'): +h.c.$}

By using \eqnref{eq:g_W_in_cf} and \eqnref{eq:W_W_onsite}, the term can be simplfied to
\eqa{
& \frac{1}{2} \int d^2 r d^2 r' V(\bsl{r}-\bsl{r}') :\rho_{cf}(\bsl{r}): :\rho_{cf}(\bsl{r}'): +h.c. \\
& \approx \frac{1}{2}\sum_{\bsl{k_1}}^{|\bsl{k_1}|\leq \Lambda_c}\sum_{\bsl{k_2}}^{|\bsl{k_2}|\leq \Lambda_c}\sum_{\eta,s}\sum_{\alpha\beta\bsl{R}}c^\dagger_{\eta,\bsl{k}_1,\beta,s}f_{\eta,\bsl{R},\alpha,s}\sum_{\eta',s'}\sum_{\alpha'\beta'}c^\dagger_{\eta',\bsl{k}_2,\beta',s'}f_{\eta',\bsl{R},\alpha',s'} \\
& \quad \times \frac{1}{N}e^{-\ii \bsl{k}_1\cdot\bsl{R}}e^{-\ii \bsl{k}_2\cdot\bsl{R}}\frac{1}{\A}\sum_{\bsl{p}}V(\bsl{p}) (\widetilde{u}^\dagger_{\eta,c,\beta}(0)\widetilde{v}_{\eta,f,\alpha}(\bsl{p}+\bsl{k}_1)) (\widetilde{u}^\dagger_{\eta',c,\beta'}(0)\widetilde{v}_{\eta',f,\alpha'}(-\bsl{p}+\bsl{k}_2))+h.c.\ .
}
Then, due to the small $\Lambda_c$, we can set $\bsl{k}_1=0$ and $\bsl{k}_2=0$ in $(\widetilde{u}^\dagger_{\eta,c,\beta}(0)\widetilde{v}_{\eta,f,\alpha}(\bsl{p}+\bsl{k}_1)) (\widetilde{u}^\dagger_{\eta',c,\beta'}(0)\widetilde{v}_{\eta',f,\alpha'}(-\bsl{p}+\bsl{k}_2))$, resulting in 
\eqa{
 \frac{1}{2} \int d^2 r d^2 r' V(\bsl{r}-\bsl{r}') :\rho_{cf}(\bsl{r}): :\rho_{cf}(\bsl{r}'): +h.c. \approx H_{int,\widetilde{J}}\ ,
 }
 where 
 \eq{
 \label{eq:H_int_J_tilde}
H_{int,\widetilde{J}} =  \frac{\Omega}{2}\sum_{\bsl{R}} \sum_{\eta,s,\alpha,\beta}\sum_{\eta',s',\alpha',\beta'} J_{-\eta'\beta'\alpha',\eta\beta\alpha} c^\dagger_{\eta,\bsl{R},\beta,s}c^\dagger_{\eta',\bsl{R},\beta',s'}  f_{\eta',\bsl{R},\alpha',s'}f_{\eta,\bsl{R},\alpha,s}+h.c.\ .
}

\subsubsection{$ \left[\frac{1}{2} \int d^2 r d^2 r' V(\bsl{r}-\bsl{r}') :\rho_{ff}(\bsl{r}): :\rho_{cf}(\bsl{r}'): +h.c. \right]$ $\&$ $\left[\frac{1}{2} \int d^2 r d^2 r' V(\bsl{r}-\bsl{r}') :\rho_{ff}(\bsl{r}): :\rho_{fc}(\bsl{r}'): +h.c. \right]$}

With \eqnref{eq:rho_ff_cc_fc_cf}, \eqnref{eq:Coulomb_V}, \eqnref{eq:W_W_onsite}, \eqnref{eq:nf_p}, and \eqnref{eq:g_W_in_cf}, we can get 
\eqa{
& \frac{1}{2} \int d^2 r d^2 r' V(\bsl{r}-\bsl{r}') :\rho_{ff}(\bsl{r}): :\rho_{cf}(\bsl{r}'): +h.c. \\
& \approx \frac{1}{2} \sum_{\bsl{R}} :\rho_f(\bsl{R}):\int d^2 r' \sum_{\eta',s'}\sum_{\beta',\alpha'}c^\dagger_{\eta',\bsl{r}',\beta',s'}f_{\eta',\bsl{R},\beta',s'} \frac{1}{\A}\sum_{\bsl{p}} V(\bsl{p}) n_f(-\bsl{p})\\
& \quad \times \frac{1}{\sqrt{\Omega}N}\sum_{\bsl{p}_2} e^{\ii \bsl{p}_2\cdot\bsl{r}'}e^{-\ii \bsl{p}_2\cdot \bsl{R}} \widetilde{u}^\dagger_{\eta',c,\beta'}(0)\widetilde{v}_{\eta',f,\alpha'}(\bsl{p}_2-\bsl{p}) + h.c.\ .
}
Since $\bsl{p}_2$ is carried by $c$ modes and thus is small due to the small $\Lambda_c$, we can set $\bsl{p}_2=0$ in $\widetilde{u}^\dagger_{\eta',c,\beta'}(0)\widetilde{v}_{\eta',f,\alpha'}(\bsl{p}_2-\bsl{p})$ as a good approximation, resulting in 
\eqa{
& \frac{1}{2} \int d^2 r d^2 r' V(\bsl{r}-\bsl{r}') :\rho_{ff}(\bsl{r}): :\rho_{cf}(\bsl{r}'): +h.c. \\
& \approx \frac{1}{2} \sum_{\bsl{R}} :\rho_f(\bsl{R}): \sum_{\eta',s'}\sum_{\beta',\alpha'}c^\dagger_{\eta',\bsl{R},\beta',s'}f_{\eta',\bsl{R},\beta',s'} [\widetilde{Y}_{\eta'}]_{\beta'\alpha'} + h.c.\ ,
}
where
\eq{
 \widetilde{Y}_{\eta'} = \frac{1}{\sqrt{\Omega}N}\sum_{\bsl{p}} V(\bsl{p}) n_f(\bsl{p}) \widetilde{u}^\dagger_{\eta',c}(0)\widetilde{v}_{\eta',f}(\bsl{p})\ .
}
Based on \eqnref{eq:sym_rep_f}, \eqnref{eq:sym_rep_c} and \eqnref{eq:n_f_sym}, we find 
\eqa{
& \widetilde{Y}_{+}=\mat{ e^{-\ii \frac{2\pi}{3}  \sigma_z} & \\ & \sigma_0} \widetilde{Y}_{+}  e^{\ii \frac{2\pi}{3}  \sigma_z}   =\mat{\sigma_x & \\ & \sigma_x} \widetilde{Y}_{+} \sigma_x  =  - \mat{\sigma_z & \\ & \sigma_z} \widetilde{Y}_{+}  \sigma_z   = \widetilde{Y}_-^*\ ,
}
leading to 
\eq{
\widetilde{Y}_\eta = 0\Rightarrow  \frac{1}{2} \int d^2 r d^2 r' V(\bsl{r}-\bsl{r}') :\rho_{ff}(\bsl{r}): :\rho_{cf}(\bsl{r}'): +h.c. \approx 0\ .
}
Similarly, we have 
\eq{
\frac{1}{2} \int d^2 r d^2 r' V(\bsl{r}-\bsl{r}') :\rho_{ff}(\bsl{r}): :\rho_{fc}(\bsl{r}'): +h.c. \approx 0\ .
}

\subsubsection{$ \left[\frac{1}{2} \int d^2 r d^2 r' V(\bsl{r}-\bsl{r}') :\rho_{cc}(\bsl{r}): :\rho_{cf}(\bsl{r}'): +h.c. \right]+\left[\frac{1}{2} \int d^2 r d^2 r' V(\bsl{r}-\bsl{r}') :\rho_{cc}(\bsl{r}): :\rho_{fc}(\bsl{r}'): +h.c. \right]$}

By using  \eqnref{eq:rho_ff_cc_fc_cf}, \eqnref{eq:Coulomb_V}, \eqnref{eq:g_star_g} and \eqnref{eq:g_W_in_cf}, we can get
\eqa{
&  \left[\frac{1}{2} \int d^2 r d^2 r' V(\bsl{r}-\bsl{r}') :\rho_{cc}(\bsl{r}): :\rho_{cf}(\bsl{r}'): +h.c. \right]+\left[\frac{1}{2} \int d^2 r d^2 r' V(\bsl{r}-\bsl{r}') :\rho_{cc}(\bsl{r}): :\rho_{fc}(\bsl{r}'): +h.c. \right] \\
& = \frac{1}{2} \int d^2 r d^2 r' V(\bsl{r}-\bsl{r}') \left\{:\rho_{cc}(\bsl{r}):\ ,\ :\rho_{cf}(\bsl{r}'): \right\} + h.c. \\
& =\frac{1}{2}\int d^2r d^2 r' \sum_{\eta,s}\sum_{\beta,\beta'}\sum_{\eta_1,s_1,\beta_1,\alpha_1} \sum_{\bsl{R}} \left\{ :c^\dagger_{\eta,\bsl{r},\beta,s}c_{\eta,\bsl{r},\beta',s}:\ ,\  c^\dagger_{\eta_1,\bsl{r}',\beta_1,s_1}f_{\eta_1,\bsl{R},\alpha_1,s_1} \right\} \\
& \quad \times \frac{\sqrt{\Omega}}{\A^2} \sum_{\bsl{p}}\sum_{\bsl{G}}\sum_{\bsl{p}_1} e^{-\ii \bsl{r}\cdot\bsl{p}} V(\bsl{p}-\bsl{G}) \widetilde{u}^\dagger_{\eta,c,\beta}(\bsl{G}) \widetilde{u}_{\eta,c,\beta'}(0) e^{\ii \bsl{r}'\cdot \bsl{p}} e^{-\ii (\bsl{p}_1-\bsl{p})\cdot\bsl{R}} U^\dagger_{\eta_1,c,\beta_1}(0) U_{\eta_1,f,\alpha_1}(\bsl{p}_1-\bsl{p}+\bsl{G})\ .
}
Clearly, both $\bsl{p}$ and $\bsl{p}_1$ are carried by $c$ modes, and are small due to the small $\Lambda_c$.
Then, we can set $\bsl{p}=\bsl{p}_1=0$ in $U^\dagger_{\eta_1,c,\beta_1}(0) U_{\eta_1,f,\alpha_1}(\bsl{p}_1-\bsl{p}+\bsl{G})$ as a good approximation, resulting in
\eqa{
& \left[\frac{1}{2} \int d^2 r d^2 r' V(\bsl{r}-\bsl{r}') :\rho_{cc}(\bsl{r}): :\rho_{cf}(\bsl{r}'): +h.c. \right]+\left[\frac{1}{2} \int d^2 r d^2 r' V(\bsl{r}-\bsl{r}') :\rho_{cc}(\bsl{r}): :\rho_{fc}(\bsl{r}'): +h.c. \right] \\
& \approx  H_{int,K}\ ,
}
where 
\eq{
\label{eq:H_int_K}
H_{int,K}=\frac{1}{2}\Omega^{3/2}\sum_{\eta,s}\sum_{\beta,\beta'}\sum_{\eta_1,s_1,\beta_1,\alpha_1}\sum_{\bsl{R}} \{ :c^\dagger_{\eta,\bsl{R},\beta,s}c_{\eta,\bsl{R},\beta',s}:, c^\dagger_{\eta_1,\bsl{R},\beta_1,s_1}f_{\eta_1,\bsl{R},\alpha_1,s_1} \} K_{\eta\beta\beta',\eta_1\beta_1\alpha_1} + h.c.
}
and
\eq{
K_{\eta\beta\beta',\eta_1\beta_1\alpha_1} = \frac{1}{\Omega} \sum_{\bsl{G}} V(\bsl{G}) \widetilde{u}_{\eta,c,\beta}^\dagger(\bsl{G}) \widetilde{u}_{\eta,c,\beta'}(0) U_{\eta_1,c,\beta_1}^\dagger(0) \widetilde{v}_{\eta,f,\alpha_1}(\bsl{G})\ .
}
Numerically, we find the biggest components of $K_{\eta\beta\beta',\eta_1\beta_1\alpha_1}$ have magnitudes being $7.054\meV$ in EUS.

\subsubsection{In sum}

In sum, we have 
\eq{
H_{int}^{TBG} \approx H_{int,U} + H_{int,V,c} + H_{int,W, fc} + H_{int,J} + H_{int,\widetilde{J}} + H_{int,K} + const.\ ,
}
where the definitions of $H_{int,U}$, $H_{int,V,c}$, $H_{int,W, fc}, H_{int,J}$, $H_{int,\widetilde{J}}$, $H_{int,K}$ can be found in \eqnref{eq:H_int_U}, \eqnref{eq:H_int_V_c}, \eqnref{eq:H_int_W_fc}, \eqnref{eq:H_int_J}, \eqnref{eq:H_int_J_tilde}, and \eqnref{eq:H_int_K}, respectively.
Among all these terms, only $H_{int,\widetilde{J}}$ and $H_{int,K}$ do not preserve the number of $f$ modes.
Moreover, according to \tabref{tab:fcd_int}, the strengths of $H_{int,\widetilde{J}}$ and $H_{int,K}$ are small compared to the onsite interaction among $f$ modes in $H_{int,U}$, as $|J|\sim U_1/4$ and $|K_{\eta\beta\beta',\eta_1\beta_1\alpha_1}|< U_1/10$.
Therefore, we neglect $H_{int,\widetilde{J}}$ and $H_{int,K}$.
We can also neglect the $const.$ in $H_{int}^{TBG}$, since it is just a shift in the total energy, leading to
\eq{
\label{eq:H_int_TBG_projected}
H_{int}^{TBG} \approx H_{int,U} + H_{int,V,c} + H_{int,W, fc}+ H_{int,J}\ ,
}

At the end of this part, we address the issue of the $\sqrt{2}$ scaling.
As discussed in \secref{sec:f_c}, the parameters values of the single-particle TBG block of MATSTG are $\sqrt{2}$ scaled compared to those of the ordinary MATBG discussed in \refcite{Song20211110MATBGHF}.
As shown in \tabref{tab:fcd_int}, the same $\sqrt{2}$ scaling does not necessarily occur to the interaction strengthes in $H_{int}^{TBG}$ of the MATSTG compared to those in \refcite{Song20211110MATBGHF}.
It is because we choose the gate distance (\eqnref{eq:parameter_values_int}) for MATSTG to be the same as that for MATBG, since there is no obvious reason for us to decrease the gate distance by a factor of $\sqrt{2}$ when switching MATBG to MATSTG.
Therefore, the relative ratios among the interaction strenghes in $H_{int}^{TBG}$ of the MATSTG are not the same as those in \refcite{Song20211110MATBGHF}, allowing $W_1$ and $W_3$ to be slightly larger than $U_1$.
Nevertheless, we should still expect $U_1$ dominates the low-energy physics since $W_1$ and $W_3$ involve $c$ modes with relatively higher energies, while $U_1$ only involves the low-energy $f$ modes.

\subsection{Details on $H_{int}^{TBG-D}$ and $H_{int}^{D}$}

Now we turn to the other two terms in \eqnref{eq:H_int_three_parts}, \ie, $H_{int}^{TBG-D}$ and $H_{int}^{D}$, which are not covered in \refcite{Song20211110MATBGHF}.
First note that 
\eq{
d^\dagger_{\eta,\bsl{r},\sigma,s}=\frac{1}{\sqrt{\A}} \sum_{\bsl{p}} e^{-\ii \bsl{p}\cdot\bsl{r}} d^\dagger_{\eta,\bsl{p},\sigma,s} = \frac{1}{\sqrt{\A}} \sum_{\bsl{p}}^{|\bsl{p}|\leq \Lambda_d} e^{-\ii \bsl{p}\cdot\bsl{r}} d^\dagger_{\eta,\bsl{p},\sigma,s} + ... = \widetilde{d}^\dagger_{\eta,\bsl{r},\sigma,s} + ...\ ,
}
where $\widetilde{d}^\dagger_{\eta,\bsl{r},\sigma,s}$ is defined under \eqnref{eq:H_int_V_d}, and ``$...$" represents the higher-energy $d$ modes.
Then, we know 
\eq{
\rho_D(\bsl{r}) = \rho_d(\bsl{r}) + ...\ ,
}
where $\rho_d(\bsl{r})$ is defined under \eqnref{eq:H_int_V_d}.
Combined with \eqnref{eq:rho_tilde_proj}, we have
\eqa{
\label{eq:H_int_TBG_D}
H_{int}^{TBG-D} & = \int d^2 r d^2 r' V(\bsl{r}-\bsl{r}') :\widetilde{\rho}(\bsl{r}): :\rho_D(\bsl{r}'): \\
& \approx \int d^2 r d^2 r' V(\bsl{r}-\bsl{r}') :\rho_{ff}(\bsl{r}): :\rho_d(\bsl{r}'): \\
& \quad + \left[\int d^2 r d^2 r' V(\bsl{r}-\bsl{r}')  :\rho_{cf}(\bsl{r}): :\rho_d(\bsl{r}'): + h.c. \right]\\
& \quad + \int d^2 r d^2 r' V(\bsl{r}-\bsl{r}') :\rho_{cc}(\bsl{r}): :\rho_d(\bsl{r}'): \ .
}
Furthermore, we have 
\eqa{
H_{int}^{D} & = \frac{1}{2} \int d^2 r d^2 r' V(\bsl{r}-\bsl{r}') :\rho_D(\bsl{r}): :\rho_D(\bsl{r}'): \\
& \approx \frac{1}{2} \int d^2 r d^2 r' V(\bsl{r}-\bsl{r}') :\rho_d(\bsl{r}): :\rho_d(\bsl{r}'): = H_{int,V,d}
}
with $H_{int,V,d}$ defined in \eqnref{eq:H_int_V_d}.

In the following, we will discuss each term in \eqnref{eq:H_int_TBG_rho}.

\subsubsection{$\int d^2 r d^2 r' V(\bsl{r}-\bsl{r}') :\rho_{ff}(\bsl{r}): :\rho_d(\bsl{r}'):$ }

With \eqnref{eq:rho_ff_cc_fc_cf}, \eqnref{eq:nf_p} and \eqnref{eq:Coulomb_V}, we have
\eqa{
& \int d^2 r d^2 r' V(\bsl{r}-\bsl{r}') :\rho_{ff}(\bsl{r}): :\rho_d(\bsl{r}'): \\
& = \int d^2 r d^2 r' V(\bsl{r}-\bsl{r}') \sum_{\bsl{R}} :\rho_{f}(\bsl{R}): n_f(\bsl{r}-\bsl{R}) :\rho_d(\bsl{r}'): \\
& = \int d^2 r d^2 r' \frac{1}{\A} \sum_{\bsl{p}} e^{-\ii \bsl{p}\cdot (\bsl{r}-\bsl{r}')} V(\bsl{p}) \sum_{\bsl{R}} :\rho_{f}(\bsl{R}): \frac{1}{\A} \sum_{\bsl{p}_1} n_f(\bsl{p}_1) e^{-\ii \bsl{p}_1\cdot (\bsl{r}-\bsl{R})} :\rho_d(\bsl{r}'): \\
& = \sum_{\bsl{R}} :\rho_{f}(\bsl{R}): \int d^2 r' \frac{1}{\A} \sum_{\bsl{p}} e^{\ii \bsl{p}\cdot \bsl{r}'} V(\bsl{p})   n_f(-\bsl{p}) e^{-\ii \bsl{p}\cdot\bsl{R}} :\rho_d(\bsl{r}'): \ .
}
Since $\bsl{p}$ is carried by $d$ modes here and $\Lambda_d$ is small, we can adopt $V(\bsl{p})   n_f(-\bsl{p}) \approx V(\bsl{p}=0)   n_f(\bsl{p}=0)$.
This approximation is rather good since if we choose $\bsl{p} = \frac{\bsl{q}_3}{6}$ with $\bsl{q}_3$ defined in \eqnref{eq:q}, the error is less than 6\%, i.e.,
\eq{
\frac{ V(\frac{\bsl{q}_3}{6})   n_f(-\frac{\bsl{q}_3}{6}) }{ V(\bsl{p}=0)   n_f(\bsl{p} = 0 )} > 94\%\ .
}
Then, by defining 
\eq{
W_{fd}= \frac{1}{\Omega}V(\bsl{p}=0)   n_f(\bsl{p}=0)\ , 
}
we have 
\eqa{
& \int d^2 r d^2 r' V(\bsl{r}-\bsl{r}') :\rho_{ff}(\bsl{r}): :\rho_d(\bsl{r}'): \\
& \approx \sum_{\bsl{R}} :\rho_{f}(\bsl{R}): :\rho_d(\bsl{R}):   V(\bsl{p}=0)   n_f(\bsl{p}=0) \\
& = \Omega \sum_{\bsl{R}} :\rho_{f}(\bsl{R}): :\rho_d(\bsl{R}):   W_{fd}  \\
& = H_{int,W,fd}\ ,
}
where $H_{int,W,fd}$ is defined in \eqnref{eq:H_int_W_fd}.
The numerical value of $W_{fd}$ is listed in \tabref{tab:fcd_int}.

\subsubsection{$\int d^2 r d^2 r' V(\bsl{r}-\bsl{r}')  :\rho_{cf}(\bsl{r}): :\rho_d(\bsl{r}'): + h.c.  $ }

With \eqnref{eq:rho_ff_cc_fc_cf}, \eqnref{eq:g_W_in_cf} and \eqnref{eq:Coulomb_V}, we can get
\eqa{
& \int d^2 r d^2 r' V(\bsl{r}-\bsl{r}')  :\rho_{fc}(\bsl{r}): :\rho_d(\bsl{r}'): + h.c.   \\
& =\int d^2 r d^2 r'  \frac{1}{\A} \sum_{\bsl{p}} e^{-\ii \bsl{p}\cdot (\bsl{r}-\bsl{r}')} V(\bsl{p})  \sum_{\eta,s} \sum_{\beta,\alpha,\bsl{R}}  :c^\dagger_{\eta,\bsl{r},\beta,s}f_{\eta,\bsl{R},\alpha,s}: \frac{1}{N\sqrt{\Omega}}\sum_{\bsl{p}_1} e^{\ii \bsl{p}_1 \cdot (\bsl{r}-\bsl{R})} \widetilde{u}_{\eta,c,\beta}^\dagger(0) \widetilde{v}_{\eta,f,\alpha}(\bsl{p}_1)   :\rho_d(\bsl{r}'): + h.c.\\
& =\int d^2 r d^2 r'  \frac{\sqrt{\Omega}}{\A^2} \sum_{\eta,s} \sum_{\beta,\alpha,\bsl{R}}:c^\dagger_{\eta,\bsl{r},\beta,s}f_{\eta,\bsl{R},\alpha,s}: \sum_{\bsl{p},\bsl{p}_1} e^{\ii (\bsl{p}_1-\bsl{p})\cdot \bsl{r}} V(\bsl{p}) e^{-\ii \bsl{p}_1 \cdot  \bsl{R} }       \widetilde{u}_{\eta,c,\beta}^\dagger(0) \widetilde{v}_{\eta,f,\alpha}(\bsl{p}_1)  e^{\ii  \bsl{p}\cdot \bsl{r}'}  :\rho_d(\bsl{r}'): + h.c.\\
& =\int d^2 r d^2 r'  \frac{\sqrt{\Omega}}{\A^2} \sum_{\eta,s} \sum_{\beta,\alpha,\bsl{R}}:c^\dagger_{\eta,\bsl{r},\beta,s}f_{\eta,\bsl{R},\alpha,s}: \sum_{\bsl{p},\bsl{p}_1} e^{\ii \bsl{p}_1\cdot \bsl{r}} V(\bsl{p}) e^{-\ii (\bsl{p}_1+\bsl{p}) \cdot  \bsl{R} }       \widetilde{u}_{\eta,c,\beta}^\dagger(0) \widetilde{v}_{\eta,f,\alpha}(\bsl{p}_1+\bsl{p})  e^{\ii  \bsl{p}\cdot \bsl{r}'}  :\rho_d(\bsl{r}'): + h.c.\ .
}
As $\bsl{p}_1$ is carried by $c$ modes and $\bsl{p}$ is carried by $d$ modes, both of them are small, and we can adopt 
\eq{
 \widetilde{u}_{\eta,c,\beta}^\dagger(0) \widetilde{v}_{\eta,f,\alpha}(\bsl{p}_1+\bsl{p}) \approx  \widetilde{u}_{\eta,c,\beta}^\dagger(0) \widetilde{v}_{\eta,f,\alpha}(0) = 0
}
as a good approximation, where the second equality comes from the orthogonality of $\widetilde{u}$ and $\widetilde{v}$ at the same momentum.
Then, we know 
\eq{
\int d^2 r d^2 r' V(\bsl{r}-\bsl{r}')  :\rho_{cf}(\bsl{r}): :\rho_d(\bsl{r}'): + h.c. \approx 0\ .
}

\subsubsection{$\int d^2 r d^2 r' V(\bsl{r}-\bsl{r}') :\rho_{cc}(\bsl{r}): :\rho_d(\bsl{r}'):$ }

With \eqnref{eq:rho_ff_cc_fc_cf}, \eqnref{eq:g_star_g} and \eqnref{eq:Coulomb_V}, we can get
\eqa{
& \int d^2 r d^2 r' V(\bsl{r}-\bsl{r}') :\rho_{cc}(\bsl{r}): :\rho_d(\bsl{r}'): \\
& = \int d^2 r d^2 r' \frac{1}{\A} \sum_{\bsl{p}} e^{-\ii \bsl{p}\cdot (\bsl{r}-\bsl{r}')} V(\bsl{p})  \sum_{\eta,s,\beta,\beta'} : c^\dagger_{\eta,\bsl{r},\beta,s} c_{\eta,\bsl{r},\beta',s}:  \sum_{\bsl{G}} e^{-\ii  \bsl{G}\cdot\bsl{r}} \widetilde{u}_{\eta,c,\beta}^\dagger(\bsl{G})\widetilde{u}_{\eta,c,\beta'}(0) :\rho_d(\bsl{r}'):\\
& = \int d^2 r d^2 r' \frac{1}{\A} \sum_{\bsl{G}}\sum_{\bsl{p}} e^{-\ii (\bsl{p}+\bsl{G})\cdot\bsl{r}} V(\bsl{p})  \sum_{\eta,s,\beta,\beta'} : c^\dagger_{\eta,\bsl{r},\beta,s} c_{\eta,\bsl{r},\beta',s}:   e^{\ii  \bsl{p}\cdot \bsl{r}'} \widetilde{u}_{\eta,c,\beta}^\dagger(\bsl{G})\widetilde{u}_{\eta,c,\beta'}(0) :\rho_d(\bsl{r}'):\ .
}
As $\bsl{p}+\bsl{G}$ is carried by $c$ modes and $\bsl{p}$ is carried by $d$ modes, both of them should be small, and thus we should only keep $\bsl{G}=0$ in summation, resulting in 
\eqa{
& \int d^2 r d^2 r' V(\bsl{r}-\bsl{r}') :\rho_{cc}(\bsl{r}): :\rho_d(\bsl{r}'): \\
& = \int d^2 r d^2 r' \frac{1}{\A} \sum_{\bsl{p}} e^{-\ii  \bsl{p}\cdot\bsl{r}} V(\bsl{p})  \sum_{\eta,s,\beta,\beta'} : c^\dagger_{\eta,\bsl{r},\beta,s} c_{\eta,\bsl{r},\beta',s}:   e^{\ii  \bsl{p}\cdot \bsl{r}'} \widetilde{u}_{\eta,c,\beta}^\dagger(0)\widetilde{u}_{\eta,c,\beta'}(0) :\rho_d(\bsl{r}'):\\
& = \int d^2 r d^2 r' \frac{1}{\A} \sum_{\bsl{p}} e^{-\ii  \bsl{p}\cdot(\bsl{r}-\bsl{r}')} V(\bsl{p})  \sum_{\eta,s,\beta} : c^\dagger_{\eta,\bsl{r},\beta,s} c_{\eta,\bsl{r},\beta,s}:   :\rho_d(\bsl{r}'):\\
& = \int d^2 r d^2 r'  V(\bsl{r}-\bsl{r}')  : \rho_c (\bsl{r}):   :\rho_d(\bsl{r}'): = H_{int,V,cd}\ ,
}
where $H_{int,V,cd}$ is defined in \eqnref{eq:H_int_V_cd}.

\subsubsection{In sum }

In sum, we have 
\eq{
H^{TBG-D}_{int} \approx H_{int,W,fd} + H_{int,V,cd}\ ,
}
where  $H_{int,W,fd}$ is defined in \eqnref{eq:H_int_W_fd}, and $H_{int,V,cd}$ is defined in \eqnref{eq:H_int_V_cd}.
Furthermore, we have
\eqa{
H_{int}^{D} \approx H_{int,V,d}
}
with $H_{int,V,d}$ defined in \eqnref{eq:H_int_V_d}.

\section{More Details on the numerical Hartree-Fock Calculations}
\label{app:Hartree_Fock}

In this section, we provide more details on the numerical Hartree-Fock Calculations.

\subsection{Hartree-Fock Hamiltonian}

We first present more details for the Hartree-Fock Hamiltonian. 

In general, given an interacting Hamiltonian of the form 
\eq{
H = \sum_{i,j}\psi^\dagger_i \psi_j t_{ij} + \frac{1}{2}\sum_{i_1,i_2,i_3,i_4} U_{i_1,i_2,i_3,i_4} \psi^\dagger_{i_1} \psi^\dagger_{i_2} \psi_{i_3} \psi_{i_4}
}
with some generic fermion annihilation operator $\psi_i$.
The Hartree-Fock approximation is to choose the ground state as a single Slater determinant:
\eq{
\ket{\Psi} = a^\dagger_1 a^\dagger_2 a^\dagger_3 ...  a^\dagger_N \ket{0}\ ,
}
where $a^\dagger_n = \sum_{i} \psi^\dagger_i (\zeta_{n})_i$ and $\zeta_{1}$, $\zeta_{2}$, ..., $\zeta_{N}$ are orthonormal vectors.
Then, the Hartree-Fock Hamiltonian is derived as 
\eq{
\label{eq:H_HF_gen}
H_{HF} = \sum_{i,j}\psi^\dagger_i \psi_j t_{ij} + \frac{1}{2}\sum_{i_1,i_2,i_3,i_4} U_{i_1 i_2 i_3 i_4} \left( \psi^\dagger_{i_1} \psi_{i_4} O_{i_2 i_3} + \psi^\dagger_{i_2} \psi_{i_3} O_{i_1 i_4} -  \psi^\dagger_{i_2} \psi_{i_4} O_{i_1 i_3} - \psi^\dagger_{i_1} \psi_{i_3} O_{i_2 i_4}\right) - E_{0}\ ,
}
where $O_{i_1 i_2} = \bra{\Psi} \psi^\dagger_{i_1} \psi_{i_2} \ket{\Psi} = \sum_{n=1}^N (\zeta_n^* \zeta_n^T)_{ij}$, and
\eq{
E_0 = \frac{1}{2}\sum_{i_1,i_2,i_3,i_4} U_{i_1 i_2 i_3 i_4} (O_{i_1 i_4}O_{i_2 i_3}-O_{i_1 i_3}O_{i_2 i_4})\ .
}
$H_{HF}$ satisfies $\bra{\Psi}H_{HF}\ket{\Psi}=\bra{\Psi}H\ket{\Psi}$.
Note that $H_{HF}$ has the same form as the mean-field Hamiltonian; in fact, the Hartree-Fock approximation is equivalent to the mean-field approximation.
$O_{i_1 i_2}$ is called the order parameter or the density matrix.
$\bra{\Psi}H_{HF}\ket{\Psi}$ is also called the Hartree-Fock energy.

Now we come back to MATSTG.
We only consider the states that are invariant under the Moir\'e lattice translations.
Moreover, similar to \refcite{Song20211110MATBGHF}, we only care about the following averaged density matrices for simplicity
\eqa{
\label{eq:Omat}
& O^{ff}_{\eta_1\alpha_1 s_1,\eta_2\alpha_2 s_2} = \frac{1}{N}\sum_{\bsl{R}} \ave{f^\dagger_{\eta_1,\bsl{R},\alpha_1,s_1}f_{\eta_2,\bsl{R},\alpha_2,s_2}}\\
& O^{cc}_{\eta_1\beta_1 s_1,\eta_2\beta_2 s_2} = \frac{1}{N}\sum_{\bsl{p}}^{|\bsl{p}|\leq \Lambda_c} \ave{c^\dagger_{\eta_1,\bsl{p},\beta_1,s_1}c_{\eta_2,\bsl{p},\beta_2,s_2}}-\frac{n_{\Lambda_c}}{2} \delta_{\eta_1 \eta_2}\delta_{\beta_1 \beta_2}\delta_{s_1 s_2}\\
& O^{dd}_{\eta\sigma_1 s_1,\eta\sigma_2 s_2} = \frac{1}{N}\sum_{\bsl{p}}^{|\bsl{p}|\leq \Lambda_d} \ave{d^\dagger_{\eta,\bsl{p},\sigma_1,s_1}d_{\eta,\bsl{p},\sigma_2,s_2}}-\frac{n_{\Lambda_d}}{2} \delta_{\sigma_1 \sigma_2}\delta_{s_1 s_2}\\
& O^{dd}_{\eta\beta_1 s_1,-\eta\beta_2 s_2} = \frac{1}{3N}\sum_{\bsl{p}}^{|\bsl{p}|\leq \Lambda_d}\sum_{\bsl{p}'}^{|\bsl{p}'|\leq \Lambda_d}\sum_{n=0,1,2} \delta_{\bsl{p}-\bsl{p}', C_3^n \eta \bsl{q}_1} \ave{d^\dagger_{\eta,\bsl{p},\sigma_1,s_1}d_{-\eta,\bsl{p}',\sigma_2,s_2}}\\
& O^{cf}_{\eta_1\beta_1 s_1,\eta_2\alpha_2 s_2} = \frac{1}{N}\sum_{\bsl{p}}^{|\bsl{p}|\leq \Lambda_c} \ave{c^\dagger_{\eta_1,\bsl{p},\beta_1,s_1}f_{\eta_2,\bsl{p},\alpha_2,s_2}}\\
& O^{fc} = [O^{cf}]^\dagger \\
& O^{df}_{\eta_1\sigma_1 s_1,\eta_2\alpha_2 s_2} = \frac{1}{N}\sum_{\bsl{p}}^{|\bsl{p}|\leq \Lambda_c} \ave{d^\dagger_{\eta_1,\bsl{p},\sigma_1,s_1}f_{\eta_2,\bsl{p}+\eta_1 \KM,\alpha_2,s_2}}\\
& O^{fd} = [O^{df}]^\dagger \\
& O^{cd}_{\eta_1\beta_1 s_1,\eta_2\sigma_2 s_2} = \frac{1}{3N}\sum_{n=0,1,2}\sum_{\bsl{p}}^{|\bsl{p}|\leq \Lambda_c\& |\bsl{p}-\eta_2 C_3^n \bsl{q}_1|\leq \Lambda_d} \ave{c^\dagger_{\eta_1,\bsl{p},\beta_1,s_1}d_{\eta_2,\bsl{p}-\eta_2 C_3^n \bsl{q}_1,\sigma_2,s_2}}\\
& O^{dc} = [O^{cd}]^\dagger \ ,
}
where $\ave{...}$ is the expectation done with respect to Hartree-Fock ground state, $n_{\Lambda_c} = \frac{1}{N}\sum_{\bsl{p}}^{|\bsl{p}|\leq \Lambda_c}$, and $n_{\Lambda_d} = \frac{1}{N}\sum_{\bsl{p}}^{|\bsl{p}|\leq \Lambda_d}$.
We note that the expressions $O^{ff}$, $O^{cc}$, $O^{fc}$ and $O^{cf}$ are the same as those in \refcite{Song20211110MATBGHF}.
We also note that for $d^\dagger_{\eta,\bsl{p},\sigma_1,s_1}$ and $d_{-\eta,\bsl{p}',\sigma_2,s_2}$ in  $O^{dd}_{\eta\beta_1 s_1,-\eta\beta_2 s_2}$, $\bsl{p}$ and $\bsl{p}'$ must be different in order to preserve the Moir\'e lattice translations, owing to the fact that $d^\dagger_{+,\bsl{p}}$ and $d^\dagger_{-,\bsl{p}}$ are around $\KM$ and $-\KM$ points, respectively, as discussed in and below \eqnref{eq:f_c_d_translation}.
Because of the same reason, we choose the $d$ and $f$ modes in $O^{df}_{\eta_1\sigma_1 s_1,\eta_2\alpha_2 s_2}$ as $d^\dagger_{\eta_1,\bsl{p},\sigma_1,s_1}$ and $f_{\eta_2,\bsl{p}+\eta_1 \KM,\alpha_2,s_2}$ to preserve the Moir\'e lattice translations.

Then, combining \eqnref{eq:H_fcd} with \eqnref{eq:Omat} and \eqnref{eq:H_HF_gen}, the Hartree-Fock Hamiltonian reads
\eqa{
\label{eq:H_HF}
& H_{HF}=\sum_{\eta}H^{eff}_{0,\eta} + H_U + H_{V,c} + H_{V,d} + H_{V,cd} + H_{W,fc} + H_{W,fd} + H_J \\
& \quad - ( E_U + E_{V,c} + E_{V,d} + E_{V,cd} + E_{W,fc} + E_{W,fd} + E_J) + const.\ ,
}
where $H^{eff}_{0,\eta}$ is in \eqnref{eq:H_fcd_SP}, $const.$ stands for a scalar that is independent of the ground state,  and the rest of the terms are discussed in the following.
Before going in to details, we define $f_{\bsl{k}}^\dagger = (...,f_{\eta,\bsl{k},\alpha,s}^\dagger,...)$, $f_{\bsl{R}}^\dagger = (...,f_{\eta,\bsl{R},\alpha,s}^\dagger,...)$, $c_{\bsl{p}}^\dagger = (...,c_{\eta,\bsl{p},\beta,s}^\dagger,...)$, and $d_{\bsl{p}}^\dagger = (...,d_{\eta,\bsl{p},\sigma,s}^\dagger,...)$.

First, we go over  $H_U$, $H_{V,c}$, $H_{W,fc}$, $H_{J}$, $E_U$, $E_{V,c}$, $E_{W,fc}$ and $E_{J}$, which are the same as the corresponding Hartree-Fock terms in \refcite{Song20211110MATBGHF} since they only involve the $f$ and $c$ modes derived from the TBG part.
For more details, one can refer to \refcite{Song20211110MATBGHF}.
\eq{
H_U = \sum_{\bsl{R}} \left\{ \rho_f(\bsl{R})\left[ U_1 (\Tr[O^{ff}]-3.5) + 6 U_2 (\Tr[O^{ff}]-4)\right] - U_1 f^\dagger_{\bsl{R}} [O^{ff}]^T f_{\bsl{R}} \right\}\ ,
}
and
\eq{
E_U=\frac{N}{2} \Tr[O^{ff}]^2 ( U_1 + 6 U_2 ) - U_1 \frac{N}{2} \Tr[O^{ff} O^{ff}]\ .
}
\eq{
H_{V,c} = \frac{1}{\Omega}V(\bsl{p}=0) \sum_{\bsl{p}}^{|\bsl{p}|\leq \Lambda_c} c^\dagger_{\bsl{p}} c_{\bsl{p}} \Tr[O^{cc}]\ ,
}
and
\eq{
E_{V,c}=\frac{N}{2\Omega} V(\bsl{p}=0)(\Tr[O^{cc}]^2+ 16 n_{\Lambda_c} \Tr[O^{cc}])\ .
}
Here we neglect the Fock channels for $H_{V,c}$ and $E_{V,c}$, same to \refcite{Song20211110MATBGHF}, since otherwise the Hartree-Fock calculations would heavily depend on the cutoff $\Lambda_c$ due to the simplified density matrices chosen in \eqnref{eq:Omat}. 
\eqa{
\label{eq:H_W_fc}
H_{W,fc} & = \sum_{\bsl{k}\in\MBZ} f^\dagger_{\bsl{k}} f_{\bsl{k}} \Tr\left[O^{cc} \mat{ W_1 \eta_0 \tau_0 s_0 & \\
& W_3 \eta_0 \tau_0 s_0}\right] - \sum_{\bsl{p}}^{|\bsl{p}|\leq \Lambda_c} c^\dagger_{\bsl{p}} \mat{ W_1 \eta_0 \tau_0 s_0 & \\ & W_3 \eta_0 \tau_0 s_0} [O^{fc}]^T f_{\bsl{p}} \\
& \quad - \sum_{\bsl{p}}^{|\bsl{p}|\leq \Lambda_c} f^\dagger_{\bsl{p}} [O^{cf}]^T \mat{ W_1 \eta_0 \tau_0 s_0 & \\ & W_3 \eta_0 \tau_0 s_0}  c_{\bsl{p}} +  \sum_{\bsl{p}}^{|\bsl{p}|\leq \Lambda_c} c^\dagger_{\bsl{p}}  \mat{ W_1 \eta_0 \tau_0 s_0 & \\ & W_3 \eta_0 \tau_0 s_0}  c_{\bsl{p}} (\Tr[O^{ff}]-4)\ ,
}
$\eta_{0,x,y,z}$ are Pauli matrices for the valley index, and
\eq{
E_{W,fc}=N \Tr[O^{ff}] \Tr\left[O^{cc} \mat{ W_1 \eta_0 \tau_0 s_0 & \\
& W_3 \eta_0 \tau_0 s_0}\right] + 2 N n_{\Lambda_c} \left(\sum_{\beta} W_{\beta}\right) (\Tr[O^{ff}]-4) - N \Tr[O^{cf} O^{fc} \mat{ W_1 \eta_0 \tau_0 s_0 & \\
& W_3 \eta_0 \tau_0 s_0}]\ .
}
\eqa{
\label{eq:H_J}
H_{J} & = -\frac{J}{2}\sum_{\bsl{k}\in\MBZ} f^\dagger_{\bsl{k}} \left[\eta_z\tau_0 s_0 (O^{cc}_{\Gamma_1\Gamma_2, \Gamma_1\Gamma_2})^T \eta_z\tau_0 s_0 + \eta_0\tau_z s_0 (O^{cc}_{\Gamma_1\Gamma_2, \Gamma_1\Gamma_2})^T \eta_0\tau_z s_0\right] f_{\bsl{k}} \\
& \quad +\frac{J}{2} \sum_{\bsl{p}}^{|\bsl{p}|\leq \Lambda_c} c^\dagger_{\bsl{p},\Gamma_{1}\Gamma_{2}} \left[\eta_z\tau_0 s_0 \Tr[O^{fc}\mat{ 0_{8\times 8} \\ \eta_z\tau_0 s_0}] + \eta_0\tau_z s_0 \Tr[O^{fc}\mat{ 0_{8\times 8} \\ \eta_0\tau_z s_0}] \right]  f_{\bsl{p}} \\
& \quad +\frac{J}{2} \sum_{\bsl{p}}^{|\bsl{p}|\leq \Lambda_c} f_{\bsl{p}}^\dagger  \left[\eta_z\tau_0 s_0 \Tr[O^{fc}\mat{ 0_{8\times 8} \\ \eta_z\tau_0 s_0}]^* + \eta_0\tau_z s_0 \Tr[O^{fc}\mat{ 0_{8\times 8} \\ \eta_0\tau_z s_0}]^* \right] c_{\bsl{p},\Gamma_{1}\Gamma_{2}}  \\
& \quad - \frac{J}{2} \sum_{\bsl{p}}^{|\bsl{p}|\leq \Lambda_c}   c^\dagger_{\bsl{p},\Gamma_{1}\Gamma_{2}}  \left[\eta_z\tau_0 s_0 (O^{ff})^T \eta_z\tau_0 s_0 + \eta_0\tau_z s_0 (O^{ff})^T \eta_0\tau_z s_0 - \eta_0\tau_0 s_0\right] c_{\bsl{p},\Gamma_{1}\Gamma_{2}} \ ,
}
and
\eqa{
E_J & = -\frac{JN}{2}\sum_{\eta\eta'}\sum_{\alpha\alpha' s s'} O^{ff}_{\eta\alpha s,\eta' \alpha' s'} O^{cc}_{\eta'(\alpha'+2)s',\eta(\alpha+2)s} (\eta \eta' + (-1)^{\alpha+\alpha'}) - J N n_{\Lambda_c} \frac{\Tr[O^{ff}]}{2} \\ 
& \quad + \frac{J N}{2} \sum_{\eta\eta',s s',\alpha\alpha'} O^{cf}_{\eta(\alpha+2)s,\eta\alpha s} O^{fc}_{\eta'\alpha's',\eta'(\alpha'+2) s'}(\eta\eta'+(-1)^{\alpha+\alpha'})\ ,
}
where $O^{cc}_{\Gamma_1\Gamma_2, \Gamma_1\Gamma_2}$ is the $8\times 8$ diagonal block of $O^{cc}$ that correspond to $c^\dagger_{\eta,\bsl{k},\beta=3,4,s}$.

Now we move onto the terms that are not covered in \refcite{Song20211110MATBGHF}. 
First, $H_{int,V,d}$ in \eqnref{eq:H_int_V_d} can be re-written as 
\eqa{
H_{int,V,d} &  = \frac{1}{2}\int d^2 r d^2 r' V(\bsl{r}-\bsl{r}') :\rho_d (\bsl{r}): :\rho_{d}(\bsl{r}'): \\
& =\frac{1}{2}\int d^2 r d^2 r' V(\bsl{r}-\bsl{r}') (\rho_d (\bsl{r})- \frac{4}{\Omega}n_{\Lambda_d}) (\rho_d (\bsl{r}')- \frac{4}{\Omega}n_{\Lambda_d})\\
& =\frac{1}{2}\int d^2 r d^2 r' V(\bsl{r}-\bsl{r}') \left[\rho_d (\bsl{r})\rho_d (\bsl{r}') - \frac{4}{\Omega}n_{\Lambda_d} (\rho_d (\bsl{r})+\rho_d (\bsl{r}'))\right]+const.\\
& = \frac{1}{2}\frac{1}{\A} \sum_{\bsl{p}} \sum_{\bsl{p}_1\bsl{p}_2\bsl{p}_3\bsl{p}_4}^{\Lambda_d} \delta_{\bsl{p}_4, \bsl{p}+\bsl{p}_1} \delta_{\bsl{p}_2, \bsl{p}+\bsl{p}_3} V(\bsl{p}) \sum_{\eta\sigma s}\sum_{\eta'\sigma's'} d^\dagger_{\eta,\bsl{p}_1,\sigma,s} d^\dagger_{\eta',\bsl{p}_2,\sigma',s'} d_{\eta',\bsl{p}_3,\sigma',s'} d_{\eta,\bsl{p}_4,\sigma,s} \\
&\quad + \frac{1}{2}\frac{1}{\A} \sum_{\bsl{p}_1\bsl{p}}^{\Lambda_d} V(\bsl{p}_1 - \bsl{p}) \sum_{\eta\sigma s} d^\dagger_{\eta,\bsl{p},\sigma,s} d^\dagger_{\eta,\bsl{p},\sigma,s}  - \frac{4 n_{\Lambda_d}}{\Omega} V(\bsl{p}=0) \sum_{\bsl{p}}^{\Lambda_d}  \sum_{\eta\sigma s} d^\dagger_{\eta,\bsl{p},\sigma,s} d^\dagger_{\eta,\bsl{p},\sigma,s} \ ,
}
which leads to the following Hartree-Fock $H_{V,d} $
\eqa{
H_{V,d} &  =  \frac{1}{2}\frac{1}{\A} \sum_{\bsl{p}} \sum_{\bsl{p}_1 \bsl{p}_2 \bsl{p}_3 \bsl{p}_4 }^{\Lambda_d} \delta_{\bsl{p}_4, \bsl{p}+\bsl{p}_1} \delta_{\bsl{p}_2, \bsl{p}+\bsl{p}_3} V(\bsl{p}) \sum_{\eta\sigma s}\sum_{\eta'\sigma' s'} \left[ d^\dagger_{\eta,\bsl{p}_1,\sigma,s} d_{\eta,\bsl{p}_4,\sigma,s} \left\langle d^\dagger_{\eta',\bsl{p}_2,\sigma',s'} d_{\eta',\bsl{p}_3,\sigma',s'} \right\rangle  \right.\\
& -  d^\dagger_{\eta,\bsl{p}_1,\sigma,s} d_{\eta',\bsl{p}_3,\sigma',s'}\left\langle d^\dagger_{\eta',\bsl{p}_2,\sigma',s'}  d_{\eta,\bsl{p}_4,\sigma,s} \right\rangle  -  d^\dagger_{\eta',\bsl{p}_2,\sigma',s'}  d_{\eta,\bsl{p}_4,\sigma,s}\left\langle  d^\dagger_{\eta,\bsl{p}_1,\sigma,s} d_{\eta',\bsl{p}_3,\sigma',s'} \right\rangle \\
& \left.+ d^\dagger_{\eta',\bsl{p}_2,\sigma',s'} d_{\eta',\bsl{p}_3,\sigma',s'} \left\langle d^\dagger_{\eta,\bsl{p}_1,\sigma,s} d_{\eta,\bsl{p}_4,\sigma,s} \right\rangle\right] + \frac{1}{2}\frac{1}{\A} \sum_{\bsl{p}_1\bsl{p}}^{\Lambda_d} V(\bsl{p}_1 - \bsl{p}) \sum_{\eta\sigma s} d^\dagger_{\eta,\bsl{p},\sigma,s} d^\dagger_{\eta,\bsl{p},\sigma,s}  \\
& - \frac{4 n_{\Lambda_d}}{\Omega} V(\bsl{p}=0) \sum_{\bsl{p}}^{\Lambda_d}  \sum_{\eta\sigma s} d^\dagger_{\eta,\bsl{p},\sigma,s} d^\dagger_{\eta,\bsl{p},\sigma,s} \\
& \xrightarrow{\text{Neglecting Fock channel}} \frac{1}{\Omega}V(\bsl{p}=0) \sum_{\bsl{p}}^{\Lambda_d} d^\dagger_{\bsl{p}} d_{\bsl{p}} \Tr[O^{dd}]\ ,
}
where we have used \eqnref{eq:Omat}.
Similarly, we get $E_{V,d}$ as
\eqa{
E_{V,d}  = \frac{N}{2\Omega} V(\bsl{p}=0)\left(   \Tr[O^{dd}]^2 +  8 n_{\Lambda_d}\Tr[O^{dd}] \right)\ .
}
Second, $H_{int,V,cd}$ in \eqnref{eq:H_int_V_cd} can be re-written as 
\eqa{
H_{int,V,cd} &  =\int d^2 r d^2 r' V(\bsl{r}-\bsl{r}') (\rho_c (\bsl{r})- \frac{8}{\Omega}n_{\Lambda_c}) (\rho_d (\bsl{r}')- \frac{4}{\Omega}n_{\Lambda_d})\\
& =\int d^2 r d^2 r' V(\bsl{r}-\bsl{r}') \left[\rho_c (\bsl{r})\rho_d (\bsl{r}') - \frac{8}{\Omega}n_{\Lambda_c} \rho_d (\bsl{r}')- \frac{4}{\Omega}n_{\Lambda_d} \rho_c (\bsl{r})\right]+const.\\
& = \frac{1}{\A} \sum_{\bsl{p}} \sum_{\bsl{p}_1\bsl{p}_2}^{\Lambda_c}\sum_{\bsl{p}_3\bsl{p}_4}^{\Lambda_d} \delta_{\bsl{p}_2, \bsl{p}+\bsl{p}_1} \delta_{\bsl{p}_3, \bsl{p}+\bsl{p}_4} V(\bsl{p}) \sum_{\eta\sigma s}\sum_{\eta'\sigma's'} c^\dagger_{\eta,\bsl{p}_1,\beta,s} d^\dagger_{\eta',\bsl{p}_3,\sigma',s'} d_{\eta',\bsl{p}_4,\sigma',s'} c_{\eta,\bsl{p}_2,\beta,s} \\
&\quad - \frac{8 n_{\Lambda_c}}{\Omega} V(\bsl{p}=0) \sum_{\bsl{p}}^{\Lambda_d}  d^\dagger_{\bsl{p}} d_{\bsl{p}}  - \frac{4 n_{\Lambda_d}}{\Omega} V(\bsl{p}=0) \sum_{\bsl{p}}^{\Lambda_c} c^\dagger_{\bsl{p}} c_{\bsl{p}} +const.\ ,
}
which leads to the following Hartree-Fock $H_{int,V,cd}$
\eqa{
H_{V,cd} & = \frac{1}{\A} \sum_{\bsl{p}} \sum_{\bsl{p}_1\bsl{p}_2}^{\Lambda_c}\sum_{\bsl{p}_3\bsl{p}_4}^{\Lambda_d} \delta_{\bsl{p}_2, \bsl{p}+\bsl{p}_1} \delta_{\bsl{p}_3, \bsl{p}+\bsl{p}_4} V(\bsl{p}) \sum_{\eta\sigma s}\sum_{\eta'\sigma's'} \left[c^\dagger_{\eta,\bsl{p}_1,\beta,s} c_{\eta,\bsl{p}_2,\beta,s} \left\langle d^\dagger_{\eta',\bsl{p}_3,\sigma',s'} d_{\eta',\bsl{p}_4,\sigma',s'} \right\rangle \right.\\
& \quad + \left\langle c^\dagger_{\eta,\bsl{p}_1,\beta,s} c_{\eta,\bsl{p}_2,\beta,s} \right\rangle d^\dagger_{\eta',\bsl{p}_3,\sigma',s'} d_{\eta',\bsl{p}_4,\sigma',s'} - \left\langle c^\dagger_{\eta,\bsl{p}_1,\beta,s} d_{\eta',\bsl{p}_4,\sigma',s'} \right\rangle d^\dagger_{\eta',\bsl{p}_3,\sigma',s'}  c_{\eta,\bsl{p}_2,\beta,s} \\
&  \quad \left. - c^\dagger_{\eta,\bsl{p}_1,\beta,s} d_{\eta',\bsl{p}_4,\sigma',s'} \left\langle d^\dagger_{\eta',\bsl{p}_3,\sigma',s'}  c_{\eta,\bsl{p}_2,\beta,s}\right\rangle \right]\\
&\quad - \frac{4 n_{\Lambda_c}}{\Omega} V(\bsl{p}=0) \sum_{\bsl{p}}^{\Lambda_d}  d^\dagger_{\bsl{p}} d_{\bsl{p}}  - \frac{4 n_{\Lambda_d}}{\Omega} V(\bsl{p}=0) \sum_{\bsl{p}}^{\Lambda_c} c^\dagger_{\bsl{p}} c_{\bsl{p}} +const.\\
& \xrightarrow{\text{Neglecting Fock channel}} \frac{1}{\Omega}V(\bsl{p}=0) \sum_{\bsl{p}}^{|\bsl{p}|\leq \Lambda_c} c^\dagger_{\bsl{p}} c_{\bsl{p}} \Tr[O^{dd}] +\frac{1}{\Omega}V(\bsl{p}=0) \sum_{\bsl{p}}^{|\bsl{p}|\leq \Lambda_d} d^\dagger_{\bsl{p}} d_{\bsl{p}} \Tr[O^{cc}]  \ ,
}
Similarly, we get $E_{V,cd}$ as
\eqa{
 E_{V,cd} & = \frac{N}{2\Omega} V(\bsl{p}=0)\left( 2 \Tr[O^{dd}]\Tr[O^{cc}] +  8 n_{\Lambda_d} \Tr[O^{cc}]+ 16 n_{\Lambda_c}  \Tr[O^{dd}] \right)
}
Third, $H_{int,W,fd}$ in \eqnref{eq:H_int_W_fd} can be re-written as 
\eqa{
H_{int,W,fd} &  = \Omega W_{fd} \sum_{\bsl{R}} (\rho_f (\bsl{R})- 4) (\rho_d (\bsl{R})- \frac{4}{\Omega}n_{\Lambda_d})\\
& =\Omega W_{fd} \sum_{\bsl{R}}\rho_f (\bsl{R})\rho_d (\bsl{R}) - 4 \Omega W_{fd} \sum_{\bsl{R}} \rho_d(\bsl{R}) - 4 n_{\Lambda_d} W_{fd} \sum_{\bsl{R}} \rho_f(\bsl{R})+const.\\
& = \frac{W_{fd}}{N} \sum_{\eta\alpha s} \sum_{\eta'\sigma
s'}\sum_{\bsl{k}_1\bsl{k}_4}^{\MBZ}\sum_{\bsl{p}_2\bsl{p}_3}^{\Lambda_d} \sum_{\bsl{G}} \delta_{\bsl{k}_1+\bsl{p}_2, \bsl{p}_3 + \bsl{k}_4 + \bsl{G} } f^\dagger_{\eta,\bsl{k}_1,\alpha,s} d^\dagger_{\eta',\bsl{p}_2,\sigma',s'} d_{\eta',\bsl{p}_3,\sigma',s'} f_{\eta,\bsl{k}_4,\alpha,s} \\
&\quad - 4 W_{fd}  \sum_{\bsl{p}}^{\Lambda_d}  d^\dagger_{\bsl{p}} d_{\bsl{p}}  - 4 n_{\Lambda_d} W_{fd} \sum_{\bsl{k}}^{\MBZ} f^\dagger_{\bsl{k}} f_{\bsl{k}} +const.\ ,
}
which leads to the following Hartree-Fock $H_{int,W,fd}$
\eqa{
\label{eq:H_W_fd}
H_{W,fd} & = \frac{W_{fd}}{N} \sum_{\eta\alpha s} \sum_{\eta'\sigma
s'}\sum_{\bsl{k}_1\bsl{k}_4}^{\MBZ}\sum_{\bsl{p}_2\bsl{p}_3}^{\Lambda_d} \sum_{\bsl{G}} \delta_{\bsl{k}_1+\bsl{p}_2, \bsl{p}_3 + \bsl{k}_4 + \bsl{G} } \left[f^\dagger_{\eta,\bsl{k}_1,\alpha,s} f_{\eta,\bsl{k}_4,\alpha,s} \left\langle d^\dagger_{\eta',\bsl{p}_2,\sigma',s'} d_{\eta',\bsl{p}_3,\sigma',s'} \right\rangle  \right.\\
& \quad \left. + \left\langle f^\dagger_{\eta,\bsl{k}_1,\alpha,s} f_{\eta,\bsl{k}_4,\alpha,s} \right\rangle d^\dagger_{\eta',\bsl{p}_2,\sigma',s'} d_{\eta',\bsl{p}_3,\sigma',s'}   -  \left\langle f^\dagger_{\eta,\bsl{k}_1,\alpha,s}  d_{\eta',\bsl{p}_3,\sigma',s'} \right\rangle d^\dagger_{\eta',\bsl{p}_2,\sigma',s'} f_{\eta,\bsl{k}_4,\alpha,s}   \right.\\
& \quad \left. - f^\dagger_{\eta,\bsl{k}_1,\alpha,s} d_{\eta',\bsl{p}_3,\sigma',s'}  \left\langle d^\dagger_{\eta',\bsl{p}_2,\sigma',s'}  f_{\eta,\bsl{k}_4,\alpha,s}  \right\rangle \right] - 4 W_{fd}  \sum_{\bsl{p}}^{\Lambda_d}  d^\dagger_{\bsl{p}} d_{\bsl{p}}  - 4 n_{\Lambda_d} W_{fd} \sum_{\bsl{k}}^{\MBZ} f^\dagger_{\bsl{k}} f_{\bsl{k}} \\
& = W_{fd} \Tr[O^{dd}] \sum_{\bsl{R}} f^\dagger_{\bsl{R}} f_{\bsl{R}} + W_{fd} (\Tr[O^{dd}]-4)\sum_{\bsl{p}}^{|\bsl{p}|\leq \Lambda_d} d^\dagger_{\bsl{p}} d_{\bsl{p}} - W_{fd} (\sum_{\bsl{p}}^{|\bsl{p}|\leq \Lambda_d}\sum_{\eta\eta'} d^\dagger_{\eta',\bsl{p}} [O^{fd}_{\eta\eta'}]^T f_{\eta,\bsl{p}+\eta' \KM}+h.c.)\ ,
}
where we have used \eqnref{eq:Omat}.
Similarly, we get $E_{W,fd}$ as
\eqa{
E_{W,fd} = N W_{fd} \left( \Tr[O^{dd}] \Tr[O^{ff}] + 4 n_{\Lambda_d} (\Tr[O^{ff}]-4) \right) - N W_{fd} \Tr[O^{df} O^{fd}]\ .
}
Comparing \eqnref{eq:H_W_fd} to \eqnref{eq:H_W_fc}, we can see $d^\dagger_{\eta,\bsl{p}}$ couples to $f_{\eta',\bsl{p}+\eta \KM}$, while $c^\dagger_{\bsl{p}}$ couples to $f_{\bsl{p}}$, showing that $d$ modes are around the $\eta \KM$ points and $c$ modes are around the $\GM$ point.

For the calculation of Hartree-Fock density matrices, we choose $\Lambda_c=\Lambda_d=\sqrt{3}$ ($|\bsl{b}_{\mathrm{M},1}|=|\bsl{b}_{\mathrm{M},2}|=\sqrt{3}$ according to \eqnref{eq:BM_basis_k_Q} as a comparison), and the iteration for the self-consistent calculation stops when the error of the Hartree-Fock ground state energy is smaller than $10^{-4}$meV in EUS.
The initial Hartree-Fock density matrices are given by the initial states, which are specified below.
To rule out the symmetry breaking induced by the artificial cutoffs, we address the momentum points in MBZ in a symmetric way.
Specifically, when we need to sum $\bsl{k}$ over $\MBZ$ for determining the density matrices in \eqnref{eq:Omat}, instead of actually summing $\bsl{k}$ over $\MBZ$, we sum $\bsl{k}$ over the completion of $\MBZ$, \ie, the union of $\MBZ$ with its all edges and corners (shown in \figref{fig:MBZ_edge_corner}), and include a factor of $1/2$ for terms with $\bsl{k}$ on the edge and $1/3$ for terms with $\bsl{k}$ at the corners.

The Hartree-Fock band structures are plotted for $\Lambda_c=\Lambda_d=2\sqrt{3}$, in order to compare with the single-particle band structure.

\begin{figure}[h]
    \centering
    \includegraphics[width=0.6\columnwidth]{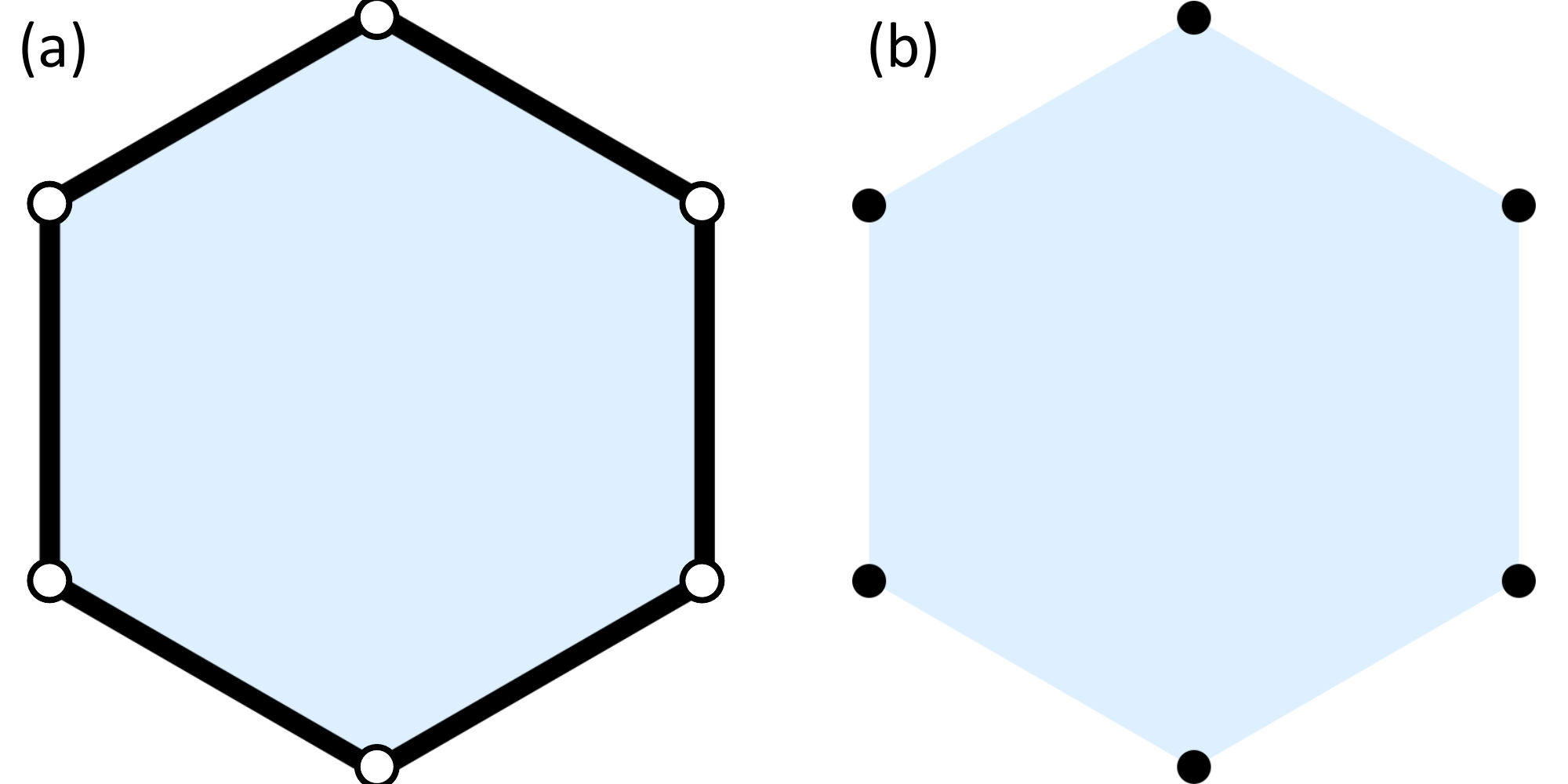}
    \caption{The edge of the MBZ is shown in (a) as the black solid line, where the cornors are excluded. 
        The corners of the MBZ are shown in (b) as the black dots.
        Only half of the edge in (a) belongs to MBZ, and only one third of the corners in (b) belongs to MBZ.
    }
    \label{fig:MBZ_edge_corner}
\end{figure}

\subsection{Initial States}
\label{app:ini_states}

Now we specify the initial states for the self-consistent Hartree-Fock calculations for $\nu=0,-1,-2$.
The choice of the initial states are inspired by the numerical results in \refcite{BAB20210628TSTGII}, which show that (i) the ground state at low-$\E$ is similar to TBG, and (ii) the ground states at high-$\E$ have zero inter-valley coherence.
Therefore, for all the considered fillings, we will include all the initial states that correspond to those used in the study of TBG in \refcite{Song20211110MATBGHF} and include representative states without inter-valley coherence.

Recall that we choose the initial states to have the form of \eqnref{eq:state_ini_HF}, where $\ket{\text{Fermi Sea}}$ stands for the hall-filled Fermi sea of the free $c$ and $d$ modes.
For the initial states without inter-valley coherence, the filling in each valley is well-defined and can be evaluated as $\nu_\eta = \Tr[\zeta_\eta \zeta_\eta^\dagger] - 2$ for the $\eta$ valley, where $\zeta_\eta$ is defined in \eqnref{eq:xi_eta}; we have $\nu_++\nu_-=\nu$.
Then, we choose certain representative initial states without inter-valley coherence for all combinations of $(\nu_+,\nu_-)$ with $\nu_+\leq \nu_-$, since the $\nu_+<\nu_-$ subspace is related to the $\nu_+>\nu_-$ subspace by the TR symmetry.

The initial states that we choose for the self-consistent calculations at $\nu=0$ are
\eq{
\left|\text{VP}^{\nu = 0}_{0} \right\rangle  =  \prod_{\bsl{R}} f^\dagger_{+,\bsl{R},1,\uparrow} f^\dagger_{+,\bsl{R},1,\downarrow} f^\dagger_{+,\bsl{R},2,\uparrow} f^\dagger_{+,\bsl{R},2,\downarrow} 
\ket{\text{Fermi Sea}}\ ,
}
\eq{
\left|\text{IVC}^{\nu = 0}_{0} \right\rangle  = \prod_{\bsl{R}} \frac{1}{4} (f^\dagger_{+,\bsl{R},1,\uparrow}-\ii f^\dagger_{-,\bsl{R},2,\uparrow})(f^\dagger_{+,\bsl{R},1,\downarrow}-\ii f^\dagger_{-,\bsl{R},2,\downarrow})(-\ii f^\dagger_{-,\bsl{R},1,\uparrow}+f^\dagger_{+,\bsl{R},2,\uparrow})(-\ii f^\dagger_{-,\bsl{R},1,\downarrow}+f^\dagger_{+,\bsl{R},2,\downarrow})\ket{\text{Fermi Sea}}\ ,
}
\eq{
\label{eq:KIVC_nu0_ini}
\left|\text{K-IVC}^{\nu = 0}_{0} \right\rangle  = \prod_{\bsl{R}} \frac{1}{4} (f^\dagger_{+,\bsl{R},1,\uparrow}+f^\dagger_{-,\bsl{R},2,\uparrow})(f^\dagger_{+,\bsl{R},1,\downarrow}+f^\dagger_{-,\bsl{R},2,\downarrow})(-f^\dagger_{-,\bsl{R},1,\uparrow}+f^\dagger_{+,\bsl{R},2,\uparrow})(-f^\dagger_{-,\bsl{R},1,\downarrow}+f^\dagger_{+,\bsl{R},2,\downarrow})\ket{\text{Fermi Sea}}\ ,
}
\eq{
\left|\text{PVP}^{1,\nu = 0}_{0} \right\rangle = \prod_{\bsl{R}} f^\dagger_{+,\bsl{R},1,\uparrow} f^\dagger_{+,\bsl{R},1,\downarrow} f^\dagger_{+,\bsl{R},2,\downarrow} f^\dagger_{-,\bsl{R},1,\downarrow} 
\ket{\text{Fermi Sea}}\ ,
}
\eq{
\left|\text{PVP}^{2,\nu = 0}_{0} \right\rangle = \prod_{\bsl{R}} f^\dagger_{+,\bsl{R},1,\uparrow} f^\dagger_{+,\bsl{R},1,\downarrow} f^\dagger_{+,\bsl{R},2,\uparrow} f^\dagger_{-,\bsl{R},2,\downarrow} 
\ket{\text{Fermi Sea}}\ ,
}
\eq{
\left|\VH^{\nu = 0}_{0} \right\rangle  = \prod_{\bsl{R}} f^\dagger_{+,\bsl{R},1,\uparrow} f^\dagger_{+,\bsl{R},1,\downarrow} f^\dagger_{-,\bsl{R},1,\uparrow} f^\dagger_{-,\bsl{R},1,\downarrow} 
\ket{\text{Fermi Sea}}\ ,
}
\eq{
\label{eq:Ch_nu0_ini}
\left|\text{Chern}^{\nu = 0}_{0} \right\rangle  = \prod_{\bsl{R}} f^\dagger_{+,\bsl{R},1,\uparrow} f^\dagger_{+,\bsl{R},1,\downarrow} f^\dagger_{-,\bsl{R},2,\uparrow} f^\dagger_{-,\bsl{R},2,\downarrow} 
\ket{\text{Fermi Sea}}\ ,
}
\eq{
\left|\text{half-Chern}^{\nu = 0}_{0} \right\rangle  = \prod_{\bsl{R}} f^\dagger_{+,\bsl{R},1,\uparrow} f^\dagger_{+,\bsl{R},1,\downarrow} f^\dagger_{-,\bsl{R},1,\downarrow} f^\dagger_{-,\bsl{R},2,\uparrow}  
\ket{\text{Fermi Sea}}\ ,
}
and
\eq{
\left|\text{$C_2\TR$-invariant}^{\nu = 0}_{0} \right\rangle  = \prod_{\bsl{R}}  f^\dagger_{+,\bsl{R},1,\downarrow} f^\dagger_{+,\bsl{R},2,\uparrow} f^\dagger_{-,\bsl{R},1,\downarrow}  f^\dagger_{-,\bsl{R},2,\uparrow}  
\ket{\text{Fermi Sea}}\ .
}
Here $\left|\text{VP}^{\nu = 0}_{0} \right\rangle$, $\left|\text{IVC}^{\nu = 0}_{0} \right\rangle$ and $\left|\text{K-IVC}^{\nu = 0}_{0} \right\rangle$ are chosen because the corresponding states are used in \refcite{Song20211110MATBGHF} for TBG.
$\left|\text{VP}^{\nu = 0}_{0} \right\rangle$ is also a representative state without inter-valley coherence for $(\nu_+,\nu_-)=(2,-2)$.
We choose $\left|\text{PVP}^{1,\nu = 0}_{0} \right\rangle$ and $\left|\text{PVP}^{2,\nu = 0}_{0} \right\rangle $ as the representative states without inter-valley coherence for $(\nu_+,\nu_-)=(1,-1)$.
We choose $\left|\VH^{\nu = 0}_{0} \right\rangle$, $\left|\text{Chern}^{\nu = 0}_{0} \right\rangle$, $\left|\text{half-Chern}^{\nu = 0}_{0} \right\rangle$ and $\left|\text{$C_2\TR$-invariant}^{\nu = 0}_{0} \right\rangle$ as the representative states without inter-valley coherence for $(\nu_+,\nu_-)=(0,0)$.

The initial states that we choose for  the self-consistent calculations at $\nu=-1$ are

\eq{
\left|\text{VP}^{\nu = -1}_{0} \right\rangle  = \prod_{\bsl{R}} f^\dagger_{+,\bsl{R},1,\uparrow} f^\dagger_{+,\bsl{R},1,\downarrow} f^\dagger_{+,\bsl{R},2,\uparrow} 
\ket{\text{Fermi Sea}}\ ,
}

\eq{
\left|\text{IVC}^{\nu = -1}_{0} \right\rangle  = \prod_{\bsl{R}} \frac{1}{2\sqrt{2}} (f^\dagger_{+,\bsl{R},1,\uparrow}-\ii f^\dagger_{-,\bsl{R},2,\uparrow})(f^\dagger_{+,\bsl{R},1,\downarrow}-\ii f^\dagger_{-,\bsl{R},2,\downarrow})(-\ii f^\dagger_{-,\bsl{R},1,\uparrow}+f^\dagger_{+,\bsl{R},2,\uparrow})\ket{\text{Fermi Sea}}\ ,
}

\eq{
\left|\text{VP+IVC}^{\nu = -1}_{0} \right\rangle  = \prod_{\bsl{R}} \frac{1}{2} (f^\dagger_{+,\bsl{R},1,\uparrow}+f^\dagger_{-,\bsl{R},2,\uparrow})(-f^\dagger_{-,\bsl{R},1,\uparrow}+f^\dagger_{+,\bsl{R},2,\uparrow})f^\dagger_{+,\bsl{R},1,\downarrow}\ket{\text{Fermi Sea}}\ ,
}
\eq{
\left|\text{PVP}^{1,\nu = -1}_{0} \right\rangle  = \prod_{\bsl{R}} f^\dagger_{+,\bsl{R},1,\uparrow} f^\dagger_{+,\bsl{R},1,\downarrow} f^\dagger_{-,\bsl{R},1,\downarrow} 
\ket{\text{Fermi Sea}}\ ,
}
\eq{
\left|\text{PVP}^{2,\nu = -1}_{0} \right\rangle  = \prod_{\bsl{R}} f^\dagger_{+,\bsl{R},2,\uparrow} f^\dagger_{+,\bsl{R},2,\downarrow} f^\dagger_{-,\bsl{R},1,\downarrow} 
\ket{\text{Fermi Sea}}\ ,
}
and
\eq{
\left|\text{PVP}^{3,\nu = -1}_{0} \right\rangle  = \prod_{\bsl{R}} f^\dagger_{+,\bsl{R},1,\downarrow} f^\dagger_{+,\bsl{R},2,\uparrow}  f^\dagger_{-,\bsl{R},1,\downarrow} 
\ket{\text{Fermi Sea}}\ .
}
Here $\left|\text{VP}^{\nu = -1}_{0} \right\rangle$, $\left|\text{IVC}^{\nu = -1}_{0} \right\rangle$ and $\left|\text{VP+IVC}^{\nu = -1}_{0} \right\rangle$ are chosen because the corresponding states are used in \refcite{Song20211110MATBGHF} for TBG.
$\left|\text{VP}^{\nu = -1}_{0} \right\rangle$ is also a representative state without inter-valley coherence for $(\nu_+,\nu_-)=(1,-2)$.
We choose $\left|\text{PVP}^{1,\nu = -1}_{0} \right\rangle $, $\left|\text{PVP}^{2,\nu = -1}_{0} \right\rangle$, and $\left|\text{PVP}^{3,\nu = -1}_{0} \right\rangle$ as the representative states without inter-valley coherence for $(\nu_+,\nu_-)=(0,-1)$.

The initial states that we choose for the self-consistent calculations at $\nu=-2$ are

\eq{
\left|\text{K-IVC}^{\nu = -2}_{0} \right\rangle  = \prod_{\bsl{R}} \frac{1}{2} (f^\dagger_{+,\bsl{R},1,\uparrow}+f^\dagger_{-,\bsl{R},2,\uparrow})(f^\dagger_{-,\bsl{R},1,\uparrow}-f^\dagger_{+,\bsl{R},2,\uparrow})\ket{\text{Fermi Sea}}\ ,
}

\eq{
\left|\text{IVC}^{\nu = -2}_{0} \right\rangle  = \prod_{\bsl{R}} \frac{1}{2} (f^\dagger_{+,\bsl{R},1,\uparrow}-\ii f^\dagger_{-,\bsl{R},2,\uparrow})(-\ii f^\dagger_{-,\bsl{R},1,\uparrow}+f^\dagger_{+,\bsl{R},2,\uparrow})\ket{\text{Fermi Sea}}\ ,
}

\eq{
\left|\text{VP}^{\nu = -2}_{0} \right\rangle  = \prod_{\bsl{R}} f^\dagger_{+,\bsl{R},1,\uparrow} f^\dagger_{+,\bsl{R},2,\uparrow}
\ket{\text{Fermi Sea}}\ ,
}

\eq{
\left|\text{VP}^{1,\nu = -2}_{0} \right\rangle  = \prod_{\bsl{R}} f^\dagger_{+,\bsl{R},1,\downarrow} f^\dagger_{+,\bsl{R},2,\uparrow}
\ket{\text{Fermi Sea}}\ ,
}
\eq{
\left|\text{VP}^{2,\nu = -2}_{0} \right\rangle  = \prod_{\bsl{R}} f^\dagger_{+,\bsl{R},1,\uparrow} f^\dagger_{+,\bsl{R},1,\downarrow}
\ket{\text{Fermi Sea}}\ ,
}
\eq{
\left|\text{valley-unpolarized}^{1,\nu = -2}_{0} \right\rangle  = \prod_{\bsl{R}} f^\dagger_{+,\bsl{R},1,\uparrow} f^\dagger_{-,\bsl{R},1,\uparrow}
\ket{\text{Fermi Sea}}\ ,
}
and
\eq{
\left|\text{valley-unpolarized}^{2,\nu = -2}_{0} \right\rangle  = \prod_{\bsl{R}} f^\dagger_{+,\bsl{R},1,\uparrow} f^\dagger_{-,\bsl{R},2,\downarrow}
\ket{\text{Fermi Sea}}\ .
}
Here $\left|\text{K-IVC}^{\nu = -2}_{0} \right\rangle$, $\left|\text{IVC}^{\nu = -2}_{0} \right\rangle$ and $\left|\text{VP}^{\nu = -2}_{0} \right\rangle$ are chosen because the corresponding states are used in \refcite{Song20211110MATBGHF} for TBG.
We choose $\left|\text{VP}^{1,\nu = -2}_{0} \right\rangle$ and $\left|\text{VP}^{2,\nu = -2}_{0} \right\rangle$ (as well as $\left|\text{VP}^{\nu = -2}_{0} \right\rangle$) as the representative states without inter-valley coherence for $(\nu_+,\nu_-)=(0,-2)$.
We choose  $\left|\text{valley-unpolarized}^{1,\nu = -2}_{0} \right\rangle$ and $\left|\text{valley-unpolarized}^{2,\nu = -2}_{0} \right\rangle$ as the representative states without inter-valley coherence for $(\nu_+,\nu_-)=(-1,-1)$.

\section{More Details on Analytical Understanding}
\label{app:ana}

In this section, we provide more details on the analytical understanding.

\subsection{One-Shot Hartree-Fock Hamiltonian}
We develop the analytical understanding by using the one-shot Hartree-Fock Hamiltonian, which is derived as the follows.
First, based on the initial state \eqnref{eq:state_ini_HF}, we can derive the initial density matrices as 
\eqa{
\label{eq:Omat_ini}
& O^{ff}_{ini} = \zeta^*\zeta^T\ ,\\
& [O^{cc}_{ini}]_{\eta \beta s,\eta' \beta' s'} = \delta_{\eta \eta'} \left[Z_{\eta}\right]_{\beta \beta'} \delta_{ss'} \text{ with } \left[Z_{\eta}\right]_{\beta \beta} = 0 \ ,\\
& O^{dd}_{ini} = 0\ ,\\
& O^{cf}_{ini} = 0\ ,\ O^{fc}_{ini} = 0 \\
& O^{df}_{ini} = 0\ ,\ O^{fd}_{ini} = 0 \\
& O^{cd}_{ini} = 0\ ,\ O^{dc}_{ini} = 0 \ ,
}
where $Z_{\eta}$ is a $4\times 4$ matrix
Then, we can substitute \eqnref{eq:Omat_ini} into the Hartree-Fork Hamiltonian \eqnref{eq:H_HF}, and the resultant Hartree-Fock Hamiltonian is the one-shot Hartree-Fock Hamiltonian, which reads
\eqa{
& H_{HF,OS}=\sum_{\eta}H^{eff}_{0,\eta} + H_{U,OS} + H_{V,c,OS} + H_{W,fc,OS} + H_{J,OS} + H_{V,d,OS} + H_{V,cd,OS} + H_{W,fd,OS}  - E_0^{OS} + const.\ ,
}
where $H^{eff}_{0,\eta}$ is in \eqnref{eq:H_fcd_SP}, 
\eq{
E_0^{OS} =  E_{U,OS} + E_{V,c,OS} + E_{W,fc,OS} + E_{J,OS} + E_{V,d,OS} + E_{V,cd,OS} + E_{W,fd,OS}
}
\eq{
H_{U,OS} = \sum_{\bsl{R}}  f^\dagger_{\bsl{R}} h_U f_{\bsl{R}} \ ,
}
\eq{
h_U = \frac{1}{2}U_1 +\nu U_1 + 6 \nu U_2 - U_1 \zeta \zeta^\dagger\ ,
}
\eq{
E_{U,OS}=\frac{N}{2} (4+\nu)^2 ( U_1 + 6 U_2 ) - U_1 \frac{N}{2} (4+\nu)\ ,
}
\eq{
H_{V,c,OS} = 0\ ,
}
\eq{
E_{V,c,OS}=0\ ,
}
\eq{
H_{W,fc,OS} = \sum_{\bsl{p}}^{|\bsl{p}|\leq \Lambda_c} c^\dagger_{\bsl{p}} h_{W,fc} c_{\bsl{p}}\ ,
}
\eq{
h_{W,fc} = \mat{ \nu W_1 \eta_0 \tau_0 s_0 & \\ & \nu W_3 \eta_0 \tau_0 s_0}\ ,
}
\eq{
E_{W,fc,OS}=2 N \nu n_{\Lambda_c}(2W_1+2W_3)+const.\ ,
}
\eq{
H_{J,OS} =  \sum_{\bsl{p}}^{|\bsl{p}|\leq \Lambda_c} c^\dagger_{\bsl{p}} h_J c_{\bsl{p}} \ ,
}
\eq{
h_J = \mat{ 0_{8\times 8} & \\
 & \eta_z\tau_0 s_0 \zeta\zeta^\dagger \eta_z\tau_0 s_0 + \eta_0\tau_z s_0 \zeta\zeta^\dagger \eta_0\tau_z s_0 - \eta_0\tau_0 s_0 }\ ,
}
\eqa{
E_{J,OS} = -\frac{J}{2}\nu N n_{\Lambda_c} + const.
}
\eqa{
H_{V,d,OS} = H_{V,cd,OS}  = 0 \ ,
}
\eqa{
& E_{V,d,OS}= const.\\ 
& E_{V,cd,OS} = const.\ ,
}
\eq{
H_{W,fd,OS}  = W_{fd} \nu \sum_{\bsl{p}}^{|\bsl{p}|\leq \Lambda_d}  d^\dagger_{\bsl{p}} d_{\bsl{p}}\ ,
}
and
\eqa{
E_{W,fd,OS} = 4 \nu N W_{fd} n_{\Lambda_d}+const.\ .
}
Here ``const." consists of scalar terms that do not depend on the density matrices.
It is clear that the dependence of $E_0^{OS}$ on the ground state is only through the filling $\nu$, which is solely determined by the $f$ modes at the one-shot level.
Therefore, the energy difference for different states with the same filling only comes from the operator part, which we will focus on in the following.

\subsection{Simple Rule for High-$\E$ States: High-$\E$ Limit }
\label{app:simple_rule_high_E}

Now we provide more details on the analytical understanding of the simple rule for high-$\E$ states, under the high-$\E$ limit.
Here high-$\E$ limit mean that we choose $|\E|$ to be infinitely large compared with all other energy quantities.
We also approximate the chemical potential as 
\eq{
\label{eq:mu_approx}
\mu = \nu(U_1+6U_2)\ ,
}
which is the correction of the chemical potential due to the density-density interaction of $f$ modes \cite{Song20211110MATBGHF}.
The validity of these simplifications will be discussed in \appref{app:simple_rule_300_nu0}.
In \appref{app:simple_rule_300_nu0}, we will demonstrate the validity of those approximations for $\nu = 0$.

Throughout this part, we choose $\nu\in \{0,-1,-2\}$.
As an effective theory, we will focus on the Hartree-Fock Hamiltonian at $\pm \KM$ and $\Gamma_{\text{M}}$.
We will first consider $\pm \KM$ and then consider $\Gamma_{\text{M}}$.

\subsubsection{$\pm\KM$}

The one-shot Hartree-Fock Hamiltonain around $\pm\KM$ reads
\eqa{
H_{HF,OS}^{\eta \KM}  = \sum_{\bsl{p}}^{|\bsl{p}|<\Lambda_d} (f^\dagger_{+,\eta \KM+\bsl{p}}, f^\dagger_{-,\eta \KM+\bsl{p}}, d^\dagger_{\eta,\bsl{p}} ) h_{HF,OS}^{\eta \KM}(\bsl{p}) 
\mat{
 f_{+,\eta \KM+\bsl{p}} \\
 f_{-,\eta \KM+\bsl{p}} \\ 
 d_{\eta,\bsl{p}} 
}
}
\eq{
h_{HF,OS}^{\KM}(\bsl{p}) =
\left(\begin{array}{c|c}
\frac{1}{2} U_1 +\nu (U_1+6U_2) - U_1 \zeta\zeta^\dagger & \begin{array}{c}  M_1 \E (\tau_0+\ii\tau_z)s_0   \\  0_{4\times 4 }   \end{array} \\
\hline
 \begin{array}{cc}  M_1 \E (\tau_0-\ii\tau_z)s_0  &  0_{4\times 4 }   \end{array} &  W_{fd} \nu + (p_x \sigma_x + p_y \sigma_y)s_0
\end{array}\right) 
}
and
\eq{
h_{HF,OS}^{-\KM}(\bsl{p}) =
\left(\begin{array}{c|c}
\frac{1}{2} U_1 +\nu (U_1+6U_2) - U_1 \zeta\zeta^\dagger & \begin{array}{c}  0_{4\times 4 } \\ M_1 \E (\tau_0-\ii\tau_z)s_0     \end{array} \\
\hline
 \begin{array}{cc} 0_{4\times 4 } &  M_1 \E (\tau_0+\ii\tau_z)s_0  \end{array} &  W_{fd} \nu + (-p_x \sigma_x + p_y \sigma_y)s_0
\end{array}\right) \ .
}
By performing $ d^\dagger_{\eta,\bsl{p}} \rightarrow d^\dagger_{\eta,\bsl{p}} e^{-\ii \tau_z \frac{\pi}{4}}s_0$, we have
\eq{
\label{eq:ht_HFOS_K}
h_{HF,OS}^{\KM}(\bsl{p}) \rightarrow  \widetilde{h}_{HF,OS}^{\KM}(\bsl{p})=
\left(\begin{array}{c|c}
\frac{1}{2} U_1 +\nu (U_1+6U_2) - U_1 \zeta\zeta^\dagger & \begin{array}{c}  \sqrt{2} M_1 \E  \tau_0 s_0   \\  0_{4\times 4 }   \end{array} \\
\hline
 \begin{array}{cc}  \sqrt{2} M_1 \E  \tau_0 s_0  &  0_{4\times 4 }   \end{array} &  W_{fd} \nu + (-p_x \sigma_y + p_y \sigma_x)s_0
\end{array}\right) 
}
and
\eq{
\label{eq:ht_HFOS_mK}
h_{HF,OS}^{-\KM}(\bsl{p}) \rightarrow \widetilde{h}_{HF,OS}^{-\KM}(\bsl{p})=
\left(\begin{array}{c|c}
\frac{1}{2} U_1 +\nu (U_1+6U_2) - U_1 \zeta\zeta^\dagger & \begin{array}{c}  0_{4\times 4 } \\ \sqrt{2} M_1 \E  \tau_0 s_0     \end{array} \\
\hline
 \begin{array}{cc} 0_{4\times 4 } &  \sqrt{2} M_1 \E  \tau_0 s_0  \end{array} &  W_{fd} \nu + (-p_x \sigma_y - p_y \sigma_x)s_0
\end{array}\right) \ ,
}
which are convenient to use.

Since we focus on $\pm\KM$ in this part, we only consider $\bsl{p}=0$ for $f^\dagger_{+,\eta \KM+\bsl{p}}$, $f^\dagger_{-,\eta \KM+\bsl{p}}$ and $d^\dagger_{\eta,\bsl{p}}$.
To proceed, let us define the following two unitary matrices: 
\eq{
\widetilde{U}_{+\KM} = \mat{ \chi_{0,1} & \chi_{1,1} & \\ & & 1 \\ \chi_{0,2} & \chi_{1,2} & } \otimes \mathds{1}_{4\times 4}
\text{ and }
\widetilde{U}_{-\KM} = \mat{  &  & 1 \\ \chi_{0,1} & \chi_{1,1} &  \\ \chi_{0,2} & \chi_{1,2} & } \otimes \mathds{1}_{4\times 4}\ ,
}
where 
\eq{
\mat{ 
\nu (U_1 + 6 U_2) & \sqrt{2} M_1 \E \\ 
\sqrt{2} M_1 \E & W_{fd} \nu
} \chi_{\gamma} = \epsilon_\gamma \chi_{\gamma} \ ,
}
$\gamma = 0,1$, $\chi_\gamma = (\chi_{\gamma,1}, \chi_{\gamma,2})$ is real, and 
\eq{
\label{eq:gamma_epsilon}
\epsilon_\gamma = \frac{\nu (U_1 + 6 U_2 + W_{fd} )}{2} + (-)^\gamma \sqrt{\left[ \frac{\nu (U_1+ 6 U_2 - W_{fd})}{2} \right]^2 + 2 M_1^2 \E^2 }\ .
}
Then, we use $\widetilde{U}_{\eta \KM}$ to unitarily transformation $\widetilde{h}_{HF,OS}^{\eta \KM}(0)$ to
\eq{
\label{eq:transformed_htilde}
\widetilde{U}_{\eta \KM}^\dagger \widetilde{h}_{HF,OS}^{\eta \KM}(0) \widetilde{U}_{\eta \KM}
=
\mat{ 
\epsilon_0 \mathds{1}_{4\times 4} & & \\
 & \epsilon_1 \mathds{1}_{4\times 4} & \\
 & & \nu(U_1+6U_2)\mathds{1}_{4\times 4}
} - U_1 
\left(\begin{array}{c|c}
\mat{ |\chi_{0,1}|^2 & \chi_{0,1}^* \chi_{1,1} \\ \chi_{1,1}^* \chi_{0,1} & |\chi_{1,1}|^2} \otimes (\zeta_\eta \zeta_\eta^\dagger - \frac{1}{2})  & \begin{array}{c}  \chi_{0,1}^* \zeta_{\eta}\zeta_{-\eta}^\dagger \\  \chi_{1,1}^* \zeta_{\eta}\zeta_{-\eta}^\dagger    \end{array} \\
\hline
 \begin{array}{cc}  \chi_{0,1} \zeta_{-\eta}\zeta_{\eta}^\dagger &  \chi_{1,1} \zeta_{-\eta}\zeta_{\eta}^\dagger    \end{array} &  \zeta_{-\eta} \zeta_{-\eta}^\dagger - \frac{1}{2}
\end{array}\right) \ .
}
We perform the transformation in \eqnref{eq:transformed_htilde} because (i) it gives a block-diagonal term (\ie, the first term) that has the three blocks with energies $\epsilon_0$, $\epsilon_1$ and $\nu(U_1+6U_2)$, and (ii) the gaps among $\epsilon_0$, $\epsilon_1$ and $\nu(U_1+6U_2)$ are of order $|\E|$ according to \eqnref{eq:gamma_epsilon}, which is much larger than $U_1$ in the high-$\E$ limit.
Therefore, in \eqnref{eq:transformed_htilde}, the elements (of the second term) that couple different blocks in the first term can only change the eigenvalues at the order of $O(|U_1/\E|)$.

In the following, we will neglect all corrections to the energies that are of order $O(|U_1/\E^2|)$.
Then, we only need to consider the following Hamiltonian 
\eq{
\label{eq:h_HF_OS_pmK}
\widetilde{U}_{\eta \KM}^\dagger \widetilde{h}_{HF,OS}^{\eta \KM}(0) \widetilde{U}_{\eta \KM}
\approx 
\mat{ 
\epsilon_0 \mathds{1}_{4\times 4} - U_1 |\chi_{0,1}|^2  (\zeta_\eta \zeta_\eta^\dagger - \frac{1}{2})& & \\
 & \epsilon_1 \mathds{1}_{4\times 4} - U_1 |\chi_{1,1}|^2  (\zeta_\eta \zeta_\eta^\dagger - \frac{1}{2}) & \\
 & & \nu(U_1+6U_2)\mathds{1}_{4\times 4} -  U_1  (\zeta_{-\eta} \zeta_{-\eta}^\dagger - \frac{1}{2})
}\ .
}
Recall that $\epsilon_1<\nu(U_1+6U_2)<\epsilon_0$ and the gaps between them are of order $O( |\E| )$, which is much larger than $U_1$.
Then, based on the expression of the chemical potential in \eqnref{eq:mu_approx}, the $\epsilon_0$ block should be fully empty, while the $\epsilon_1$ block should be fully occupied.
Eventually, we know that the occupied states of \eqnref{eq:h_HF_OS_pmK} are all eigenstates of 
\eq{
\label{eq:OS_K_contribution_1}
[\epsilon_1-\nu(U_1+6U_2)] \mathds{1}_{4\times 4} - U_1 |\chi_{1,1}|^2  (\zeta_\eta \zeta_\eta^\dagger - \frac{1}{2}) \text{ for both $\eta=\pm$, }
}
and all negative-energy states of 
\eq{
\label{eq:OS_K_contribution_2}
- U_1 (\zeta_{\eta} \zeta_{\eta}^\dagger - \frac{1}{2})\text{ for both $\eta=\pm$. }
}
We have subtracted the chemical potential in \eqnref{eq:OS_K_contribution_1} and \eqnref{eq:OS_K_contribution_2} compared to the corresponding block in \eqnref{eq:h_HF_OS_pmK}.
We label the total energy of all those occupied states as $E_{\pm \KM}$.

We want to minimize $E_{\pm \KM}$.
Recall that \eqnref{eq:OS_K_contribution_1} should be fully occupied for both $\eta=\pm$.
To express the remaining contribution to $E_{\pm \KM}$, we use $\lambda_{i}$ $(i=1,2,...,8)$ to label the eight eigenvalues of 
\eq{
\mat{
\zeta_{+} \zeta_{+}^\dagger & \\
 & \zeta_{-} \zeta_{-}^\dagger
}\ ,
}
since all negative-energy states of \eqnref{eq:OS_K_contribution_2} are all negative-energy states of
\eq{
- U_1 \left[
\mat{
\zeta_{+} \zeta_{+}^\dagger & \\
 & \zeta_{-} \zeta_{-}^\dagger
} - \frac{1}{2}\right]\ .
}
We choose $\lambda_1\geq \lambda_2\geq ... \geq \lambda_8$ without loss of generality, and choose $n$ to be the largest integer that gives $\lambda_n\geq 1/2$.
Then, according to \eqnref{eq:OS_K_contribution_1} and \eqnref{eq:OS_K_contribution_2}, we have
\eqa{
\label{eq:E_pmK_1}
E_{\pm \KM} & = 8 [\epsilon_1-\nu(U_1+6U_2)] - U_1  |\chi_{1,1}|^2  \sum_{\eta} \Tr[\zeta_\eta \zeta_\eta^\dagger - \frac{1}{2}] - U_1 \sum_{i=1}^n (\lambda_i -\frac{1}{2}) + O(|U_1|^2/|\E|) \\
& = 8 [\epsilon_1-\nu(U_1+6U_2)] - U_1  |\chi_{1,1}|^2  \nu - U_1 \sum_{i=1}^n (\lambda_i -\frac{1}{2})+ O(|U_1|^2/|\E|)\ ,
}
where we have used
\eq{
\sum_{\eta} \Tr[\zeta_\eta \zeta_\eta^\dagger] = \Tr[\zeta \zeta^\dagger ] = 4+\nu\ .
}

To proceed, let us derive the constraints on $\lambda_i$.
First, as $\mat{
\zeta_{+} \zeta_{+}^\dagger & \\
 & \zeta_{-} \zeta_{-}^\dagger
}$ is positive semi-definite, $\lambda_i\geq 0$.
Second, $\lambda_i\leq 1$.
To see this, recall that $\zeta$ defined in \eqnref{eq:xi} is a $8\times (4+\nu)$ matrix whose columns ($\zeta_{1},...,\zeta_{4+\nu}$) are orthonormal.
Then, there exists $4-\nu$ $8$-component vectors, $\bar{\zeta}_1$,...,$\bar{\zeta}_{4-\nu}$, such that $\zeta_{1},...,\zeta_{4+\nu}$ and $\bar{\zeta}_1$,...,$\bar{\zeta}_{4-\nu}$ form an orthonormal basis of $\dsC^8$.
Let us define the $\bar{\zeta}=\mat{ \bar{\zeta}_1 & ... & \bar{\zeta}_{4-\nu}}$ as a $8\times (4-\nu)$ matrix, whose columns are orthonormal and which satisfies $\bar{\zeta}^\dagger \zeta = 0 $ and $\zeta \zeta^\dagger + \bar{\zeta} \bar{\zeta}^\dagger = \mathds{1}_{8\times 8}$.
Then, we have 
\eq{
\mat{\zeta_+ \zeta^\dagger_+ & \zeta_+ \zeta^\dagger_- \\ \zeta_- \zeta^\dagger_+ & \zeta_- \zeta^\dagger_- } + \mat{\bar{\zeta}_+ \bar{\zeta}^\dagger_+ & \bar{\zeta}_+ \bar{\zeta}^\dagger_- \\ \bar{\zeta}_- \bar{\zeta}^\dagger_+ & \bar{\zeta}_- \bar{\zeta}^\dagger_- } = \mathds{1}_{8\times 8} \Rightarrow \zeta_\eta \zeta^\dagger_\eta + \bar{\zeta}_\eta \bar{\zeta}^\dagger_\eta = \mathds{1}_{4\times 4}\ .
}
Combined with the fact that $ \bar{\zeta}_\eta \bar{\zeta}^\dagger_\eta$ is also positive semi-definite, we can get $\lambda_i\leq 1$.
Third, 
\eq{
\sum_{i=1}^8 \lambda_i =\sum_{\eta} \Tr[\zeta_\eta \zeta_\eta^\dagger] = 4 + \nu\ .
}
In sum, we know $\lambda_i\in[0,1]$ and $\sum_{i=1}^8 \lambda_i = 4 + \nu$.

With the constraints on $\lambda_i$, we have
\eq{
\label{eq:sum_n_lambda}
\sum_{i=1}^n \lambda_i = 4 + \nu \leq \frac{4+\nu}{2}\ .
}
To see this, we first consider $n>4+\nu$, which gives 
\eq{
\sum_{i=1}^n (\lambda_i-\frac{1}{2}) = \sum_{i=1}^n  \lambda_i-\frac{n}{2} \leq \sum_{i=1}^8  \lambda_i-\frac{n}{2} = 4+\nu - \frac{n}{2} = \frac{4+\nu}{2} + \frac{4+\nu-n}{2} < \frac{4+\nu}{2}\ .
}
For $n<4+\nu$, we have
\eq{
\sum_{i=1}^n (\lambda_i-\frac{1}{2}) \leq \sum_{i=1}^n  \frac{1}{2} < \frac{4+\nu}{2}\ .
}
For $n=4+\nu$, we have
\eq{
\sum_{i=1}^n (\lambda_i-\frac{1}{2}) = \sum_{i=1}^n  \lambda_i-\frac{n}{2} \leq \sum_{i=8}^n  \lambda_i-\frac{n}{2} = \frac{4+\nu}{2}\ .
}
Therefore, we proved  \eqnref{eq:sum_n_lambda} and we know the equality in \eqnref{eq:sum_n_lambda} only happens when $n= 4+\nu$.

\eqnref{eq:sum_n_lambda} and \eqnref{eq:E_pmK_1} give
\eq{
E_{\pm \KM} \geq  8 [\epsilon_1-\nu(U_1+6U_2)] - U_1 |\chi_{1,1}|^2  \nu - U_1 \frac{4+\nu}{2} + O(|U_1|^2/|\E|) \ .
}
Then, we know that the lowest $E_{\pm \KM}$ is achieved if and only if $\sum_{i=1}^n \lambda_i = 4 + \nu$, which only appears for $n= 4+\nu$.
Owing to $\sum_{i=1}^8 \lambda_i = 4 + \nu$, we have
\eqa{
& \sum_{i=1}^n \lambda_i = 4 + \nu \\
& \Leftrightarrow \sum_{i=1}^n \lambda_i = 4 + \nu\ \&\ n= 4+\nu \\
& \Leftrightarrow \lambda_1 = \lambda_2 = ... = \lambda_{4+\nu}=1 \\
& \Leftrightarrow \lambda_1 = \lambda_2 = ... = \lambda_{4+\nu}=1\ \&\ \lambda_{4+\nu+1} = \lambda_{4+\nu+2} = ... = \lambda_{8}\ = 0\ .
}
Therefore, the lowest $E_{\pm \KM}$ is achieved if and only if 
\eq{
\label{eq:mat_sim_xi_pm}
\mat{
\zeta_{+} \zeta_{+}^\dagger & \\
 & \zeta_{-} \zeta_{-}^\dagger
} 
\cong
\text{diag}(\underbrace{1,1,...,1}_{4+\nu},\underbrace{0,0,...,0}_{4-\nu})\ ,
}
where $\cong$ stands for matrix similarity defined by unitary transformations in $\U(8)$.
\eqnref{eq:mat_sim_xi_pm} suggests that $\mat{
\zeta_{+} \zeta_{+}^\dagger & \\
 & \zeta_{-} \zeta_{-}^\dagger
} $ is a projection matrix.
Then, we know 
\eq{
\Tr\left[ \mat{
\zeta_{+} \zeta_{+}^\dagger & \\
 & \zeta_{-} \zeta_{-}^\dagger
} \mat{
\zeta_{+} \zeta_{+}^\dagger & \\
 & \zeta_{-} \zeta_{-}^\dagger
}  \right] =\Tr\left[ \mat{
\zeta_{+} \zeta_{+}^\dagger & \\
 & \zeta_{-} \zeta_{-}^\dagger
}  \right] = \Tr\left[ \zeta \zeta^\dagger \right] =  \Tr\left[ \zeta \zeta^\dagger \zeta \zeta^\dagger \right]\ ,
}
which results in 
\eq{
\Tr[  \zeta_{+} \zeta_{+}^\dagger \zeta_{+} \zeta_{+}^\dagger ]  + \Tr[  \zeta_{-} \zeta_{-}^\dagger \zeta_{-} \zeta_{-}^\dagger ]
=
\Tr[  \zeta_{+} \zeta_{+}^\dagger \zeta_{+} \zeta_{+}^\dagger ]  + \Tr[  \zeta_{-} \zeta_{-}^\dagger \zeta_{-} \zeta_{-}^\dagger ] + \Tr[  \zeta_{+} \zeta_{-}^\dagger \zeta_{-} \zeta_{+}^\dagger ]  + \Tr[  \zeta_{-} \zeta_{+}^\dagger \zeta_{+} \zeta_{-}^\dagger ]\ ,
}
which results in 
\eq{
\Tr[  \zeta_{+} \zeta_{-}^\dagger \zeta_{-} \zeta_{+}^\dagger ] = 0  \Rightarrow  \zeta_{+} \zeta_{-}^\dagger = 0 \ .
}
Combined with 
\eq{
\zeta_{+} \zeta_{-}^\dagger = 0\Rightarrow \mat{
\zeta_{+} \zeta_{+}^\dagger & \\
 & \zeta_{-} \zeta_{-}^\dagger
} = \zeta \zeta^\dagger \Rightarrow \mat{
\zeta_{+} \zeta_{+}^\dagger & \\
 & \zeta_{-} \zeta_{-}^\dagger
} 
\cong
\text{diag}(\underbrace{1,1,...,1}_{4+\nu},\underbrace{0,0,...,0}_{4-\nu})\ ,
}
we know $\zeta_{+} \zeta_{-}^\dagger = 0$ is equivalent to \eqnref{eq:mat_sim_xi_pm}.
Therefore, in the high-$\E$ limit, the lowest $E_{\pm \KM}$ is achieved if and only if $\zeta_{+} \zeta_{-}^\dagger = 0$ (\ie, the intervalley coherence is zero), if we neglect all corrections to the energies that are of order $O(|U_1/\E^2|)$.

\subsubsection{$\Gamma_{\text{M}}$}

Now let us turn to the $\Gamma_{\text{M}}$ point.
The one-shot Hartree-Fock Hamiltonian at $\Gamma_{\text{M}}$ reads
\eq{
\label{eq:h_OS_Gamma}
\mat{
\frac{1}{2}U_1 + \nu (U_1 + 6 U_2) - U_1 \zeta \zeta^\dagger & \widetilde{\gamma} \mathds{1}_8 & \\
\widetilde{\gamma} \mathds{1}_8 & \nu W_1 & \\
 & & \nu W_3 + h_{\Gamma_{1}\Gamma_{2}}\ ,
}
}
where $\widetilde{\gamma} = \gamma + B_\gamma \E^2$, 
\eq{
h_{\Gamma_{1}\Gamma_{2}} = \widetilde{M} \eta_0\sigma_x s_0 - \frac{J}{2} \left( \eta_z \sigma_0 s_0 \zeta \zeta^\dagger \eta_z\sigma_0 s_0 + \eta_0 \sigma_z s_0 \zeta \zeta^\dagger \eta_0\sigma_z s_0  - \mathds{1}_8 \right)\ ,
}
and $\widetilde{M} = M + B_M \E^2$.
Owing to
\eq{
\frac{1}{2}U_1 + \nu (U_1 + 6 U_2) - U_1 \zeta \zeta^\dagger
\cong
\mat{ 
(-\frac{1}{2}U_1 + \nu (U_1 + 6 U_2))\mathds{1}_{(4+\nu)}   & \\
 & (\frac{1}{2}U_1 + \nu (U_1 + 6 U_2))\mathds{1}_{(4-\nu)}  
}\ ,
}
the eigenvalues of 
\eq{
\mat{
\frac{1}{2}U_1 + \nu (U_1 + 6 U_2) - U_1 \zeta \zeta^\dagger & \widetilde{\gamma} \mathds{1}_8  \\
\widetilde{\gamma} \mathds{1}_8 & \nu W_1  
}
}
does not depend on the $\zeta$ as long as $\nu$ is given.
Therefore, we will focus on $h_{\Gamma_{1}\Gamma_{2}}$.

Since we consider the high-$\E$ limit, we have $|\widetilde{M}|\gg J$.
Then, the energy difference between different states given by $h_{\Gamma_{1}\Gamma_{2}}$ should be of order $J$, which is generally much smaller than the energy difference at $\pm\KM$ which is of the order $U_1$.
Therefore, we should only focus on the states with lowest $E_{\pm \KM}$, $\ie$ states with zero intervalley coherence.
In other words, the discussion at $\pm\KM$ already suggests that only states without intervalley coherence are favored at large $\E$.

Now we show that  $h_{\Gamma_{1}\Gamma_{2}}$ further picks out the favored high-$\E$ states among all states without intervalley coherence.
Since we now only care about the states without intervalley coherence (\ie, $\zeta_+\zeta_-^\dagger=0$), we have 
\eq{
\zeta \zeta^\dagger
=
\mat{
\zeta_{+} \zeta_{+}^\dagger & \\
 & \zeta_{-} \zeta_{-}^\dagger
}\ .
}
In general, $\zeta_\eta \zeta_\eta^\dagger$ has the following form
\eq{
\label{eq:xi_eta_sim}
\zeta_\eta \zeta_\eta^\dagger = \sum_{\mu\nu\in\{0,x,y,z\}} \left(y_\eta\right)_{\mu\nu} \sigma_\mu s_\nu\ ,
}
where $\left(y_\eta\right)_{\mu\nu}$ are the real coefficients.
Owing to the spin-charge $U(2)$ symmetries in each valley, namely $U(2)\times U(2)$, we can always first rotate $\sum_{\nu\in\{0,x,y,z\}} \left(y_\eta\right)_{z\nu} \sigma_z s_\nu$ to $\sum_{\nu\in\{0,z\}} \left(y_\eta\right)_{z\nu} \sigma_z s_\nu$, and then rotate $\sum_{\nu\in\{0,x,y,z\}} \left(y_\eta\right)_{0\nu} \sigma_0 s_\nu$ to $\sum_{\nu\in\{0,x,z\}} \left(y_\eta\right)_{0\nu} \sigma_0 s_\nu$.
Therefore, we have 
\eq{
\left(y_\eta\right)_{zx} = \left(y_\eta\right)_{zy} = \left(y_\eta\right)_{0y} = 0
}
up to $U(2)\times U(2)$.
With this observation, we have the following expression
\eq{
\zeta_\eta \zeta_\eta^\dagger
=a_{0,\eta} + a_{\eta}\sigma_0 s_z + c_\eta \sigma_0 s_x + \frac{b_{2-\eta}+b_{3-\eta}}{2}\sigma_z s_0 + \frac{b_{2-\eta}-b_{3-\eta}}{2}\sigma_z s_z +\sum_{\mu\in\{0,z\},\nu\in\{0,x,y,z\}} \left(y_\eta\right)_{\mu\nu} \sigma_\mu s_\nu
}
up to $U(2)\times U(2)$.
Then, 
\eq{
h_{\Gamma_{1}\Gamma_{2}} = \mat{ \widetilde{h}_{+} & \\ & \widetilde{h}_{-} }
}
up to $U(2)\times U(2)$, where
\eq{
\widetilde{h}_{\eta} = \widetilde{M} \sigma_x s_0 -J \left[a_{0,\eta} + a_{\eta}\sigma_0 s_z + c_\eta \sigma_0 s_x + \frac{b_{2-\eta}+b_{3-\eta}}{2}\sigma_z s_0 + \frac{b_{2-\eta}-b_{3-\eta}}{2}\sigma_z s_z\right] + \frac{J}{2}.
}
According to \eqnref{eq:h_OS_Gamma}, the eigenstates of $h_{\Gamma_{1}\Gamma_{2}}$ with energies lower than $\nu (U_1 + 6 U_2- W_3)$ are occupied.
Before proceeding, we list some useful constraints derived from $\zeta \zeta^\dagger$ being a projection matrix of rank $4+\nu$.
First, $\zeta_\eta\zeta_\eta^\dagger \zeta_\eta\zeta_\eta^\dagger = \zeta_\eta\zeta_\eta^\dagger$ gives
\eqa{
& a_{0,\eta} = \frac{m_\eta}{4}\\
& a_{0,\eta}^2 + a_\eta^2 + c_\eta^2 + \frac{b^2_{2-\eta}+b^2_{3-\eta}}{2} + \sum_{\mu\in\{0,z\},\nu\in\{0,x,y,z\}} \left(y_\eta\right)_{\mu\nu}^2   = \frac{m_\eta}{4}\ ,
}
where $m_\eta  = \Tr[ \zeta_\eta\zeta_\eta^\dagger ]\in \dsZ$ and $m_+ + m_- = 4+\nu$.
Then, owing to the fact that the diagonal elements of $\zeta_\eta\zeta_\eta^\dagger$ are in $[0,1]$, we have
\eqa{
& a_{0,\eta} + a_{\eta} + b_{2-\eta}, a_{0,\eta} + a_{\eta} - b_{2-\eta}, a_{0,\eta} - a_{\eta} + b_{3-\eta}, a_{0,\eta} - a_{\eta} - b_{3-\eta} \in [0,1] \\
& \Rightarrow  
\left\{
\begin{array}{l}
 a_\eta + b_{2-\eta}\in [-\frac{m_\eta}{4}, 1 - \frac{m_\eta}{4}] \ \&\ a_\eta - b_{2-\eta}\in [-\frac{m_\eta}{4}, 1 - \frac{m_\eta}{4}]  \\
  -a_\eta + b_{3-\eta}\in [-\frac{m_\eta}{4}, 1 - \frac{m_\eta}{4}]\ \&\ -a_\eta - b_{3-\eta}\in [-\frac{m_\eta}{4}, 1 - \frac{m_\eta}{4}] 
\end{array}
\right. \\
& \Rightarrow  
\left\{
\begin{array}{l}
 a_\eta \in [ -\frac{m_\eta}{4} , 1 - \frac{m_\eta}{4} ] \ \&\ \pm b_{2-\eta}\in [-\frac{m_\eta}{4} - a_\eta, 1 - \frac{m_\eta}{4} - a_\eta ]  \\
  a_\eta \in [-1 + \frac{m_\eta}{4} , \frac{m_\eta}{4}]\ \&\ \pm b_{3-\eta}\in [-\frac{m_\eta}{4} + a_\eta, 1 - \frac{m_\eta}{4} + a_\eta] 
\end{array}
\right. \\
& \Rightarrow |a_\eta|\in \left[0,\min(\frac{m_\eta}{4},1-\frac{m_\eta}{4})\right] \ \&\ |b_{2-\eta}|\in \left[0,\min(\frac{m_\eta}{4}+a_\eta,1-\frac{m_\eta}{4}-a_\eta)\right]\ \&\ |b_{3-\eta}|\in \left[0,\min(\frac{m_\eta}{4}-a_\eta,1-\frac{m_\eta}{4}+a_\eta)\right]\ .
}
Then, since we only care about $\nu\in\{ -4,-3,-2,-1,0 \}$, we have
\eqa{
& |a_\eta| = |b_{2-\eta}| = |b_{3-\eta}| = 0 \ ,\ \text{for } m_\eta = 0 \\
& |a_\eta| \in [0,\frac{m_\eta}{4} ] \ ,\  |b_{2-\eta}| \in [0, \frac{m_\eta}{4} + a_\eta ]\ ,\ |b_{3-\eta}| \in [ 0 , \frac{m_\eta}{4} -a_\eta] \ ,\ \text{for } m_\eta = 1 \\
& |a_\eta| \in [0,\frac{1}{2} ] \ ,\  |b_{2-\eta}| \in [0, \frac{1}{2} - |a_\eta| ]\ ,\ |b_{3-\eta}| \in [ 0 , \frac{1}{2} -|a_\eta|] \ ,\ \text{for } m_\eta = 2 \\
& |a_\eta| \in [0, 1 - \frac{m_\eta}{4} ] \ ,\  |b_{2-\eta}| \in [0, 1 - \frac{m_\eta}{4} - a_\eta ]\ ,\ |b_{3-\eta}| \in [ 0 , 1- \frac{m_\eta}{4} + a_\eta ] \ ,\ \text{for } m_\eta = 3 \\
& |a_\eta| = |b_{2-\eta}| = |b_{3-\eta}| = 0 \ ,\ \text{for } m_\eta = 4\ ,
}
which leads to 
\eq{
b_{2-\eta}^2 + b_{3-\eta}^2 \leq \min(\frac{m_\eta}{4}, 1- \frac{m_\eta}{4})\ .
}
In sum, the constraints that we will use are summarized as
\eqa{
\label{eq:constraints}
& a_{0,\eta} = \frac{m_\eta}{4}\ ,\ m_\eta \in \dsZ_{\geq 0}\ ,\ m_+ + m_- = 4+\nu \\
& a_{0,\eta}^2 + a_\eta^2 + c_\eta^2 + \frac{b^2_{2-\eta}+b^2_{3-\eta}}{2} + \sum_{\mu\in\{0,z\},\nu\in\{0,x,y,z\}} \left(y_\eta\right)_{\mu\nu}^2   = \frac{m_\eta}{4}\\ 
& |a_\eta|\in \left[0,\min(\frac{m_\eta}{4},1-\frac{m_\eta}{4})\right] \ ,\ |b_{2-\eta}|\in \left[0,\min(\frac{m_\eta}{4}+a_\eta,1-\frac{m_\eta}{4}-a_\eta)\right]\ ,\ |b_{3-\eta}|\in \left[0,\min(\frac{m_\eta}{4}-a_\eta,1-\frac{m_\eta}{4}+a_\eta)\right]\\\
& b_{2-\eta}^2 + b_{3-\eta}^2 \leq \min(\frac{m_\eta}{4}, 1- \frac{m_\eta}{4})\ .
}

We label the four eigenvalues of $\widetilde{h}_{\eta}$ as $E_{\eta,1}^{\Gamma_{\text{M}}}\leq E_{\eta,2}^{\Gamma_{\text{M}}} \leq E_{\eta,3}^{\Gamma_{\text{M}}} \leq E_{\eta,4}^{\Gamma_{\text{M}}}$.
Owing to $|\widetilde{M}|\gg J$, we can solve the eigenvalues perturbatively to $O(M_J^{-2})$, where $M_J=|\widetilde{M}/J|$.
To do so, we use the following unitary matrix
\eq{
\widetilde{U}_{\eta} = \frac{1}{\sqrt{2}} 
\mat{
1 & 1 \\
1 & -1
}
\otimes s_0 \ ,
}
\eq{
\widetilde{U}_{\eta}^\dagger \frac{\widetilde{h}_\eta}{J} \widetilde{U}_{\eta}
=
\frac{2-m_\eta}{4}  + 
\mat{
- M_J - a_\eta s_z - c_\eta s_x & \mat{ -b_{2-\eta} & \\ & -b_{3-\eta}} \\
 \mat{ -b_{2-\eta} & \\ & -b_{3-\eta}} & M_J -a_\eta s_z - c_\eta s_x 
}\ ,
}
where we used the fact that $\widetilde{M}<0$ and $J>0$.
Then, we can project the the off-diagonal $b_{2-\eta}$ and $b_{3-\eta}$ terms to the diagonal block, and get two effective Hamiltonians from $\widetilde{U}_{\eta}^\dagger \frac{\widetilde{h}_\eta}{J} \widetilde{U}_{\eta}$ as
\eq{
\frac{2-m_\eta}{4}  \pm M_J - a_\eta s_z -c_\eta s_x \pm \frac{1}{2 M_J} \mat{ -b_{2-\eta} & \\ & -b_{3-\eta}}^2 + O(M_J^{-2})
}
leading to 
\eq{
\label{eq:EJ_approx}
E_{\eta,i}^{\GM}/J = \frac{2-m_\eta}{4} + (-1)^{\lceil i/2 \rceil}\left[M_J+\frac{1}{4} (b_{2-\eta}^2+ b_{3-\eta}^2)M_J^{-1}\right] + (-1)^i \left[\sqrt{a_\eta^2 + c_\eta^2}+O(M_J^{-1})\right] +O(M_J^{-2})\ ,
}
where $i=1,2,3,4$ and $\lceil i/2 \rceil$ is the smallest interger that is no smaller than $i/2$.
Since we have $|\widetilde{M}|\gg |\nu(U_1+6 U_2-W_3)|$ in the high-$\E$ limit, $E_{\eta,i}^{\GM}$ with $i=1,2$ are the only occupied levels, leading to
\eq{
\label{eq:E_Gamma}
E_{\Gamma_{\text{M}}}/J = - 4 M_J  - \frac{1}{2}(b_1^2+b_2^2+b_3^2+b_4^2) M_J^{-1} + E_{\nu,\Gamma_{\text{M}}}/J + O(M_J^{-2}) \ ,
}
where $E_{\nu=0,\Gamma_{\text{M}}}$ contains the other contribution that does not rely on $\zeta$ as long as $\nu$ is fixed.
To lower $E_{\Gamma_{\text{M}}}$, we just need to maximize $b_1^2+b_2^2+b_3^2+b_4^2$.
In the following, we will do it for $\nu=0,-1,-2$ separately.

For $\nu = 0$, we have three cases distinguished by the values of $m_{\pm}$, \ie, there exists $\eta_0\in\{+,-\}$ such that $(m_{\eta_0}, m_{-\eta_0}) = (4,0), (3,1), (2,2)$, which respectively leads to 
\eq{
\sum_{\eta} \min(\frac{m_\eta}{4}, 1- \frac{m_\eta}{4}) = 0, \frac{1}{2}, 1\ .
}
Owing to \eqnref{eq:constraints}, we then have
\eq{
E_{\Gamma_{\text{M}}}/J \geq - 4 M_J -\frac{1}{2} M_J^{-1} + E_{\nu=0,\Gamma_{\text{M}}}/J + O(M_J^{-2})\ .
}
Then, by exploiting \eqnref{eq:constraints}, 
\eqa{
& E_{\Gamma_{\text{M}}}/J = - 4 M_J -\frac{1}{2} M_J^{-1} + E_{\nu=0,\Gamma_{\text{M}}}/J + O(M_J^{-2}) \text{, \ie, minimizing $E_{\Gamma_{\text{M}}}$ for $\nu=0$ states without intervalley coherence} \\
& \Leftrightarrow 
\left\{ 
\begin{array}{l}
  m_+ = m_- = 2  \\
   b_1^2+b_2^2+b_3^2+b_4^2 = 1 
\end{array}
\right. \\
& \Leftrightarrow 
\left\{ 
\begin{array}{l}
   m_+ = m_- = 2  \\
   b_1^2+b_2^2=b_3^2+b_4^2 = \frac{1}{2} 
\end{array}
\right. \\
& \Leftrightarrow 
\left\{ 
\begin{array}{l}
 m_+ = m_- = 2  \\
 |b_1|=|b_2|=|b_3|=|b_4| = \frac{1}{2} \\
 a_\eta^2 + c_\eta^2 + \sum_{\mu\in\{0,z\},\nu\in\{0,x,y,z\}} \left(y_\eta\right)_{\mu\nu}^2   = 0 \ \forall \eta\in\{+,-\}
\end{array}
\right.\ .
}
Then, combined with \eqnref{eq:constraints} and \eqnref{eq:xi_eta_sim}, it means that $E_{\Gamma_{\text{M}}}$ is minimized if and only if $\zeta \zeta^\dagger$ is (up to $U(2)\times U(2)$) spin-diagonal with each of the 4 valley-spin blocks being $\frac{1}{2}(1\pm \sigma_z)$.

For $\nu=-1$, we have two cases distinguished by the values of $m_{\pm}$, \ie, there exists $\eta_0\in\{+,-\}$ such that $(m_{\eta_0}, m_{-\eta_0}) = (3,0), (2,1)$, which respectively leads to 
\eq{
\sum_{\eta} \min(\frac{m_\eta}{4}, 1- \frac{m_\eta}{4}) = \frac{1}{4}, \frac{3}{4} \ .
}
Owing to \eqnref{eq:constraints}, we then have
\eq{
E_{\Gamma_{\text{M}}}/J \geq - 4 M_J - \frac{3}{8} M_J^{-1} + E_{\nu=-1,\Gamma_{\text{M}}}/J + O(M_J^{-2})\ .
}
Then, by exploiting \eqnref{eq:constraints}, 
\eqa{
& E_{\Gamma_{\text{M}}}/J = - 4 M_J -\frac{3}{8} M_J^{-1} + E_{\nu=0,\Gamma_{\text{M}}}/J + O(M_J^{-2}) \text{, \ie, minimizing $E_{\Gamma_{\text{M}}}$ for $\nu=-1$ states without intervalley coherence} \\
& \Leftrightarrow 
\left\{ 
\begin{array}{l}
 m_{\eta_0} =2,\  m_{-\eta_0} = 1  \\
 b_1^2+b_2^2+b_3^2+b_4^2 = \frac{3}{4} 
\end{array}
\right. \\
& \Leftrightarrow 
\left\{ 
\begin{array}{l}
 m_{\eta_0} =2,\  m_{-\eta_0} = 1  \\
 b_{2-\eta_0}^2+b_{3-\eta_0}^2= \frac{1}{2}\ ,\ b_{2+\eta_0}^2+b_{3+\eta_0}^2 = \frac{1}{4} 
\end{array}
\right. \\
& \Leftrightarrow 
\left\{ 
\begin{array}{l}
 m_{\eta_0} =2 \ ,\  |b_{2-\eta_0}|=|b_{3-\eta_0}|= \frac{1}{2}\ ,\ a_{\eta_0}^2 + c_{\eta_0}^2 + \sum_{\mu\in\{0,z\},\nu\in\{0,x,y,z\}} \left(y_{\eta_0}\right)_{\mu\nu}^2   = 0 \\
m_{-\eta_0} = 1  \ ,\ |a_{-\eta_0}|\in [0,\frac{1}{4}]\ ,\ \frac{1}{4}\leq (\frac{1}{4}+a_{\eta_0})^2+(\frac{1}{4}-a_{-\eta_0})^2\ ,\ b_{2+\eta_0}^2+b_{3+\eta_0}^2 = \frac{1}{4} 
\end{array}
\right.
\\
& \Leftrightarrow 
\left\{ 
\begin{array}{l}
 m_{\eta_0} =2 \ ,\  |b_{2-\eta_0}|=|b_{3-\eta_0}|= \frac{1}{2}\ ,\ a_{\eta_0}^2 + c_{\eta_0}^2 + \sum_{\mu\in\{0,z\},\nu\in\{0,x,y,z\}} \left(y_{\eta_0}\right)_{\mu\nu}^2   = 0 \\
m_{-\eta_0} = 1  \ ,\ a_{-\eta_0} = \pm \frac{1}{4}, |b_{2-\eta_0}|= a_{-\eta_0} + \frac{1}{4}, |b_{3-\eta_0}| = -a_{-\eta_0} + \frac{1}{4} \ ,\  c_{-\eta_0}^2 + \sum_{\mu\in\{0,z\},\nu\in\{0,x,y,z\}} \left(y_{-\eta_0}\right)_{\mu\nu}^2   = 0
\end{array}
\right.\ .
}
Then, combined with \eqnref{eq:constraints} and \eqnref{eq:xi_eta_sim}, it means that $E_{\Gamma_{\text{M}}}$ is minimized if and only if $\zeta \zeta^\dagger$ is (up to $U(2)\times U(2)$)  spin-diagonal with 1 valley-spin block being zero and each of the remaining 3 valley-spin blocks being $\frac{1}{2}(1\pm \sigma_z)$.

For $\nu=-2$, we have two cases distinguished by the values of $m_{\pm}$, \ie, there exists $\eta_0\in\{+,-\}$ such that $(m_{\eta_0}, m_{-\eta_0}) = (2,0), (1,1)$, which both leads to 
\eq{
\sum_{\eta} \min(\frac{m_\eta}{4}, 1- \frac{m_\eta}{4}) = \frac{1}{2} \ .
}
Owing to \eqnref{eq:constraints}, we then have
\eq{
E_{\Gamma_{\text{M}}}/J \geq - 4 M_J -\frac{1}{4} M_J^{-1} + E_{\nu=-2,\Gamma_{\text{M}}}/J + O(M_J^{-2})\ .
}
Then, by combining \eqnref{eq:constraints} with the discussions for $\nu=0$ and $\nu=-1$, we can get that $E_{\Gamma_{\text{M}}}$ is minimized if and only if $\zeta \zeta^\dagger$ is (up to $U(2)\times U(2)$) spin-diagonal with 2 valley-spin block being zero and each of the remaining 2 valley-spin blocks being $\frac{1}{2}(1\pm \sigma_z)$.

Therefore, we obtain the statement that for states without intervalley coherence, $E_{\Gamma_{\text{M}}}$ is the lowest if and only of $\zeta \zeta^\dagger$ (up to $U(2)\times U(2)$) has a spin-valley diagonal form with $4+\nu$ blocks being $(\sigma_0\pm\sigma_z)/2$ and $4-\nu$ blocks being zero.

\subsubsection{In sum}

Let us summarize the whole procedure.
In the high-$\E$ limit and given $\nu\in \{0,-1,-2\}$, we require the ground states to first minimize the total energy of the occupied levels of \eqnref{eq:h_HF_OS_pmK}, which make them have zero intervalley-coherence, and then minimize the total energy of the occupied levels of \eqnref{eq:h_OS_Gamma}.
Then, we arrive at the \propref{prop:high_field_states}.

We numerate all initial states that satisfy \propref{prop:high_field_states} for $\nu=0,-1,-2$.
All the states we found are (up to the symmetries of the total interacting Hamiltonian \eqnref{eq:H_tot_sym}) included in \appref{app:ini_states}.

Explicitly, at $\nu=0$, the states that satisfy \propref{prop:high_field_states} are $\left|\VH^{\nu = 0}_{0} \right\rangle$, $\left|\text{Chern}^{\nu = 0}_{0} \right\rangle$, $\left|\text{half-Chern}^{\nu = 0}_{0} \right\rangle$, $\left|\text{$C_2\TR$-invariant}^{\nu = 0}_{0} \right\rangle$ and their symmetry-related states.

At $\nu=-1$, the states that satisfy \propref{prop:high_field_states} are $\left|\text{PVP}^{1,\nu = -1}_{0} \right\rangle$, $\left|\text{PVP}^{2,\nu = -1}_{0} \right\rangle$ and $\left|\text{PVP}^{3,\nu = -1}_{0} \right\rangle$ and their symmetry-related states.

At $\nu=-2$, the states that satisfy \propref{prop:high_field_states} are $\left| \text{VP}^{1,\nu = -2}_{0} \right\rangle$, $\left|\text{VP}^{2,\nu = -2}_{0} \right\rangle $, $\left|\text{valley-unpolarized}^{1,\nu = -2}_{0} \right\rangle$, $\left|\text{valley-unpolarized}^{2,\nu = -2}_{0} \right\rangle$ and their symmetry-related states.

According to \secref{sec:NHF_cal}, these inital states, after performing the self-consistent calculations, give the high-$\E$ low-energy states with very similar energies (similar for a fixed $\nu$).

\subsection{Simple Rule for High-$\E$ States: $\E = 300\meV$ in EUS}
\label{app:simple_rule_300_nu0}

(Recall that EUS is the unit system in which $\AA$ is the length unit and $\meV$ is the energy unit, as discussed at the beginning of \secref{sec:BM_TSTG}.)

In \appref{app:simple_rule_high_E}, we analytically derive \propref{prop:high_field_states} by looking at $\pm \KM$ and $\Gamma_{\text{M}}$ in the high-$\E$ limit, \ie, assuming an infinitely large $\E$.
Furthermore, we assume $\mu = \nu (U_1 + 6 U_2)$.
However, as shown in \figref{fig:bands_SP}, we can only claim the validity of the $f-c-d$ model in $\E \in [0,300\meV]$ (EUS).
Therefore, in this part, we will discuss the validity of \appref{app:simple_rule_high_E} for $\E=300\meV$ (EUS).

First, we note that $|\widetilde{M}|\gg |\nu(U_1+6 U_2-W_3)|$ need to hold in order to use \eqnref{eq:E_Gamma}.
For $\E=300\meV$ (EUS), we have $|\widetilde{M}|\approx 0.14$.
However, for $\nu=-1$, we have $|\nu(U_1+6 U_2-W_3)|\approx 0.12$, which is close to $|\widetilde{M}|$; for $\nu=-2$, we have $|\nu(U_1+6 U_2-W_3)|\approx 0.24$, which is larger than $|\widetilde{M}|$.
On the other hand, the same issue does not occur for $\nu=0$ since $|\nu(U_1+6 U_2-W_3)| = 0$.
Therefore, the simplifications in \appref{app:simple_rule_high_E} are not all valid for $\nu=-1,-2$.

Now let us focus on $\nu =0$, for which $ \mu=\nu (U_1 + 6 U_2) = 0$ is exactly correct.
We discuss $\pm \KM$ first.
We have $\epsilon_\gamma = (-)^\gamma |\sqrt{2} M_1 \E|$ and $\chi_\gamma = \frac{1}{\sqrt{2}} (1,(-)^\gamma \sgn{M_1\E})$ according to \eqnref{eq:gamma_epsilon}.
Then, \eqnref{eq:transformed_htilde} becomes
\eq{
\widetilde{U}_{\eta \KM}^\dagger \widetilde{h}_{HF,OS}^{\eta \KM}(0) \widetilde{U}_{\eta \KM}
=
\mat{ 
\epsilon_0 \mathds{1}_{4\times 4} & & \\
 & \epsilon_1 \mathds{1}_{4\times 4} & \\
 & & 0\mathds{1}_{4\times 4}
} - U_1 
\left(\begin{array}{c|c}
\mat{ \frac{1}{2} & \frac{1}{2} \\ \frac{1}{2} & \frac{1}{2}} \otimes (\zeta_\eta \zeta_\eta^\dagger - \frac{1}{2})  & \begin{array}{c}  \frac{1}{\sqrt{2}} \zeta_{\eta}\zeta_{-\eta}^\dagger \\  \frac{1}{\sqrt{2}} \zeta_{\eta}\zeta_{-\eta}^\dagger    \end{array} \\
\hline
 \begin{array}{cc}  \frac{1}{\sqrt{2}} \zeta_{-\eta}\zeta_{\eta}^\dagger &  \frac{1}{\sqrt{2}} \zeta_{-\eta}\zeta_{\eta}^\dagger    \end{array} &  \zeta_{-\eta} \zeta_{-\eta}^\dagger - \frac{1}{2}
\end{array}\right) \ .
}
Since $P_{ii}\in[0,1]$ and $|P_{i\neq j}| \in [0,\frac{1}{2}]$ for any hermitian projectoin matrix $P$, the elements that couple different $\epsilon_0 \mathds{1}_{4\times 4}$, $\epsilon_1 \mathds{1}_{4\times 4}$ and $0\mathds{1}_{4\times 4}$ blocks have amplitudes no larger than $\frac{1}{2\sqrt{2}} U_1$, while the gaps among those blocks are no smaller than $|\sqrt{2} M_1 \E|$.
The energy contributions of the elements that couple different $\epsilon_0 \mathds{1}_{4\times 4}$, $\epsilon_1 \mathds{1}_{4\times 4}$ and $0\mathds{1}_{4\times 4}$ blocks are of the order $\left|\frac{U_1}{4 M_1 \E}\right|^2\approx 0.3$, which can be neglected.
Then, it is legitimate to only consider the \eqnref{eq:h_HF_OS_pmK}, which eventually leads to the fact that only states without inter-valley coherent should be considered.

Now turn to $\Gamma_{\text{M}}$.
In \eqnref{eq:EJ_approx}, the terms that we neglect compared to the largest-order term are of order $M_J^{-3}\approx 0.29$, which is also reasonable.
Then, the later derivation based on \eqnref{eq:EJ_approx} in \appref{app:simple_rule_high_E} should all be valid, leading to \propref{prop:high_field_states}.
Therefore, the derivation in \appref{app:simple_rule_high_E} should be valid for $\nu=0$ even if $\E=300\meV$ (EUS).

The numerical evidence for the validity for $\nu=0$ and $\E=300\meV$ (EUS) is that if we only compare the energies of the occupied levels of the one-shot Hamiltonian at $\pm \KM$ and $\Gamma_{\text{M}}$ for the initial states in \appref{app:ini_states} at $\nu=0$ and $\E=300\meV$ (EUS), we can get the right ground states, as shown in \tabref{tab:nu0_sim}.

\begin{table}[H]
    \centering
    \begin{tabular}{|c|c|c|c|c|}
    \hline
    \text{Initial States} & $\Gamma_{\text{M}}$ & $\KM$ & $- \KM$ & Total\\
    \hline
        $\left|\text{K-IVC}^{\nu = 0}_{0} \right\rangle$ & -1.30093 & -1.15152 & -1.15152 & -3.60397 \\
        \hline
        $\left|\text{IVC}^{\nu = 0}_{0} \right\rangle$ & -1.27088 & -1.15152 & -1.15152 & -3.57392 \\
        \hline
        $\left|\text{VP}^{\nu = 0}_{0} \right\rangle$ & -1.27088 & -1.32885 & -1.32885 & -3.92858 \\
        \hline
        $\left|\text{PVP}^{1,\nu = 0}_{0} \right\rangle $ & -1.28591 & -1.32885 & -1.32885 & -3.9436 \\
        \hline
        $\left|\text{PVP}^{2,\nu = 0}_{0} \right\rangle $ & -1.28591 & -1.32885 & -1.32885 & -3.9436\\
        \hline
        $\left|\VH^{\nu = 0}_{0} \right\rangle$ & -1.30093 & -1.32885 & -1.32885 & -3.95863 \\
        \hline
        $\left|\text{Chern}^{\nu = 0}_{0} \right\rangle$ & -1.30093 & -1.32885 & -1.32885 & -3.95863 \\
        \hline
        $\left|\text{half-Chern}^{\nu = 0}_{0} \right\rangle$ & -1.30093 & -1.32885 & -1.32885 & -3.95863 \\
        \hline
        $\left|\text{$C_2\TR$-invariant}^{\nu = 0}_{0} \right\rangle$ & -1.30093 & -1.32885 & -1.32885 & -3.95863 \\
        \hline
    \end{tabular}
    \caption{This table shows the energies of the occupied levels of the one-shot Hamiltonian at $\pm \KM$ and $\Gamma_{\text{M}}$ for the initial states in \appref{app:ini_states} at $\nu=0$ and $\E=300\meV$ (EUS).
    The first column specifies the initial states.
    The second, third and fourth columns specifies the energies of the occupied levels of the one-shot Hamiltonian at $\Gamma_{\text{M}}$, $\KM$ and $-\KM$, respectively.
    The fifth column shows the total of the second, third and fourth columns.
    The states with the lowest total are the lowest four, which are the high-$\E$ ground states found in the numerical calculations described in \appref{app:Hartree_Fock}. 
    }
    \label{tab:nu0_sim}
\end{table}

\subsection{One-Shot Hartree-Fock Energies for High-$\E$ States}

\label{app:one-shot_sym_energy}

The competing energies of the high-$\E$ Hartee-Fock ground states with fixed $\nu$ can also be understood analytically at one-shot level.

At $\nu=0$, the one-shot Hartree-Fock Hamiltonains for different high-$\E$ initial states are related with each other:
\eq{
\label{eq:VH_OS_nu0}
H^{\VH,OS}= H^{\VH,OS}_{+,\uparrow}+H^{\VH,OS}_{+,\downarrow}+H^{\VH,OS}_{-,\uparrow}+H^{\VH,OS}_{-,\downarrow}-E_0^{OS}\ ,
}
\eq{
\label{eq:Chern_OS_nu0}
H^{\Chern,OS}= H^{\VH,OS}_{+,\uparrow}+H^{\VH,OS}_{+,\downarrow}+ C_2\overline{\TR} H^{\VH,OS}_{-,\uparrow}(C_2\overline{\TR})^{-1} + C_2\overline{\TR} H^{\VH,OS}_{-,\downarrow} (C_2\overline{\TR})^{-1}-E_0^{OS}\ ,
}

\eq{
H^{\text{half-Chern},OS}= H^{\VH,OS}_{+,\uparrow}+H^{\VH,OS}_{+,\downarrow}+ C_2\overline{\TR} H^{\VH,OS}_{-,\uparrow}(C_2\overline{\TR})^{-1} + H^{\VH,OS}_{-,\downarrow}-E_0^{OS}\ ,
}

\eq{
H^{C_2\TR\text{-invaraint},OS}= C_2\overline{\TR} H^{\VH,OS}_{+,\uparrow}(C_2\overline{\TR})^{-1} +H^{\VH,OS}_{+,\downarrow}+ C_2\overline{\TR} H^{\VH,OS}_{-,\uparrow} (C_2\overline{\TR})^{-1}+ H^{\VH,OS}_{-,\downarrow}-E_0^{OS}\ ,
}
where 
\eq{
H^{\VH,OS}_{\eta,s} = H_{0,\eta,s} + (-\frac{1}{2}U_1)\sum_{\bsl{R}_M} f^\dagger_{\eta,\bsl{R}_M,s} \tau_z f_{\eta,\bsl{R}_M,s} + (-\frac{J}{2})\sum_{\bsl{k}}^{\Lambda_c} c^\dagger_{\eta,\bsl{k},\Gamma_1\Gamma_2,s} \tau_z c_{\eta,\bsl{k},\Gamma_1\Gamma_2,s} \ ,
}
$H_{0,\eta,s}$ is the spin-s part of $H_{0,\eta}$ which is specified in \eqnref{eq:H_fcd_SP}, and $\overline{\TR}$ is the spinless time-reversal symmetry.

At $\nu=-1$, the one-shot Hartree-Fock Hamiltonians for different high-$\E$ initial states are related:
\eq{
H^{\text{PVP}_1,OS}= H^{\text{PVP}_1,OS}_{+,\uparrow}+H^{\text{PVP}_1,OS}_{+,\downarrow}+H^{\text{PVP}_1,OS}_{-,\uparrow}+H^{\text{PVP}_1,OS}_{-,\downarrow}-E_0^{OS}\ ,
}
\eq{
H^{\text{PVP}_2,OS}= C_2\overline{\TR} H^{\text{PVP}_1,OS}_{+,\uparrow} (C_2\overline{\TR})^{-1}+ C_2\overline{\TR} H^{\text{PVP}_1,OS}_{+,\downarrow} (C_2\overline{\TR})^{-1}+H^{\text{PVP}_1,OS}_{-,\uparrow}+H^{\text{PVP}_1,OS}_{-,\downarrow}-E_0^{OS}\ ,
}
\eq{
H^{\text{PVP}_3,OS}= C_2\overline{\TR} H^{\text{PVP}_1,OS}_{+,\uparrow} (C_2\overline{\TR})^{-1}+  H^{\text{\text{PVP}}_1,OS}_{+,\downarrow} +H^{\text{PVP}_1,OS}_{-,\uparrow}+H^{\text{PVP}_1,OS}_{-,\downarrow}-E_0^{OS}\ ,
}
where 
\eqa{
H^{\text{PVP}_1,OS}_{+,\uparrow} & = H_{0,+,\uparrow} + \sum_{\bsl{R}_M} f^\dagger_{+,\bsl{R}_M,\uparrow} (-\frac{1}{2}U_1-6U_2-\frac{1}{2}U_1+\frac{1}{2}U_1 \tau_z) f_{+,\bsl{R}_M,\uparrow} + (-W_1) \sum_{\bsl{k}}^{\Lambda_c} c^\dagger_{+,\bsl{k},\Gamma_3,\uparrow} \tau_z c_{+,\bsl{k},\Gamma_3,\uparrow} \\
& \quad + \sum_{\bsl{k}}^{\Lambda_c} c^\dagger_{+,\bsl{k},\Gamma_1\Gamma_2,\uparrow} (-W_3 +\frac{J}{2}\tau_z) c_{+,\bsl{k},\Gamma_1\Gamma_2,\uparrow} + (-W_{fd}) \sum_{\bsl{p}}^{\Lambda_d} d^\dagger_{+,\bsl{p},\uparrow} d_{+,\bsl{p},\uparrow}\ ,
}
\eqa{
H^{\text{PVP}_1,OS}_{+,\downarrow} & = H_{0,+,\downarrow} + \sum_{\bsl{R}_M} f^\dagger_{+,\bsl{R}_M,\downarrow} (-\frac{1}{2}U_1-6U_2-\frac{1}{2}U_1+\frac{1}{2}U_1 \tau_z) f_{+,\bsl{R}_M,\downarrow} + (-W_1) \sum_{\bsl{k}}^{\Lambda_c} c^\dagger_{+,\bsl{k},\Gamma_3,\downarrow} \tau_z c_{+,\bsl{k},\Gamma_3,\downarrow}  \\
& \quad+ \sum_{\bsl{k}}^{\Lambda_c} c^\dagger_{+,\bsl{k},\Gamma_1\Gamma_2,\downarrow} (-W_3 +\frac{J}{2}\tau_z) c_{+,\bsl{k},\Gamma_1\Gamma_2,\downarrow} + (-W_{fd}) \sum_{\bsl{p}}^{\Lambda_d} d^\dagger_{+,\bsl{p},\downarrow} d_{+,\bsl{p},\downarrow}  \ ,
}
\eqa{
H^{\text{PVP}_1,OS}_{-,\uparrow} & = H_{0,-,\uparrow} + \sum_{\bsl{R}_M} f^\dagger_{-,\bsl{R}_M,\uparrow} (-\frac{1}{2}U_1-6U_2) f_{-,\bsl{R}_M,\uparrow} + (-W_1) \sum_{\bsl{k}}^{\Lambda_c} c^\dagger_{-,\bsl{k},\Gamma_3,\uparrow} \tau_z c_{-,\bsl{k},\Gamma_3,\uparrow}  \\
& \quad+ \sum_{\bsl{k}}^{\Lambda_c} c^\dagger_{-,\bsl{k},\Gamma_1\Gamma_2,\uparrow} (-W_3 +\frac{J}{2}) c_{-,\bsl{k},\Gamma_1\Gamma_2,\uparrow} + (-W_{fd}) \sum_{\bsl{p}}^{\Lambda_d} d^\dagger_{-,\bsl{p},\uparrow} d_{-,\bsl{p},\uparrow} \ ,
}
and
\eqa{
H^{\text{PVP}_1,OS}_{-,\downarrow} & = H_{0,-,\downarrow} + \sum_{\bsl{R}_M} f^\dagger_{-,\bsl{R}_M,\downarrow} (-\frac{1}{2}U_1-6U_2-\frac{1}{2}U_1-\frac{1}{2}U_1 \tau_z) f_{-,\bsl{R}_M,\downarrow} + (-W_1) \sum_{\bsl{k}}^{\Lambda_c} c^\dagger_{-,\bsl{k},\Gamma_3,\downarrow} \tau_z c_{-,\bsl{k},\Gamma_3,\downarrow}  \\
& \quad+ \sum_{\bsl{k}}^{\Lambda_c} c^\dagger_{-,\bsl{k},\Gamma_1\Gamma_2,\downarrow} (-W_3 -\frac{J}{2}\tau_z) c_{-,\bsl{k},\Gamma_1\Gamma_2,\downarrow} + (-W_{fd}) \sum_{\bsl{p}}^{\Lambda_d} d^\dagger_{-,\bsl{p},\downarrow} d_{-,\bsl{p},\downarrow} \ .
}

At $\nu=-2$, the one-shot Hartree-Fock Hamiltonains for different high-$\E$ initial states are related:
\eq{
H^{\text{VP}_1,OS}= H^{\text{VP}_1,OS}_{+,\uparrow}+H^{\text{VP}_1,OS}_{+,\downarrow}+H^{\text{VP}_1,OS}_{-,\uparrow}+H^{\text{VP}_1,OS}_{-,\downarrow}-E_0^{OS}\ ,
}
\eq{
H^{\text{VP}_2,OS}= C_2\overline{\TR} H^{\text{VP}_1,OS}_{+,\uparrow} (C_2\overline{\TR})^{-1} +H^{\text{VP}_1,OS}_{+,\downarrow}+H^{\text{VP}_1,OS}_{-,\uparrow}+H^{\text{VP}_1,OS}_{-,\downarrow}-E_0^{OS}\ ,
}
\eq{
H^{\text{VUP}_1,OS}= C_2\overline{\TR} H^{\text{VP}_1,OS}_{+,\uparrow} (C_2\overline{\TR})^{-1} + \overline{\TR} H^{\text{VP}_1,OS}_{-,\downarrow} \overline{\TR}^{-1} + \overline{\TR} H^{\text{VP}_1,OS}_{+,\uparrow} \overline{\TR}^{-1} +H^{\text{VP}_1,OS}_{-,\downarrow}-E_0^{OS}\ ,
}
\eq{
H^{\text{VUP}_2,OS}= C_2\overline{\TR} H^{\text{VP}_1,OS}_{+,\uparrow} (C_2\overline{\TR})^{-1} + \overline{\TR} H^{\text{VP}_1,OS}_{-,\downarrow} \overline{\TR}^{-1} +  H^{\text{VP}_1,OS}_{-,\uparrow}  + C_2 H^{\text{VP}_1,OS}_{+,\downarrow} C_2^{-1}-E_0^{OS}\ ,
}
where 
\eqa{
H^{\text{VP}_1,OS}_{+,\uparrow} & = H_{0,+,\uparrow} + \sum_{\bsl{R}_M} f^\dagger_{+,\bsl{R}_M,\uparrow} (-\frac{3}{2}U_1-12U_2-\frac{1}{2}U_1+\frac{1}{2}U_1 \tau_z) f_{+,\bsl{R}_M,\uparrow} + (-2W_1) \sum_{\bsl{k}}^{\Lambda_c} c^\dagger_{+,\bsl{k},\Gamma_3,\uparrow} \tau_z c_{+,\bsl{k},\Gamma_3,\uparrow} \\
& \quad + \sum_{\bsl{k}}^{\Lambda_c} c^\dagger_{+,\bsl{k},\Gamma_1\Gamma_2,\uparrow} (-2W_3 +\frac{J}{2}\tau_z) c_{+,\bsl{k},\Gamma_1\Gamma_2,\uparrow} + (-2W_{fd}) \sum_{\bsl{p}}^{\Lambda_d} d^\dagger_{+,\bsl{p},\uparrow} d_{+,\bsl{p},\uparrow}\ ,
}
\eqa{
H^{\text{VP}_1,OS}_{+,\downarrow} & = H_{0,+,\downarrow} + \sum_{\bsl{R}_M} f^\dagger_{+,\bsl{R}_M,\downarrow} (-\frac{3}{2}U_1-12U_2-\frac{1}{2}U_1-\frac{1}{2}U_1 \tau_z) f_{+,\bsl{R}_M,\downarrow} + (-2W_1) \sum_{\bsl{k}}^{\Lambda_c} c^\dagger_{+,\bsl{k},\Gamma_3,\downarrow} \tau_z c_{+,\bsl{k},\Gamma_3,\downarrow}  \\
& \quad+ \sum_{\bsl{k}}^{\Lambda_c} c^\dagger_{+,\bsl{k},\Gamma_1\Gamma_2,\downarrow} (-2W_3 -\frac{J}{2}\tau_z) c_{+,\bsl{k},\Gamma_1\Gamma_2,\downarrow} + (-2W_{fd}) \sum_{\bsl{p}}^{\Lambda_d} d^\dagger_{+,\bsl{p},\downarrow} d_{+,\bsl{p},\downarrow}  \ ,
}
\eqa{
H^{\text{VP}_1,OS}_{-,\uparrow} & = H_{0,-,\uparrow} + \sum_{\bsl{R}_M} f^\dagger_{-,\bsl{R}_M,\uparrow} (-\frac{3}{2}U_1-12U_2) f_{-,\bsl{R}_M,\uparrow} + (-2W_1) \sum_{\bsl{k}}^{\Lambda_c} c^\dagger_{-,\bsl{k},\Gamma_3,\uparrow} \tau_z c_{-,\bsl{k},\Gamma_3,\uparrow}  \\
& \quad+ \sum_{\bsl{k}}^{\Lambda_c} c^\dagger_{-,\bsl{k},\Gamma_1\Gamma_2,\uparrow} (-2W_3 +\frac{J}{2}) c_{-,\bsl{k},\Gamma_1\Gamma_2,\uparrow} + (-2W_{fd}) \sum_{\bsl{p}}^{\Lambda_d} d^\dagger_{-,\bsl{p},\uparrow} d_{-,\bsl{p},\uparrow} \ ,
}
and
\eqa{
H^{\text{VP}_1,OS}_{-,\downarrow} & = H_{0,-,\downarrow} + \sum_{\bsl{R}_M} f^\dagger_{-,\bsl{R}_M,\downarrow} (-\frac{3}{2}U_1-12U_2) f_{-,\bsl{R}_M,\downarrow} + (-2W_1) \sum_{\bsl{k}}^{\Lambda_c} c^\dagger_{-,\bsl{k},\Gamma_3,\downarrow} \tau_z c_{-,\bsl{k},\Gamma_3,\downarrow}  \\
& \quad+ \sum_{\bsl{k}}^{\Lambda_c} c^\dagger_{-,\bsl{k},\Gamma_1\Gamma_2,\downarrow} (-2W_3 +\frac{J}{2}) c_{-,\bsl{k},\Gamma_1\Gamma_2,\downarrow} + (-2W_{fd}) \sum_{\bsl{p}}^{\Lambda_d} d^\dagger_{-,\bsl{p},\downarrow} d_{-,\bsl{p},\downarrow} \ .
}

Combined with the fact that the dependence of $E_0^{OS}$ on the states is only through the filling $\nu$, all the listed high-$\E$ initial states with same $\nu$ have exactly the same Hartree-Fock energies.
The exact degeneracy will be broken in the self-consistent calculation.
It is because the density matrix obtained from the self-consistent calculation will have nonzero $O^{fc}$ (defined in \eqnref{eq:Omat}), while $O^{fc}=0$ for the initial states shown in \eqnref{eq:Omat_ini}.
To be concrete, let us consider the VH state and the Chern state for $\nu=0$, whose exact same energies at one-shot level require 
\eq{
 H^{\VH,OS}_{+,\uparrow}+H^{\VH,OS}_{+,\downarrow} =  H^{\Chern,OS}_{+,\uparrow}+H^{\Chern,OS}_{+,\downarrow}
}
according to \eqnref{eq:Chern_OS_nu0}, which further requires
\eq{
\Tr[O^{fc}_{\VH}\mat{ 0_{8\times 8} \\ \eta_z\tau_0 s_0}] = \Tr[O^{fc}_{\Chern}\mat{ 0_{8\times 8} \\ \eta_z\tau_0 s_0}]
}
according to \eqnref{eq:H_J}.
As $\Tr[O^{fc}_{\VH}\mat{ 0_{8\times 8} \\ \eta_z\tau_0 s_0}]$ is not necessarily equal to $\Tr[O^{fc}_{\Chern}\mat{ 0_{8\times 8} \\ \eta_z\tau_0 s_0}]$ beyond one-shot level, the energies of the VH state and the Chern state for $\nu=0$ are not necessarily the same in the self-consistent calculation.
However, the self-consistent calculation shows that the degeneracy breaking effect is very small.

\subsection{Treating $\E$ as a Perturbation at $\nu=0$}

\label{app:E_perturbatively}

In this part, we will treat $\E$ perturbatively at $\nu=0$ in order to analytically answer two questions:
(i) why there is a phase transition as we gradually increase $\E$? 
(ii) and what are the Chern numbers for the high-$\E$ ground states?

Let us discuss (i) first.
To answer this question, we again try to develop effective models at $\pm \KM$ and at $\Gamma_{\text{M}}$.
Since we consider $\E$ as a perturbation, the low-energy modes at $\pm \KM$ should be the $d$ modes, since the $f$ modes have energies $\pm \frac{1}{2} U_1$ at $\eta \KM$.
Then, based on \eqnref{eq:ht_HFOS_K} and \eqnref{eq:ht_HFOS_mK}, we can project $\E$ to the $d$ modes via the second order perturbation and get the following effective Hamiltonian at $\eta \KM$
\eqa{
\label{eq:eff_pmK_OS_small_E}
 \frac{8 M_1^2 \E^2}{U_1} (\zeta_\eta \zeta_\eta^\dagger-\frac{1}{2}\mathds{1}_4) - p_x \sigma_y s_0  + \eta  p_y  \sigma_x s_0  \ .
}
The effective energy at $\pm \KM$ is given by occupying the lowest 4 bands in each valley.
Then, by using \eqnref{eq:KIVC_nu0_ini} and \eqnref{eq:Ch_nu0_ini}, the effective energy at $\pm \KM$ reads
\eqa{
& E^{\text{K-IVC},\nu =0}_{\pm\KM} = -4 \sum_{\bsl{p}}^{\Lambda_d} |\bsl{p}| \text{ for the K-IVC state,}\\
& E^{\text{Ch},\nu =0}_{\pm\KM} = -4 \sum_{\bsl{p}}^{\Lambda_d} \sqrt{|\bsl{p}|^2+  \frac{16 M_1^4 \E^4}{U_1^2}} \text{ for the Chern state.}
}
We mention that our \eqnref{eq:eff_pmK_OS_small_E} is similar to  Eq.\,(29) in \refcite{Vishwanath20211122AMTG} , though \refcite{Vishwanath20211122AMTG} was only able to derive their Eq.\,(29) for Chern-diagonal states.

On the other hand, around $\Gamma_{\text{M}}$, the matrix form of the one-shot Hartree-Fock Hamiltonian in the basis $(f_{\bsl{p}}^\dagger, c_{\bsl{p},\Gamma_3}^\dagger, c_{\bsl{p},\Gamma_1\Gamma_2}^\dagger)$ reads
\eq{
h_{\Gamma_{\text{M}}}^{OS}(\bsl{k})=\mat{ 
\frac{1}{2}U_1  - U_1 \zeta \zeta^\dagger & \widetilde{\gamma} +  v_\star' (p_x \eta_z \tau_x \tau_0 + p_y \eta_0 \tau_y s_0) & \\
h.c. & 0 \mathds{1}_{8\times 8} &  v (k_x \eta_z \tau_0 s_0 + k_y \ii \eta_0 \tau_z s_0) \\
 & h.c. & \widetilde{M} \eta_0\tau_x s_0 - \frac{J}{2}\left[ \eta_z\tau_0 s_0 \zeta \zeta^\dagger \eta_z\tau_0 s_0 + \eta_0\tau_z s_0 \zeta \zeta^\dagger \eta_0\tau_z s_0 - \eta_0\tau_0 s_0\right]
}\ ,
}
where $f_{\bsl{k}}^\dagger = (...,f_{\eta,\bsl{k},\alpha,s}^\dagger,...)$, $c_{\bsl{p},\Gamma_3}^\dagger = (...,c_{\eta,\bsl{p},\beta,s}^\dagger,...)$ with $\beta =1,2$, $c_{\bsl{p},\Gamma_1\Gamma_2}^\dagger = (...,c_{\eta,\bsl{p},\beta,s}^\dagger,...)$ with $\beta =3,4$, $\widetilde{\gamma} = \gamma+B_\gamma \E^2$, $\widetilde{M} = M+B_M \E^2$, and we neglect $v_{\star}''$ and $B_{v''}\E^2$ since $v_{\star}''$ is small and we consider a perturbative $\E$.
Note that 
\eq{
U(\theta) h_{\Gamma_{\text{M}}}^{OS}(\bsl{k}) U^\dagger(\theta)= \left. h_{\Gamma_{\text{M}}}^{OS}(\bsl{k}) \right|_{\zeta\rightarrow U_{\bar{A}}(\theta)\zeta} \text{ when $v_\star' = 0$ }\ ,
}
where $U(\theta)$ is a chiral U(4) operation~\cite{Song20211110MATBGHF} with the form 
\eq{
U(\theta) = 
\mat{
U_{\bar{A}}(\theta) & & \\
 & U_{\bar{A}}(\theta) & \\
  & & U_{\bar{B}}(\theta)
} \text{ with } U_{\bar{A}}(\theta) = \exp\left[\ii \sum_{\mu\nu} \theta^{\mu\nu} \bar{A}_{\mu\nu} \right] \text{ and } U_{\bar{B}}(\theta) = \exp\left[\ii \sum_{\mu\nu} \theta^{\mu\nu} \bar{B}_{\mu\nu} \right]\ ,
}
and
\eqa{
& \bar{A}_{\mu\nu} = (\eta_0\tau_0 s_\nu,\eta_x\tau_x s_\nu,\eta_y\tau_x s_\nu,\eta_z\tau_0 s_\nu)_\mu \\
& \bar{B}_{\mu\nu} = (\eta_0\tau_0 s_\nu,-\eta_x\tau_x s_\nu,-\eta_y\tau_x s_\nu,\eta_z\tau_0 s_\nu)_\mu \ .
}
Since $e^{\ii \eta_y \tau_x s_0 \frac{\pi}{4}} \zeta_{\text{K-IVC},\nu = 0} =  \zeta_{\text{VH},\nu = 0}$ derived from the initial states in \appref{app:ini_states} and we know VH and Ch states have the same energies at the one-shot level, the energy difference between Ch and K-IVC states around $\Gamma_{\text{M}}$ relies on $v_\star'$.
Then, we focus on 
\eq{
\mat{ 
\frac{1}{2}U_1  - U_1 \zeta \zeta^\dagger & \widetilde{\gamma} +  v_\star' (p_x \eta_z \tau_x \tau_0 + p_y \eta_0 \tau_y s_0)  \\
h.c. & 0 \mathds{1}_{8\times 8}  
}\ .
}
Filling the lowest 8 bands would give the effective energy at $\Gamma_{\text{M}}$, resulting in 
\eqa{
& E^{\text{K-IVC},\nu =0}_{\Gamma_{\text{M}}} = -\sum_{\bsl{k}}^{\Lambda_c} \left[\sqrt{U_1^2 + 16 (|v'_{\star}\bsl{p}|-|\widetilde{\gamma}|)^2}+\sqrt{U_1^2 + 16 (|v'_{\star}\bsl{p}|+|\widetilde{\gamma}|)^2}\right] \text{ for the K-IVC state,}\\
& E^{\text{Ch},\nu =0}_{\Gamma_{\text{M}}} = -\sum_{\bsl{k}}^{\Lambda_c} \sum_{z=\pm}\left|\sqrt{U_1^2 + 16  |v'_{\star}\bsl{p}|^2 }+ z \sqrt{U_1^2 + 16  \widetilde{\gamma}^2 }\right| \text{ for the Chern state.}
}
Then, we know $E^{\text{K-IVC},\nu =0}_{\Gamma_{\text{M}}}\leq E^{\text{Ch},\nu =0}_{\Gamma_{\text{M}}}$, since 
\eq{
\sqrt{a^2 + (b-c)^2} + \sqrt{a^2 + (b+c)^2}\geq 2 \sqrt{a^2 + b^2}= \left|\sqrt{a^2 + b^2}-\sqrt{a^2 + c^2}\right| +\left|\sqrt{a^2 + b^2}+\sqrt{a^2 + c^2}\right|  \text{ for $a>0$ and $b\geq c\geq 0 $}
}
derived from
\eq{
\frac{\partial}{\partial c }\left( \sqrt{a^2 + (b-c)^2} + \sqrt{a^2 + (b+c)^2} \right) = \sqrt{1- \frac{a^2}{a^2 + (b+c)^2}} - \sqrt{1- \frac{a^2}{a^2 + (b-c)^2}} > 0\ \forall a>0, b\geq c>0\ .
}

In sum, the total effective energy is 
\eqa{
& E^{\text{K-IVC},\nu =0}_{eff} = E^{\text{K-IVC},\nu =0}_{\Gamma_{\text{M}}} + E^{\text{K-IVC},\nu =0}_{\pm\KM} \\
& E^{\text{Ch},\nu =0}_{eff} = E^{\text{Ch},\nu =0}_{\Gamma_{\text{M}}} + E^{\text{Ch},\nu =0}_{\pm\KM}\ .
}
Clearly, at $\E=0$, we have $E^{\text{K-IVC},\nu =0}_{eff}< E^{\text{Ch},\nu =0}_{eff} $ since $E^{\text{K-IVC},\nu =0}_{\Gamma_{\text{M}}}<E^{\text{Ch},\nu =0}_{\Gamma_{\text{M}}}$ and $E^{\text{K-IVC},\nu =0}_{{\pm\KM}}=E^{\text{Ch},\nu =0}_{{\pm\KM}}$.
Moreover, at $\E=\E_c$ ($\approx 294.816$meV in EUS) that satisfies $\gamma+B_\gamma \E_c^2=0$, we have $E^{\text{K-IVC},\nu =0}_{eff} > E^{\text{Ch},\nu =0}_{eff} $ since $E^{\text{K-IVC},\nu =0}_{\Gamma_{\text{M}}}=E^{\text{Ch},\nu =0}_{\Gamma_{\text{M}}}$ and $E^{\text{K-IVC},\nu =0}_{{\pm\KM}}>E^{\text{Ch},\nu =0}_{{\pm\KM}}$, demonstrating the existence of the transition (as increasing $\E$ from $\E=0$ to $\E=\E_c$).

Now we turn to the question (ii).
Before answering the question, let us first specify our convention for the Berry connection.
Given a isolated band with the cell-periodic part of its Bloch state being $\ket{u_{\bsl{k}}}$, the Berry connection is defined as 
\eq{
\bsl{A}(\bsl{k}) = -\ii \bra{u_{\bsl{k}}}\nabla_{\bsl{k}} \ket{u_{\bsl{k}}}\ .
}
The berry curvature is just the curl of the Berry connection.

With the convention specified, let us answer the question (ii).
As discussed in \appref{app:simple_rule_high_E}, $\left|\VH^{\nu = 0}_{0} \right\rangle$, $\left|\text{Chern}^{\nu = 0}_{0} \right\rangle$, $\left|\text{half-Chern}^{\nu = 0}_{0} \right\rangle$, $\left|\text{$C_2\TR$-invariant}^{\nu = 0}_{0} \right\rangle$ and their symmetry-related states are the $\nu=0$ states that are energetically favored at high $\E$.
If we plot the Hartree-Fock band structures of $\left|\VH^{\nu = 0}_{0} \right\rangle$, $\left|\text{Chern}^{\nu = 0}_{0} \right\rangle$, $\left|\text{half-Chern}^{\nu = 0}_{0} \right\rangle$ and $\left|\text{$C_2\TR$-invariant}^{\nu = 0}_{0} \right\rangle$, we find that their Hartree-Fock band structures remain gapped even for small $\E$.
Therefore, we are allowed to use determine the Chern numbers of $\left|\VH^{\nu = 0}_{0} \right\rangle$, $\left|\text{Chern}^{\nu = 0}_{0} \right\rangle$, $\left|\text{half-Chern}^{\nu = 0}_{0} \right\rangle$ and $\left|\text{$C_2\TR$-invariant}^{\nu = 0}_{0} \right\rangle$  at small $\E$
Since the one-shot Hartree-Fock Hamiltonians of $\left|\VH^{\nu = 0}_{0} \right\rangle$, $\left|\text{Chern}^{\nu = 0}_{0} \right\rangle$, $\left|\text{half-Chern}^{\nu = 0}_{0} \right\rangle$ and $\left|\text{$C_2\TR$-invariant}^{\nu = 0}_{0} \right\rangle$ are related with each other, let us consider the VH state first.
Recall that the one-shot Hartree-Fock Hamiltonian of VH state is spin-valley diagonal as shown in \eqnref{eq:VH_OS_nu0}.
Moreover, owing to the $U(2)\times U(2)$ symmetry and TR symmetry of the VH state, we have
\eqa{
 & H^{\VH,OS}_{+,\downarrow} = \left. H^{\VH,OS}_{+,\uparrow} \right|_{\text{flipping spin}} \\
 & H^{\VH,OS}_{-,s} = \overline{\TR} H^{\VH,OS}_{+, s} \overline{\TR}^{-1}\ ,
}
where $\overline{\TR}$ is the spinless TR operation.
Thus, we only need to study the Chern number of $H^{\VH,OS}_{+,\uparrow}$ at small $\E$.

Since we are considering the small $\E$, we can study the TBG part and the $d$ modes separately.
We first determine the Chern number of the TBG part of $H^{\VH,OS}_{+,\uparrow}$, following \refcite{Song20211110MATBGHF}.
The TBG part of $H^{\VH,OS}_{+,\uparrow}$ has the following matrix form
\eq{
\mat{ 
-\frac{1}{2}U_1 \tau_z & \gamma +  v_\star' (p_x  \tau_x  + p_y  \tau_y ) & \\
h.c. & 0 \mathds{1}_{2\times 2} &  v (k_x  \tau_0   + k_y \ii  \tau_z  ) \\
 & h.c. & \widetilde{M}  \tau_x  - \frac{J}{2}\tau_z 
}
}
in the basis $(f_{\bsl{p}}^\dagger, c_{\bsl{p},\Gamma_3}^\dagger, c_{\bsl{p},\Gamma_1\Gamma_2}^\dagger)$, where $f_{\bsl{k}}^\dagger = (...,f_{\eta,\bsl{k},\alpha,s}^\dagger,...)$, $c_{\bsl{p},\Gamma_3}^\dagger = (...,c_{\eta,\bsl{p},\beta,s}^\dagger,...)$ with $\beta =1,2$, and $c_{\bsl{p},\Gamma_1\Gamma_2}^\dagger = (...,c_{\eta,\bsl{p},\beta,s}^\dagger,...)$ with $\beta =3,4$.
By projecting the $f$ modes to $c$ modes via second order perturbation (which is allowed since $f$ modes have high energies ($\pm U_1/2$) and are topologically trivial), we get
\eq{
\mat{ 
 \frac{2}{U_1} \gamma^2 \tau_z &  v (k_x  \tau_0   + k_y \ii  \tau_z  ) \\
 h.c. & \widetilde{M}  \tau_x  - \frac{J}{2}\tau_z 
}\ ,
}
where we have neglected the $k^2$ term.
Since the gap stays open as we tune $(\frac{2}{U_1} \gamma^2 , -\frac{J}{2}, M)$ to $(m>0, -m,0)$ based on \tabref{tab:f-c-d_values}-\ref{tab:fcd_int}, we can determine its Chern number by 
\eq{
\mat{ 
 m \tau_z &  v (k_x  \tau_0   + k_y \ii  \tau_z  ) \\
 h.c. & -m\tau_z 
}\ ,
}
which gives $\ch=1 $ owing to $v>0$.

Now we turn to the $d$ modes.
Around $\KM$, $H^{\VH,OS}_{+,\uparrow}$ has the following matrix form
\eq{
\mat{ 
\frac{1}{2}U_1 \sigma_z & M_1 (\tau_0 + \ii \sigma_z) \E \\
M_1 (\tau_0 - \ii \sigma_z) \E  & p_x \sigma_x + p_y \sigma_y
}\ .
}
Again, by projecting the $f$ modes to $d$ modes via second order perturbation, we get
\eq{
\frac{4}{U_1}M_1\E^2 \sigma_z + p_x \sigma_x + p_y \sigma_y\ ,
}
giving $\ch = -1/2$.
Therefore, we have $\ch=1/2$ for $H^{\VH,OS}_{+,\uparrow}$.
Owing to the $U(2)\times U(2)$ symmetry and TR symmetry of the VH state, we have 
\eq{
\ch = \frac{1}{2}\eta \text{ for } H^{\VH,OS}_{\eta,s}\ .
}

The Chern number is not an integer for $H^{\VH,OS}_{\eta,s}$ because $H^{\VH,OS}_{\eta,s}$ is built from three Dirac cones of at $\eta$ valley.
If such up both valleys (and include the trivial high-energy completion), we should have well-defined interger Chern numbers.
By using the relations between different high-$\E$ states below \eqnref{eq:VH_OS_nu0}, we have 
\eqa{
& \ch = \frac{1}{2}+\frac{1}{2} - \frac{1}{2} - \frac{1}{2} =0 \text{ for } \left|\VH^{\nu = 0}_{0} \right\rangle \\ 
& \ch = \frac{1}{2}+\frac{1}{2} + \frac{1}{2} + \frac{1}{2} = 2 \text{ for } \left|\text{Chern}^{\nu = 0}_{0} \right\rangle\\ 
& \ch = \frac{1}{2}+\frac{1}{2} + \frac{1}{2} - \frac{1}{2} = 1 \text{ for } \left|\text{$C_2\TR$-invariant}^{\nu = 0}_{0} \right\rangle \\ & 
\ch = -\frac{1}{2}+\frac{1}{2} + \frac{1}{2} - \frac{1}{2} = 0\text{ for }\left|\text{$C_2\TR$-invariant}^{\nu = 0}_{0} \right\rangle \ .
}
If we include the symmetry related states, we have
\eqa{
& \ch =  0 \text{ for VH states} \\ 
& \ch =\pm 2 \text{ for Chern states}\\ 
& \ch = \pm 1 \text{ for half-Chern states} \\ 
& \ch = 0\text{ for  $C_2\TR$-invariant states} \ .
}

\section{Local TR-odd $C_2$-even $\U(2)\times \U(2)$-Invariant Perturbation on high-$\E$ states at $\nu=0$}
\label{app:stablize_Chern}

In this section, we present general symmetry arguments on how local TR-odd and $C_2$-even perturbations affect high-$\E$ states at $\nu=0$ to the leading order.
We also assume the local perturbation to preserve $\U(2)\times \U(2)$ symmetry.

For the VH states, if we keep the tensor-product nature of the states, we have the symmetry rep as
\eq{\ket{\VH}=(\ket{\VH,1},\ket{\VH,2})}
with 
\eqa{
& C_2\ket{\VH}=\ket{\VH} \sigma_x \\
& \TR\ket{\VH}=\ket{\VH}\ ,
}
leading to 
\eq{
\bra{\VH}H_\delta\ket{\VH}=0\ .
}

For the Chern states, if we keep the tensor-product nature of the states, we have the symmetry rep as
\eq{\ket{\Chern}=(\ket{\Chern,1},\ket{\Chern,2})}
with
\eqa{
& C_2\ket{\Chern}=\ket{\Chern}  \\
& \TR\ket{\Chern}=\ket{\Chern} \sigma_x\ ,
}
leading to
\eq{
\bra{\Chern}H_\delta\ket{\Chern}=b \sigma_z
}

For the half Chern states, if we keep the tensor-product nature of the states, we have the symmetry rep as
\eq{\ket{\text{half-Chern}}=(\ket{\text{half-Chern},1},\ket{\text{half-Chern},2},\ket{\text{half-Chern},3},\ket{\text{half-Chern},4})}
with
\eqa{
& C_2\ket{\text{half-Chern}}=\ket{\text{half-Chern}}\tau_x\sigma_0  \\
& \TR\ket{\text{half-Chern}}=\ket{\text{half-Chern}} \tau_0\sigma_x \ ,
}
leading to
\eq{
\label{eq:half_chern_perturbation}
\bra{\text{half-Chern}}H_\delta\ket{\text{half-Chern}}=b_{1} \tau_0\sigma_z + b_{2} \tau_x\sigma_z\ .
}

For the $C_2\TR$-invariant states, if we keep the tensor-product nature of the states, we have the symmetry rep as
\eq{
\ket{C_2\TR\text{-invariant}}=(\ket{C_2\TR\text{-invariant},1},\ket{C_2\TR\text{-invariant},2}
} 
with
\eqa{
& C_2\ket{C_2\TR\text{-invariant}}=\ket{C_2\TR\text{-invariant}}\sigma_x  \\
& \TR\ket{C_2\TR\text{-invariant}}=\ket{C_2\TR\text{-invariant}} \sigma_x \ ,
}
leading to
\eq{
\bra{C_2\TR\text{-invariant}}H_\delta\ket{C_2\TR\text{-invariant}}=0\ .
}

Here among the four types of states (\ie, VH, Chern, half-Chern and $C_2\TR$), we neglect the mixing between different types of states induced by $H_{\delta}$, since it is exponentially small due to the local nature of the perturbation $H_\delta$.
Moreover, the off-diagonal terms in \eqnref{eq:half_chern_perturbation} are exponentially small for the same reason.
As a result, we see that $H_{\delta}$ can shift the energy of certain Chern states by $-|b|$ energy, favoring it.

\end{widetext}
\bibliography{bibfile_references,alternative_ref}

\end{document}